\documentclass{jpp}
\usepackage{graphicx}

\usepackage{epstopdf, epsfig}
\usepackage{float}
\usepackage{mathrsfs}

 \usepackage{amssymb}
 \usepackage{bigints}
 \usepackage{amsmath}
 \usepackage{empheq}
 \usepackage{framed}
 \usepackage{color}
 \usepackage{enumerate}
 \usepackage{ulem}
 \usepackage{bm}
 \usepackage{appendix}
 \usepackage{multicol}
 \numberwithin{equation}{section}
 \usepackage{upgreek}
 \usepackage{textgreek}

\usepackage{stmaryrd} 
\usepackage{xfrac}    

\usepackage{hyperref}

\newcommand{\comment}[1]{}

\newcommand*{\cB}{\mathcal{B}}

\newcommand*{\cL}{\mathcal{L}}
\newcommand*{\cM}{\mathcal{M}}
\newcommand*{\cN}{\mathcal{N}}
\newcommand*{\cQ}{\mathcal{Q}}

\newcommand*{\cP}{\mathcal{P}}

\newcommand*{\cT}{\mathcal{T}}

\newcommand*{\scD}{\mathscr{D}}

\newcommand*{\scH}{\mathscr{H}}

\newcommand*{\scS}{\mathscr{S}}

\newcommand*{\ep}{\epsilon}

\newcommand*{\B}{\bm{B}}

\newcommand*{\J}{\bm{J}}

\newcommand*{\BDB}{\B\cdot \dl B}

\newcommand*{\BxDpsi}{\B\times \dl \psi}

\newcommand*{\dlBT}{\dl\B}

\newcommand*{\JxB}{\bm{J}\times \bm{B}}

\renewcommand*{\u}{\bm{u}}

\newcommand*{\dl}{\bm{\nabla}}
\newcommand*{\del}{\partial}
\newcommand*{\BD}{\bm{B}\cdot\bm{\nabla}}

\newcommand*{\dlr}{\bm{\nabla_\perp}}

\newcommand*{\bh}{\bm{\hat{b}}}
\newcommand*{\nh}{\bm{\hat{n}}}
\renewcommand*{\th}{\bm{\hat{t}}}

\newcommand*{\xh}{\bm{\hat{x}}}
\newcommand*{\yh}{\bm{\hat{y}}}
\newcommand*{\zh}{\bm{\hat{z}}}

\newcommand*{\rhoh}{\bm{\hat{\rho}}}
\newcommand*{\Rh}{\bm{\hat{R}}}

\newcommand*{\wh}{\bm{\hat{\omega}}}
\newcommand*{\zth}{\bm{\hat{\zeta}}}

\newcommand*{\lbr}{\left(}
\newcommand*{\rbr}{\right)}

\newcommand*{\zb}{\overline{z}}

\newcommand*{\Ib}{\overline{I}}

\newcommand*{\Rb}{\overline{R}}

\newcommand*{\etab}{\overline{\eta}}

\newcommand*{\psibar}{\overline{\psi}}

\newcommand*\reallywidetilde[1]{\ThisStyle{%
  \setbox0=\hbox{$\SavedStyle#1$}%
  \stackengine{-.1\LMpt}{$\SavedStyle#1$}{%
    \stretchto{\scaleto{\SavedStyle\mkern.2mu\sim}{.5467\wd0}}{.7\ht0}%
  }{O}{c}{F}{T}{S}%
}}

\def\test#1{$%
  \reallywidetilde{#1}\,
  \scriptstyle\reallywidetilde{#1}\,
  \scriptscriptstyle\reallywidetilde{#1}
$\par}
\makeatletter
\newsavebox{\@brx}
\newcommand{\llangle}[1][]{\savebox{\@brx}{\(\m@th{#1\langle}\)}%
  \mathopen{\copy\@brx\mkern2mu\kern-0.9\wd\@brx\usebox{\@brx}}}
\newcommand{\rrangle}[1][]{\savebox{\@brx}{\(\m@th{#1\rangle}\)}%
  \mathclose{\copy\@brx\mkern2mu\kern-0.9\wd\@brx\usebox{\@brx}}}
\makeatother

\newcommand{\normdet}{\ensuremath{\bar{\scD}_3}}

\shorttitle{Optical analogy for vacuum QA} 

\title{Optical analogy for stellarators: Ridges as caustics and coils as singularities}

\shortauthor{W. Sengupta et al.}
\author{Wrick Sengupta\aff{1}\corresp{\email{wsengupta@princeton.edu}}, Stefan Buller\aff{1}, Rogerio Jorge \aff{2}, John Kappel \aff{3}, Andrew Brown \aff{1}, Richard Nies \aff{4}, Pedro F. Gil\aff{5}, Nikita Nikulsin\aff{6}, Per Helander \aff{6}, Amitava Bhattacharjee\aff{1}}

 \affiliation{
 \aff{1} Department of Astrophysical Sciences, Princeton University, Princeton, NJ 08543, USA
 \aff{2} Department of Physics, University of Wisconsin-Madison, Madison, Wisconsin 53706, USA
 \aff{3} Institute for Research in Electronics and Applied Physics, University of Maryland, College Park, MD 20742,
USA
 \aff{4} Rudolf Peierls Centre for Theoretical Physics, Parks Road, Oxford, OX1 3PU, UK
 \aff{5}Max Planck Institute for Plasma Physics, 85748 Garching, Germany
 \aff{6} Stellarator Theory Division, Max Planck Institute for Plasma Physics, 17491 Greifswald, Germany
 }

\begin{document}

\maketitle

\begin{abstract}

A common feature of most numerically optimized stellarator geometries is the presence of sharp ridges on outer flux surfaces, irrespective of the rotational transform. Despite their importance, an analytical theory for their existence has been lacking. In this work, we demonstrate that ridges are not artifacts but mathematical necessities. We develop such a theory for devices with quasisymmetry (QS). We demonstrate that QS exhibits close connections with the theory of geometrical optics, following Parker's ``optical analogy" (E.N. Parker, \textit{Geophys. Astrophys. Fluid Dyn}, 1989). By mapping vacuum QS to the eikonal equation of geometrical optics, we derive the conditions for ridge formation, identified as field line caustics where magnetic field lines focus. Furthermore, we prove a geometric theorem for stellarator coil design: both ridges and filamentary coils must lie on the zero-determinant manifold of the magnetic gradient tensor. This topological constraint unifies the description of plasma ridges and external coils, providing a precise criterion for identifying valid coil locations and explaining the efficacy of the magnetic gradient lengthscale (J. Kappel et al.,  \textit{Plasma Phys. Control. Fusion}, 2024) as a coil optimization parameter. We demonstrate that as the device becomes more compact, sharp ridges naturally form on the inboard side in quasiaxisymmetry. We support our analytical theory with extensive numerical evidence. 
\end{abstract}

\section{Introduction \label{sec:intro}}

Modern optimized stellarators have reached a stage where their overall collisionless particle confinement, including fast particles, can rival that of tokamaks \citep{landremanPaul2021,goodman2023preciseQI}. In recent years, significant breakthroughs have been made in understanding concepts such as quasisymmetry (QS) and omnigeneity \citep{boozer2004_RMP,helander2014}, leading to substantial improvements in stellarator confinement. With the promise of low recirculating power, lack of current-driven instabilities and disruptions, and controllability of magnetic configurations through non-axisymmetric shaping, modern stellarators offer an encouraging path to a fusion power plant (FPP) \citep{boozer2008ITER2DEMO,boozer2021fast_path_fusion,goodman2024QI_PRXenergy,NASEM2019}. Some of the outstanding challenges to making a stellarator FPP a reality at this point include costs associated with typical large aspect ratios, designing 3D divertors and coils, and optimizing for stability and turbulence.

Stellarators typically have a much larger aspect ratio than tokamaks. For example, the tokamak-based DEMO reactor is designed with an aspect ratio of 2.6 and a major radius of 5.5 m \citep{tobita2009compactDEMO}. In contrast, the stellarator-based HELIAS concept can have an aspect ratio greater than 10 and a major radius of around 20 m \citep{beidler2001helias}. 
A compact stellarator is economically advantageous. However, it also introduces various technical challenges arising from its intrinsically three-dimensional (3D) geometry and tight aspect ratio \citep{hirshman1999_compact_stellarators,boozer2008ITER2DEMO,zarnstorff2001}. Compact stellarator-tokamak hybrids \citep{yamazaki1985tokastar,moroz1996sphericalStell,moroz1998helical_post_stell_1998}, as well as recent novel configurations \citep{henneberg2024compact,Schuett_henneberg2024compactQA}, offer attractive possibilities. 

A key component of a functional FPP will be a successful divertor design \citep{NASEM2021}. Divertors are crucial in fusion devices for efficient control of power, particle exhaust, and impurities. Compared to tokamak divertors, the three-dimensional (3D) geometry of stellarators offers a rich range of possibilities for separatrix structures. As a result, the stellarator divertor problem remains largely open \citep{bader2017_resilient_div_hsx}. Important concepts such as helical and island divertors have been proposed and experimentally demonstrated in the LHD \citep{ohyabu1994_LHD_div}. Similar approaches were tested in approximately quasi-isodynamic (QI) devices such as W7-AS and W7-X \citep{grigull2003_W7AS_island_divertor,pedersen2022_W7X_island_divertor}. However, the non-resonant divertor (NRD) concept \citep{bader2017_resilient_div_hsx,Baderetal2018,boozer_2015,punjabi2020simulation_nonresonant_div,garcia2024resilient,garcia2023explorationCTH}, which is promising for quasisymmetric stellarators with sizable bootstrap current, requires further exploration. 

In many numerically optimized stellarators, especially in small-aspect-ratio geometries, sharp ridges naturally arise in the outermost surface as it folds along certain curves \citep{bader2017_resilient_div_hsx,boozer_2015,nuhrenberg2006critical,1992Strumberger,strumberger1996sol}, creating regions of high principal curvature of the flux surface \citep{bader2017_resilient_div_hsx,Baderetal2018}. Sharply bending magnetic field lines requires significant energy. Thus, field lines typically follow ridges \citep{boozer_2015,nuhrenberg2006critical,bader2017_resilient_div_hsx}. The ridges are along the symmetry direction in tokamaks and stellarators with exact helical symmetry. The magnetic field lines along them are the so-called X-lines, with their rotational transform equal to zero or a rational number, and define a separatrix. The field lines outside of the separatrix can be diverted.
However, in stellarators, ridges do not necessarily need a resonant rotational transform and could thus be a key to designing NRDs.

The shape of the boundary also constrains the shape of the coils. 
In particular, near the inboard surfaces where the ridges are typically prominent in compact QA devices, the modular coils tend to crowd in regions between the sharp ridges and can develop sharp zig-zag patterns as illustrated in Figure 1 from \citep{kappel_Landreman2024magnetic}. Coils specifically designed with inboard shaping in mind, such as banana coils \citep{henneberg2024compact,Schuett_henneberg2024compactQA}, offer improvements over regular modular coils.

Ridges have also been linked to regions of high loss of energetic particles \citep{lion2023systems} and gaps in the Alfvén continuum \citep{paul2025shearAlfvenQS}. In particular, continuum optimization \citep{paul2025shearAlfvenQS} on the Wistell-A stellarator configuration \citep{bader2020WistellA} significantly reduces high-frequency gap widths by smoothing out these sharp ridge-like features. Ridges represent sharp gradients in the equilibrium magnetic surface geometry, thereby significantly impacting turbulent fluxes \citep{goodman2024QI_PRXenergy}. However, a thorough and precise characterization of ridge structure is currently lacking. This paper aims to take a first step to address this gap. 

In this work, we initiate a thorough analytical study of equilibrium 3D magnetic field lines near sharp ridges that can potentially serve as divertors for NRDs in compact quasisymmetric stellarators. In particular, we focus on developing a theory of the nonlinear interconnections among 3D flux-surface shaping, quasisymmetry (QS), force balance, and the coils that generate the fields. Unfortunately, analytical tools such as near-axis expansions (NAE) \citep{landreman2018a,landreman2018b,landreman2019,rodriguez_sengupta2023_constructing_QSspace_NAE,jorge2020construction} that have successfully built intuition for QS are unsuitable for studying ridges. There are several reasons. First, sharp ridges are localized far from the magnetic axis and a polynomial expansion in distance from the axis is not readily justified. Second, 3D magnetic ridges on irrational flux surfaces are not closed unlike 3D X points near magnetic islands. Thus, the near-axis expansion assuming a closed magnetic field line serving as an axis is not suitable. Third, as discussed in \citet{plunk2018,plunk2020_near_axisymmetry_MHD,henneberg2024compact,Schuett_henneberg2024compactQA,schuett2025optimization}, the solution space for QA stellarators with strong shaping is mostly restricted to outer surfaces that do not overlap with the NAE space \citep{landreman2022mapping}. This conclusion is also supported by a semi-analytical approach to compact QA stellarators \citep{nikulsin2025a}, which observes relatively smooth ridge-like structures on the inboard side. Compactness is crucial here, since near-axisymmetric but large-aspect-ratio stellarators \citep{sengupta2024QSHBS,nikulsin_sengupta2024GS,brown2025Palumbo} are essentially smoothly deformed tokamaks, where each poloidal section is rigidly shifted by a $\phi$-dependent displacement, with ridges that are weakly non-axisymmetric.

The point of departure of the current work is the formulation of the vacuum quasiaxisymmetry problem as an eikonal equation, a well-known equation in the theory of classical mechanics and geometrical optics \citep{goldstein2002,arnol2013classical_mechanics,courantHilbertvol2}. Our approach to understanding the interconnections of geometry and field strength via the eikonal equation is inspired by Parker's ``optical analogy" \citep{parker1989spontaneousI_optical_analogy,parker1989tangentialII,parker1981dissipation}, which observes that magnetic field lines and magnetic field strength are analogous to rays and refractive index of a medium. The sharp ridges then appear naturally as a result of focusing of the field lines (rays) along \textit{ray caustics} due to the lensing effect arising from the variation of the field strength (refractive index). The connection of ridges to caustics and focal surfaces is consistent with the textbook mathematical definition of ridges \citep{porteous2001geometric,koenderink1990solid}. 

Finally, the optical analogy highlights the fundamental role of the eigenvalues of the $\dlBT$ tensor in both ridges and coils. Here, we show that quasisymmetric ridges and every filamentary coil must both lie on a Det$\dlBT=0$ surface. The importance of the determinant in connection to magnetic nulls was pointed out recently by \cite{israeli_Smiet_2025topological}. The $L_{\nabla B}$ metric, previously used to estimate coil distance from flux surfaces \citep{kappel_Landreman2024magnetic}, can be derived from our formalism. Therefore, coil complexity and ridges can both be studied through the lens of the $\dlBT$ tensor.

In Section \ref{sec:ridges_caustics}, we introduce the concept of ridges, showcase the difference between ridges in different quasisymmetric configurations such as QA and QH, and provide the formal mathematical tools necessary to describe ridges. We then formulate QS in terms of the eikonal equation and establish the optical analogy in Section \ref{sec:OA_QS}. 
Our main results, stemming from the application of the optical analogy to understanding ridges, are presented in Section \ref{sec:Prop_near_ridges}. In Section \ref{sec:Exact_results} we prove two exact non-perturbative results that show that quasisymmetric ridges and coils must lie on zero-determinant manifold of the $\dlBT$ tensor. We further explore eigenvalues of the $\dlBT$ tensor and its connection with ridges, coils and the $L_{\nabla B}$ metric in Section \ref{sec:OA_LgradB}. Finally, we present numerical verifications of the key analytical insights in Section \ref{sec:Numerical} and summarize our results in Section \ref{sec:conclusions}.

\section{Geometry of flux surfaces, curvatures, ridges and coils}
\label{sec:ridges_caustics}

In this work, we focus on sharp ridges that arise naturally in most optimized stellarators and appear as creases or sharp flux surface corners. Due to their sharp features, they collimate field lines, which tend to align with them. Unlike X-points, they typically do not close on themselves.

Although ubiquitous in most 3D stellarator equilibria, especially when optimized for stability, a thorough study of these structures is complicated by several factors, including the nonlinearity of the underlying equilibrium equations, the enforcement of QS, proximity to current coils, and the 3D geometry of flux surfaces. In particular, there are no clear answers on what aspect of equilibrium physics, coils, geometry, or their combination, prompts these ridges to exist at certain locations in certain devices.

In this Section, we showcase a few typical examples to demonstrate how the geometry of a flux surface $\psi$, particularly the Gaussian curvature, can strongly influence the flux expansion $|\dl\psi|$, the locations of the ridges, and the deviations of coils from planarity. Later in the paper, we present analytical and numerical evidence to explain these observations.


\subsection{\textbf{Ridges in QA and QH}}
\label{sec:ridgeQA_QH}

\begin{figure}
\centering
\includegraphics[width=0.32\linewidth]{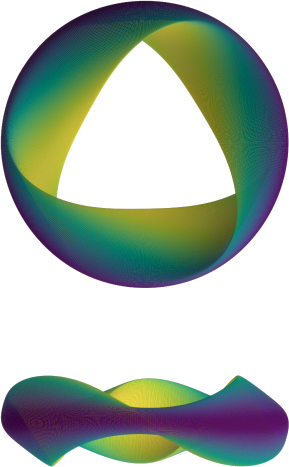}
\includegraphics[width=0.3\linewidth]{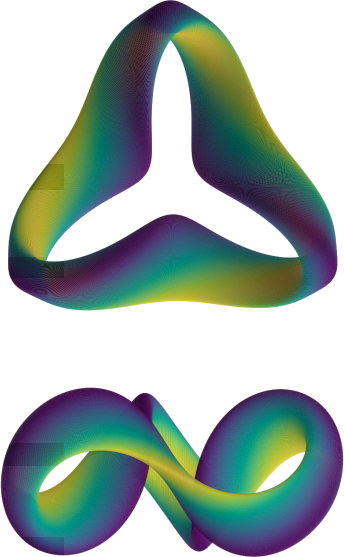}
\caption{Flux surface shapes of a compact QA (Left) and a compact QH (Right) device with three field periods, obtained using adjoint methods. Figures taken from  \citep{nies_exploration_2024}. Sharp ridges on the outermost surface exist for both of these configurations. The corresponding field strength contours in Boozer coordinates are shown in \autoref{fig:RIchard_typical_QA} and \autoref{fig:RIchard_typical_QH} display excellent quasisymmetry. }
\label{fig:RIchard_typical_3DQA_QH}
\end{figure}

\begin{figure}
\centering
\includegraphics[width=\linewidth]{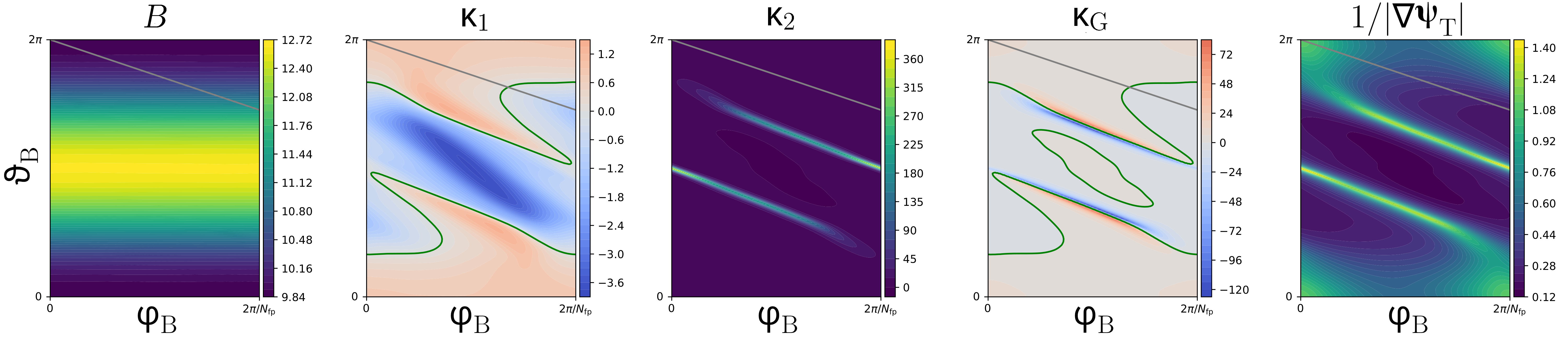}
\caption{Sharp ridges on the outermost surface of an QA device with three field periods and average rotational transform of 0.82, obtained using adjoint methods \citep{nies_exploration_2024}. From left to right, the panels show $B$, principal curvatures $\kappa_1,\kappa_2$, Gaussian curvature $K_G=\kappa_1\kappa_2$ and $1/|\dl\psi|$ employing Boozer coordinates $(\vartheta_B,\varphi_B)$. The color plots are shown on the actual device for better spatial visualization. On each plot, we show the field line with a thin black line starting from the top-left corner. We indicate the zeros of $\kappa_2,K_G$ using thin green lines.}
\label{fig:RIchard_typical_QA}

\centering
\includegraphics[width=\linewidth]{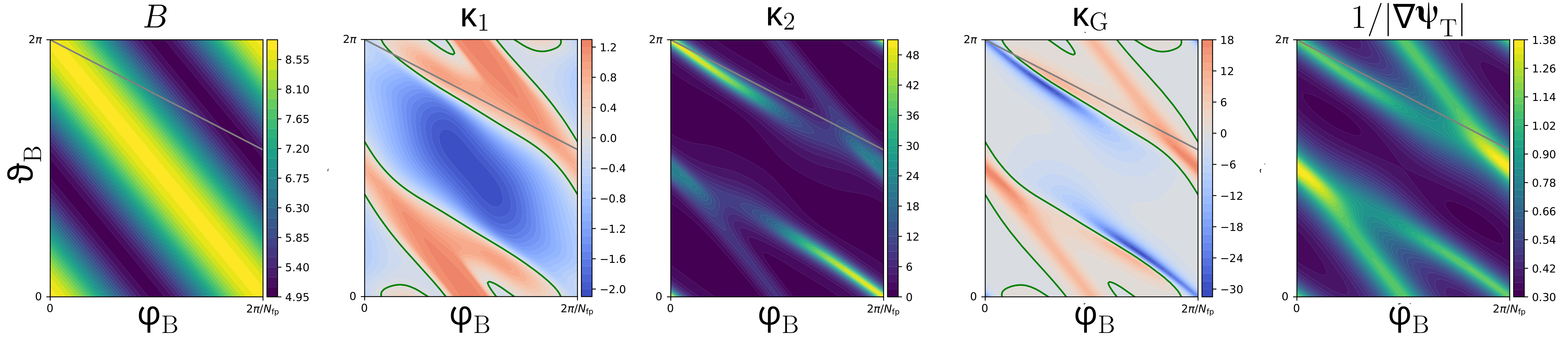}
\caption{A compact QH device ($n_{\text{fp}}=3$, $A=3.6$) obtained using Adjoint methods. See the caption for Figure \ref{fig:RIchard_typical_QA} for details. }
\label{fig:RIchard_typical_QH}
\end{figure}

The intuitive definition of ridges is that they are creases on surfaces associated with large normal curvature of the flux surface.  
We give the formal mathematical definition of ridges in Appendix \ref{app:ridge_caustics}, according to which ridges are a collection of points on a surface where a principal curvature is an extremum along the same principal direction.  

To develop physical intuition about the ridges in a quasisymmetric stellarator, we show in Figure \ref{fig:RIchard_typical_3DQA_QH} for a compact QA and a compact QH configuration with three field periods, obtained using adjoint methods \citep{nies_Paul2022adjointQS,nies_exploration_2024}. Figures \ref{fig:RIchard_typical_QA} and \ref{fig:RIchard_typical_QH} show the relevant physics and geometry aspects associated with ridges for these configurations. 

First, we focus on the commonalities of the QA and the QH ridges.
We use Boozer coordinates $(\vartheta_B,\varphi_B)$, which render both the field lines and contours of $B$ straight. The first panels from the left in the two figures show the excellent quality of QS, as evidenced by the straight $B$ contours in Boozer coordinates. The intuitive definition of sharp ridges correspond to large values of the principal curvature $\kappa_2$ in the third panel. They are almost straight because of the high alignment with the magnetic field lines indicated by thin gray lines in the panels. 

We note that the maximum of the principal curvature $\kappa_2$ also aligns quite well with the zero of the second principal curvature $\kappa_1$ indicated by the thin green line. Consequently, the Gaussian curvature $K_G = \kappa_1 \kappa_2$, shown in the fourth panel, has null points along the ridge. Also, $K_G$ attains large positive and even larger negative values immediately in the vicinity of the ridge. Since $K_G$ must average to zero on a toroidal surface, the asymmetry of the positive and the negative values requires further understanding. Further, the point with the deepest local minimum of the principal curvature $\kappa_1$ aligns very well with the maximum $B$ for both QA and QH. Finally, in the last panel, we show how the ridges also correspond to the minimum of $|\dl\psi|$. If the ridges were X-points, then $|\dl\psi|$ would have vanished identically.

Now we focus on the important differences in the QA and QH configurations. Firstly, in the second panel of the QH case, we find two pairs of ridge-like structures highlighted by the principal curvature $\kappa_1$. We see in the last panel of Figure \ref{fig:RIchard_typical_QH} that $1/|\dl\psi|$ shows a similar double ridge structure. The standard ridge is the sharper one that must be the one aligned with the field line, while the second, less prominent one is aligned with the quasihelical symmetry direction. If the ridges are not sharp enough, the distinction between the two kinds might not be obvious. A possible example of this could be HSX (see Figure 2 in \citep{bader2017_resilient_div_hsx}). Secondly, since the ridges align with the field lines near $B$ maxima, the ridges in QA are localized to the inboard side. In contrast, the QH ridges are not confined to the inboard and can rotate helically around the torus.

\subsection{Non-planarity of coils and curvature of outer flux-surface }
\label{sec:coils_Gaussian_K}

Coil complexity is a major cost driver for stellarators. Planar coils are easier to build, but to reproduce magnetic fields that lie on a complex 3D flux-surface shape in a stellarator using only modular coils, non-planarity cannot be avoided. The inboard side of the stellarators is typically where the coils get crowded \citep{kappel_Landreman2024magnetic}. Surface geometry plays a fundamental role in determining the non-planarity as shown in recent works \cite{rodriguezSengupta2026WindingCoil,warmer_pavone2026Nonplanarity}. In particular, field lines bunch in regions where the
field is stronger, which happens typically close to a ridge and spread out in between because the ridges are of finite extent on generic irrational flux surfaces. As a result, the current coils, which must be transversal to the field lines due to constraints from Biot-Savart, develop a clear S-like excursion.

\begin{figure}
\centering
\includegraphics[width=0.49\linewidth]{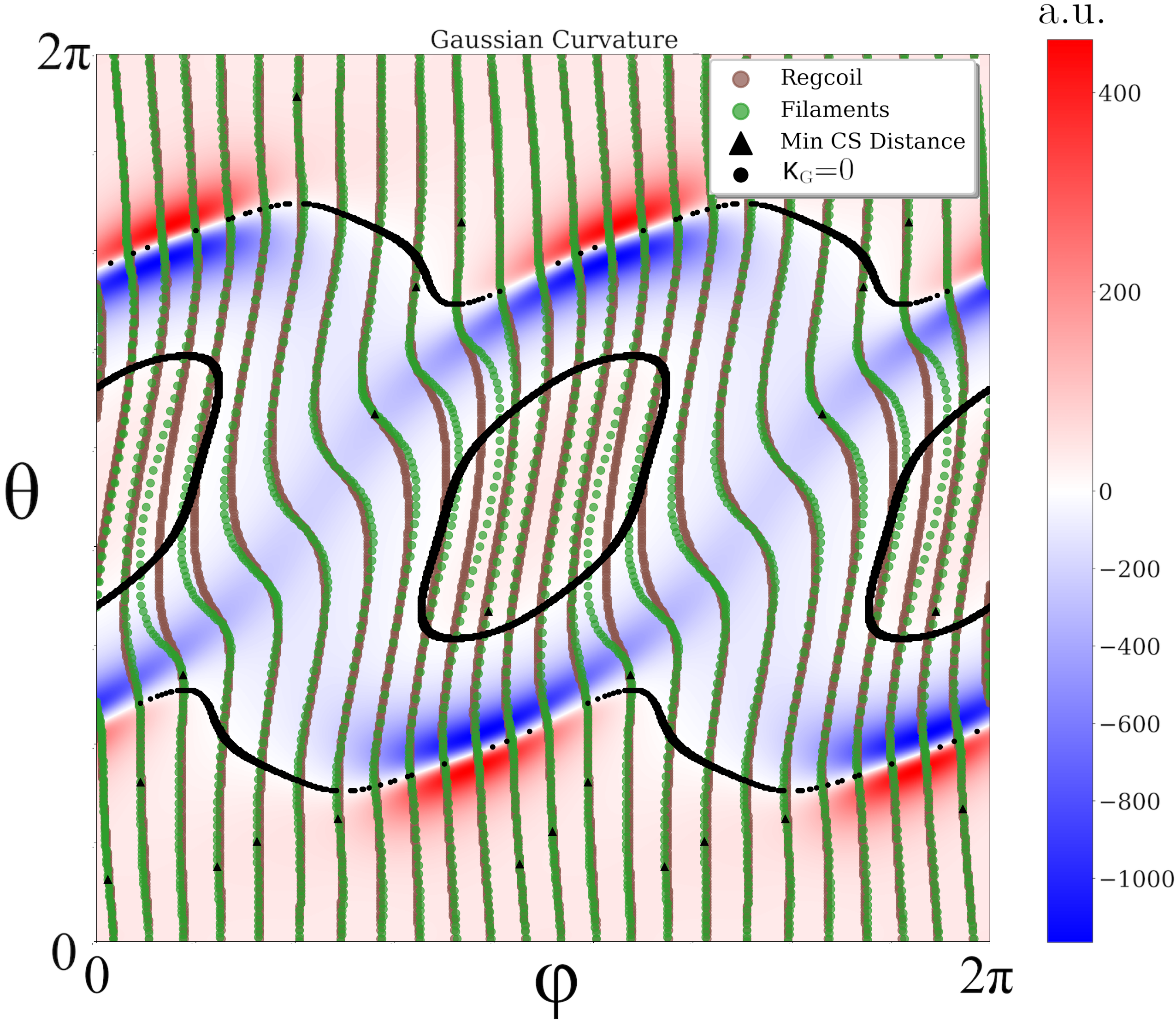}
\includegraphics[width=0.49\linewidth]{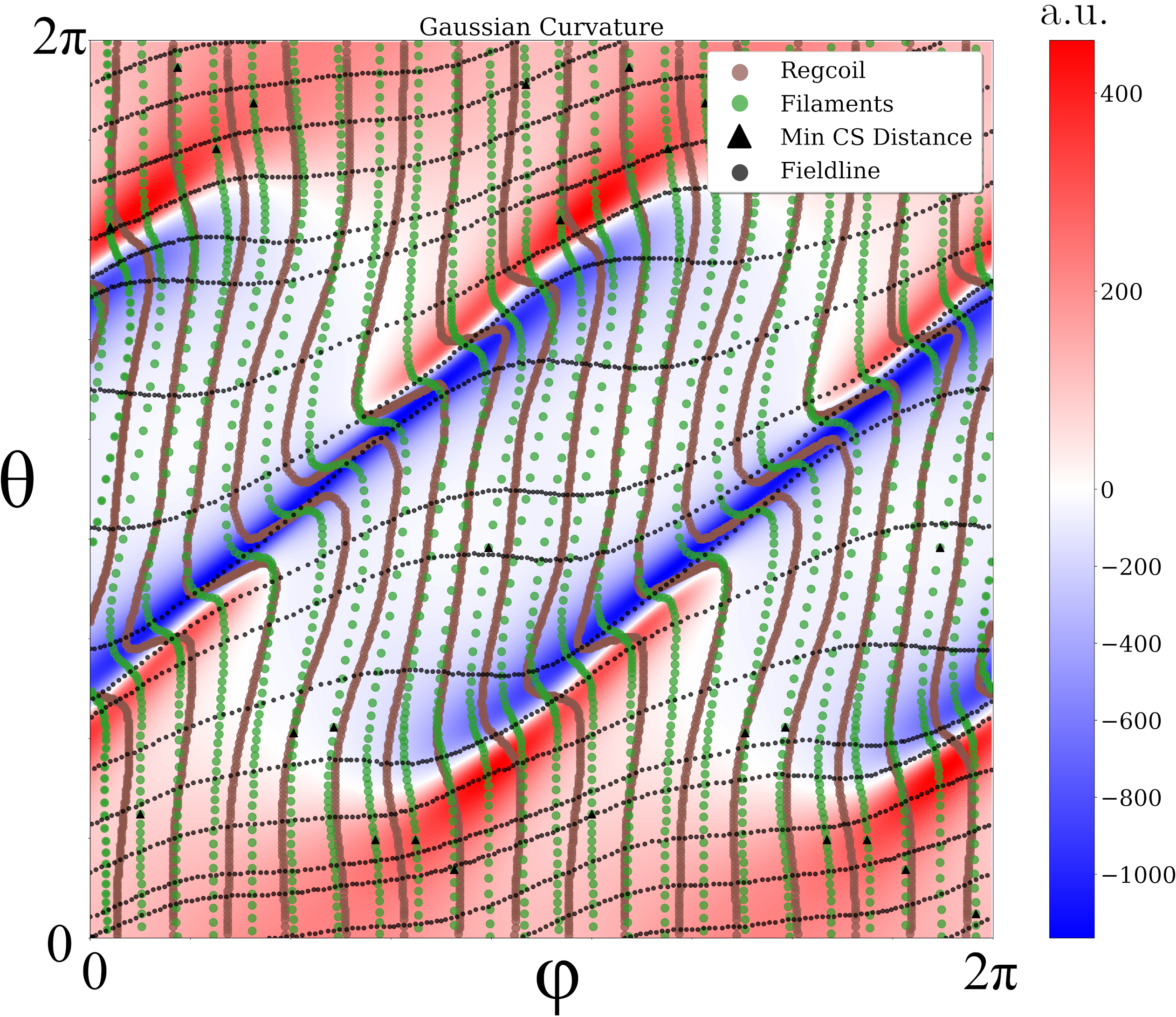}
\caption{Gaussian curvature ($K_G$) of the flux surface for two EPOS QA configurations \citep{Pedrogil2026designingEPOScoils} is shown in red-blue color with black dots denoting $K_G=0$ points. Filamentary coils projected onto the flux surface (in green dots) become non-planar and typically show a S-shaped curve in the $K_G\leq 0$ region on the inboard side. Compared to filamentary coils the REGCOIL generated coils (brown dots) when projected to the flux surface show a more pronounced non-planarity because of the restriction to lie on winding surfaces. The minimum coil-surface (CS) distance points are shown in black triangles. Field lines (black dotted lines) are seen to crowd near the ridges.}
\label{fig:Pedro_EPOS_coils_K}
\end{figure}

We show in Fig. \ref{fig:Pedro_EPOS_coils_K} this typical S-shaped non-planarity observed in filamentary coils and winding surface-based coils in the inboard side, where the Gaussian curvature $K_G\leq 0$. The non-planarity is worse in winding-surface-based coils because, unlike 3D filaments, the winding-surface coils cannot move radially.  In contrast, the $K_G>0$ regions have almost planar filaments. Thus, the $K_G=0$ curves, which are parabolic \citep{koenderink1990solid,porteous2001geometric}, seem to separate the almost planar and highly non-planar sections of filaments. Ridges align quite well with parabolic lines and are regions where the magnetic field lines are pinned down. 

In the rest of the paper, we focus on addressing the following questions. Why are ridges aligned with a) the magnetic field lines, b) the parabolic lines with $K_G=0$ c) the minimum $|\dl\psi|$, or equivalently maximum flux expansion regions, and finally d) what $K_G\leq 0$ implies for coil complexity and how it correlates with the Kappel-Landreman proxy $L_{\nabla B}$. Our analytical results are valid for both vacuum QA and QH, but we restrict ourselves to vacuum QA for the numerical analysis.

\section{Quasisymmetry and the optical analogy}\label{sec:OA_QS}

Equilibrium 3D magnetic fields, typically described by ideal magnetohydrostatics (MHS), with nested flux surfaces and nonzero rotational transform, are often imposed as a requirement for stellarator design. QS is imposed as a constraint on top of the ideal MHS equilibrium condition. Although currently no rigorous mathematical theory of 3D equilibrium magnetic fields with nested surfaces exists, imposing QS is known \citep{rodriguez2021islands,sengupta2024QSHBS} to ameliorate some of the difficulties of maintaining nestedness associated with ideal MHS.

We further restrict ourselves to vacuum fields. Besides simplicity, vacuum fields are highly relevant to stellarator optimization, since plasma pressure is typically a small fraction of the magnetic pressure. Plasma pressure and currents can always be added as perturbations in our treatment.

\subsection{\textbf{Eikonal formulation for QS}}
\label{sec:HJB}

We begin with curl-free or vacuum magnetic fields that are conveniently described using a magnetic scalar potential such that 
\begin{align}
\B=\dl \Phi, \quad \Phi=G_0 \varphi_B,
\label{eq:BisGradPhi}
\end{align}
Here, $G_0$ is a constant and $\varphi_B$ is the Boozer toroidal angle. Since $\B$ must be divergence-free,
\begin{align}
\nabla^2 \Phi=0. \label{eq:Laplace_eqn_Phi}
\end{align}

We then impose quasisymmetry (QS). QS in its various equivalent forms has been discussed in several references \citep{helander2014,boozer_2015,rodriguez2022_thesis,rodriguez2020a,burby2020,sengupta2023periodic}, therefore, we shall keep our discussion to a minimum. QS ensures integrability of guiding center trajectories. However, for this work, the key feature of a quasisymmetric vacuum field is that it must satisfy a nonlinear overdetermined system of PDEs.

Following \cite{burby2020,rodrigGBC}, we use the \textit{poloidal flux} $\psi$ to label the nested QA flux surfaces, which will be our convention for the rest of the paper. For QH, one can choose an analogous helical flux. From $\BD\psi=0$, we have the orthogonality condition
\begin{align}
\nabla \Phi \cdot \nabla \psi=0.
\label{eq:ortho_psiphi}
\end{align}

To make analytical progress while avoiding the nonlinear overdetermination problem of quasisymmetry, we now take a different approach. We now employ Boozer coordinates $(\psi,\vartheta_B,\varphi_B)$ with poloidal flux $\psi$, rotational transform $\iota(\psi)$, Boozer toroidal angle $\varphi_B$ and a poloidal or helical Boozer angle $\vartheta_B$ depending on QA or QH \citep{landreman2018a}. The covariant and the contravariant forms of $\B$ are then given by
\begin{align}
\B= \dl \Phi, \quad \B = \dl \psi \times \dl \alpha , \quad \Phi=G_0  \varphi_B, \quad \alpha \equiv \frac{1}{\iota}\vartheta_B-\varphi_B.
\label{eq:Bconcov_Boozer}
\end{align}
From the two-term form of QS in a vacuum with vanishing toroidal current,
\begin{align}
\B\times \dl \psi \cdot \dl B = G_0 \BDB \quad \implies \quad \dl \psi \times \dl \vartheta_B \cdot \dl B=0,
\label{eq:2term}
\end{align}
it follows that the hidden symmetry of QA is manifest in the Boozer coordinates with
\begin{align}
B=B(\vartheta_B,\psi).
\label{eq:QA_manifests}
\end{align}

Hence, a quasisymmetric vacuum field satisfies the following the eikonal equation \citep{Born_Wolf2013principles}
\begin{align}
|\dl\Phi|^2 = B^2(\vartheta_B,\psi).
\label{eq:HJB_QA}
\end{align}
Although the exact form of $B$ is not readily available, progress can be made due to the fact that in almost all numerically optimized designs with typical aspect ratios, the change of $B$ from the magnetic axis to the outermost surface is a small fraction ($\lessapprox 1/10 $) of its constant value on the axis. Thus, to build intuition, the exact knowledge of $B$ is not as crucial as the knowledge of the type of QS. Our exact results in Section \ref{sec:Exact_results} do not need these approximations. 

With the eikonal equation in hand, we are in a position to discuss the optical analogy, which will be our main tool to understand how $B$ and geometry interact and possible singularities of the vacuum QA system.

\subsection{\textbf{The Optical Analogy: Parker’s approach}}
\label{sec:OAnalogy}

The optical analogy is primarily motivated by E.N. Parker’s observation \citep{parker1981dissipation,parker1989tangentialII} that one can study vacuum fields and force-free fields in close analogy with optics. The analogous quantities are indicated in Table \ref{tab:OA}.

\begin{table}
  \begin{center}
\def~{\hphantom{0}}
  \begin{tabular}{c | c | c}
    Properties \qquad  &  Optics \qquad & Curl-free field \\
    \hline
     Arclength $\ell$ \qquad&   along rays \qquad   & along $\B$\\
     Refractive index \qquad & $n$ & $B$\\
     Wavefronts \qquad & $\scS$ & $\Phi$\\ 
     Rays \qquad & $\bm{k}=\dl\scS$\qquad & $\B=\dl\Phi$\\
     Eikonal equation \qquad & $|\dl\scS|^2= n^2$ & $|\dl\Phi|^2=B^2$\\
     Optical path $\phi(\bm{x};\bm{x}_0)$\quad & $\scS(\bm{x}_0)+\int_{\bm{x}_0}^{\bm{x}}\bm{k}\cdot d\bm{\ell} $&  \quad $\Phi(\bm{x}_0)+\int_{\bm{x}_0}^{\bm{x}}\bm{B}\cdot d\bm{\ell} $\\
     Ray equation \quad & $  \quad\bm{k}\cdot \dl(\bm{k}\cdot \dl \bm{x})=n\dl n \quad  $ &  \quad $\BD(\BD \bm{x})=B\dl B $\vspace{0.1in}
  \end{tabular}
  \caption{Analogy between optics and curl-free (vacuum) fields }
  \label{tab:OA}
  \end{center}
\end{table}

The essential idea is that $B$ is analogous to the refractive index $n$, while $\B$ is analogous to rays. Parker then showed that for force-free and curl-free fields, given $\B$, we can use Fermat’s principle of extremizing the optical pathlength given by $\int n d\ell\leftrightarrow \int \B\cdot \bm{d\ell}$ to obtain the fieldline trajectory. Here, $\ell$ is the arclength along the field line and $\bm{k}$ is a vector along the ray with magnitude $n$. Parker also showed that Fermat’s principle is consistent with the usual variational principle, which involves minimizing the volumetric magnetic field energy $\int d^3\bm{x} B^2$. We provide mathematical justifications of Parker’s observations in the following, focusing only on vacuum fields.

For vacuum fields with $\BD \bm{x}=\del \Phi/\del \mathbf{x}$, or equivalently in index notation $B_i = \del \Phi/\del x^i, i=(1,2,3)$, where $x^i$ are Cartesian coordinates, the field lines are obtained by solving the ODEs
\begin{align}
B \frac{dx^i}{d\ell}=\frac{\del\Phi}{\del x^i}.
\label{eq:fieldline_ODE}
\end{align}
Operating both sides of \eqref{eq:fieldline_ODE} with the $\BD$ operator, we get
\begin{align}
B \frac{d}{d\ell}\lbr B \frac{d\bm{x}(\ell)}{d\ell}\rbr=\BD\B = \dl\lbr \frac{ B^2}{2}\rbr,
\label{eq:ray_eqn}
\end{align}
where we have used the vector identity $\BD\B= \mu_0\J\times\B+\dl B^2/2$, with vacuum permeability $\mu_0$ and $\J=0$ for vacuum. Clearly, this works for force-free fields as well since $\JxB=0$ in that case.

In geometrical optics, \eqref{eq:ray_eqn} is known as the ray equation \citep{Born_Wolf2013principles}. A change of variables results in the following useful alternative form of the ray equation
\begin{align}
\frac{d^2\bm{x}}{d\cT^2} = \dl\lbr \frac{ B^2}{2}\rbr, \quad \cT= \int\frac{d\ell}{B}, \quad \BD =\frac{d}{d\cT}.
\label{eq:ray_eqn_alt}
\end{align}
The close resemblance of \eqref{eq:ray_eqn_alt} to Newton’s second law of motion with $\cT$ playing the role of time and $-B^2/2$ the potential, is not coincidental since optics and mechanics have similar mathematical structure \citep{luneburg1964mathematical_optics,synge1937geometrical,goldstein2002,arnol2013classical_mechanics}. Leveraging the classical mechanics connection, we now present a canonical Hamiltonian form for the rays \citep{kravtsov2012caustics}
\begin{align}
\frac{d\bm{x}}{d\cT}= \bm{p} , \quad \frac{d\bm{p}}{d\cT}= \dl\lbr \frac{ B^2}{2}\rbr.
\label{eq:canonical_Hamiltonian_Krav}
\end{align}
The momentum $\bm{p}=\dl\Phi$ is simply the magnetic field $\B$.

Parker used the optical analogy to understand the formation of singular current sheets in space and astrophysical plasmas. It is the nonlinear nature of the eikonal equation that naturally leads to the focusing of field lines in certain spatial locations, leading to a lack of smoothness in the solution. The classical theory of the eikonal equation and its singular solutions is to be found in the first two chapters of \citep{courantHilbertvol2} and \citep{garabedianPDE}. 
In Section \ref{sec:Prop_near_ridges} and \ref{sec:Exact_results}, we connect these ideas to ridges and coils.

\section{Equilibrium properties near magnetic ridges \label{sec:Prop_near_ridges}}
The behavior of magnetic field lines and equilibrium quantities of interest such as the field strength and $|\dl\psi|$ near fixed points such as the standard O-X points generated by a closed magnetic field line has been extensively discussed in classic references such as \citep{Mercier1964,Solovev1970} and recent works on NAEs \citep{landreman2018a,jorge_sengupta2020near,sengupta2024NAE_finite_beta,rodriguez2021islands,rodriguez2022_thesis,guinchard2025application,burby2021_NAE_normal_forms}. Sharp magnetic ridges share several properties with X points such as a field line threading the ridge and $|\dl\psi|$ being small as discussed in Section \ref{sec:ridgeQA_QH}. However, almost no results currently exist that describe the behavior of magnetic field lines near generic 3D magnetic ridges. We now show how the optical analogy can be utilized to fill this gap. 

In Section \ref{sec:ridgeQA_QH}, we pointed out several distinctive features of vacuum QA configurations near ridges. Building on the optical analogy, we now demonstrate how the magnetic fields near the maximum of $B$ can lead to the formation of ridges in regions where the Gaussian curvature is non-positive. We also show why  $|\dl\psi|\to 0$ near ridges. In Section \ref{sec:OA_QS_curvature}, we trace the failure of QA near ridges to field line curvature tending to zero on the ridges.

\subsection{\textbf{Magnetic field lines and field strength near magnetic ridges} \label{sec:BnB_ridges}}

We now provide a simple proof that the magnetic field lines must be tangential to the magnetic ridges. Let the ridge (which acts locally as the boundary of the plasma) be defined by two smooth surfaces, $\cP$ and $\cQ$, which face the plasma on one side and intersect each other along a curve $\bm{R}$, the ridge. The angle between the normals of $\cP$ and $\cQ$ is usually less than 180 degrees on the plasma-facing side. Since $\B$ is tangent to $\cP$ and $\cQ$, its component normal to these surfaces vanishes. As a result, $\B$ vanishes in two different directions orthogonal to $\bm{R}$. Unless $\B$ also vanishes in the direction of the ridge, the latter coincides with a field line.

In other words, the field cannot cross $\bm{R}$ unless its component parallel to $\bm{R}$ vanishes. On the other hand, if the parallel component vanishes, then the field can cross $\bm{R}$. An example of such behavior is a tokamak with an X-point but no toroidal field.

It also follows from continuity alone that the gradient of any flux function $\psi$ vanishes at a sharp ridge $\bm{R}$. The latter can again be thought of as an intersection between two surfaces, $\cP$ and $\cQ$. The vector $\dl \psi$ is orthogonal to both surfaces and therefore points in different directions on $\cP$ and $\cQ$. It can only be continuous at $\bm{R}$ by vanishing there.

A similar continuity based argument can be used to prove that a ridge in a field with QS is straight if it is not aligned with the symmetry direction. Starting from the two-term form of QS $|\dl \psi|\B\times \nh \cdot \dl B=G_0 \BDB $, where $\nh$ is a unit vector normal to the flux surface, we observe that $\dl \psi$ vanishing on the ridge forces $\BDB$ to vanish since $G_0$ is nonzero. Thus, QS implies that $B=|\B|$ is  constant in two different directions: along the field and in the symmetry direction. At the ridge, $\dlBT$ thus vanishes in any direction tangential to $\cP$ and $\cQ$, i.e. in three independent directions, which implies that $\dlBT=0$ on $\bm{R}$. Since the curvature vector of the field line following $\bm{R}$ is equal to $\dlr(\ln |\B|)$, it must be straight.

These results hold for sharp ridges of non-zero length. If $\bm{R}$ is only sharp in one point, then its curvature vanishes in that point, making $\bm{R}$ locally straight. The differential geometric ramifications of these results are discussed in Appendix \ref{app:Diff_geo_proof}.

\subsection{\textbf{Flux surface geometry near ridges \label{sec:geom_ridges}}}

We have argued that continuity implies that a quasisymmetric $B$ should be approximately constant $B_0$ near a sharp ridge, which implies that the ridge must be locally straight. We elaborate in this Section the far-reaching impact on the behavior of local field lines and flux surfaces due to this simple observation. 

Employing the standard Cartesian coordinates $(X,Y,Z)$ with unit vectors $(\xh,\yh,\zh)$, we now explore the implications of the eikonal equation with a constant $B$ on the geometry. Let $\Phi=B_0 \varphi$ be the scalar potential such that $\B/B_0=\nabla \varphi$. The eikonal equation reads
\begin{align}
\varphi_{,X}^2+\varphi_{,Y}^2+\varphi_{,Z}^2=1.
\label{eq:eikonal_B0}
\end{align}
For constant $B$, the ray equation \eqref{eq:canonical_Hamiltonian_Krav} implies that the rays are the constant momentum vector $\bm{p}\equiv \nabla \varphi$, which can be chosen to be
\begin{align}
\bm{p}=p_0 \xh +q_0 \yh \pm \sqrt{1-p_0^2-q_0^2}\zh.
\label{eq:ray_lines}
\end{align}
Here, $(p_0,q_0)$ are two constant parameters that serve as initial conditions for the ray tracing. The wavefront $\varphi$ is given by the 3 parameter $(\varphi_0,p_0,q_0)$ family of planes $\varphi(X,Y,Z)=u(X,Y,Z;p_0,q_0,\varphi_0)$ with
\begin{align}
u(X,Y,Z;p_0,q_0,\varphi_0)=\varphi_0+ X p_0  + Y q_0  \pm Z \sqrt{1-p_0^2-q_0^2}.
\label{eq:complete_int_B0}
\end{align}
In the simplest form, as discussed in \citep{courantHilbertvol2}, constant $B$ implies that in the optical analogy picture, rays \eqref{eq:ray_lines} are straight lines and the elementary wavefronts \eqref{eq:complete_int_B0} of the complete integral are planes. We then utilize the concept of a \textit{complete integral} \citep{courantHilbertvol2,garabedian1996,forsyth_v5_pt4} and \textit{general solution} through formation of envelopes. Treating \eqref{eq:complete_int_B0} as a complete integral that depends on three arbitrary constant parameters $(p_0,q_0,\varphi_0)$, we can obtain a general solution as envelopes of these elementary planes such that the rays are still straight lines, but the wavefronts themselves need not be planar, for example, they could be locally spherical.

To obtain less trivial solutions of the eikonal equation \eqref{eq:HJB_QA}, we now allow small $\ep$ deviations from a constant $B$. The constants $(p_0,q_0,\varphi_0)$ appearing in the elementary solution may then be promoted to functions of two ray labels $(\psi,\alpha)$, chosen to be constant along the rays. We identify these labels with Clebsch variables, so that $\B=\nabla\psi\times\nabla\alpha$ is tangent to the ray/field-line direction. We then obtain the envelope by eliminating the three quantities $p_0,q_0,\varphi_0$ from the three simultaneous equations
\begin{align}
\varphi(X,Y,Z)=u(X,Y,Z;p_0(\psi,\alpha),q_0(\psi,\alpha), \varphi_0(\psi,\alpha)), \quad \del_\psi u=0, \quad \del_\alpha u=0
\label{eq:envelopes_psi_alpha_B0}
\end{align}
The envelope equations, in principle, allow the determination of $\psi,\alpha, \varphi$ as functions of only $(X,Y,Z)$ through the standard methods of ray tracing. For $\ep\neq 0$, an explicit expression for the complete integral exists only for a handful of special cases, but nonetheless, the envelope idea persists, and can be constructed perturbatively if needed.

The rays, which are tangential to the flux surfaces, are all straight lines as $\ep\to0$ given by \eqref{eq:ray_lines}. A stringent geometrical constraint on the flux surfaces arises from this fact alone. From differential geometry (e.g., Chapter IV from \cite{hilbertCohnVossen}), we know that any straight line on a surface must be an asymptotic curve since the normal curvature vanishes identically. Asymptotic curves can exist only on surfaces with non-positive $K_G$. A surface containing a one-parameter family of straight line segments is a \textit{ruled surface}, which has $K_G\leq 0$. These line segments are asymptotic curves and acts as local rulings wherever the surface is regular. The important special case with zero $K_G$ everywhere is called \textit{developable} since it can be developed into a plane. Since a torus must have both positive and negative $K_G$, the straight field lines must be restricted to the hyperbolic $K_G\leq 0$ part of the torus, which for QA is the inboard side. For $\ep\neq 0$, the field lines remain approximately asymptotic curves even though they are not strictly straight lines anymore. We discuss this more in Section \ref{sec:OA_QS_curvature}.

Developable ruled surfaces may possess singular space curves where neighboring rulings approach each other, such as \textit{edges of regression} or \textit{cuspidal edges}. Near such curves, the surface normal curvature can be arbitrarily large. Although a generic non-developable ruled surface does not possess such an edge of regression, it has a related regular curve, the \textit{line of striction}, defined as the locus of closest approach of neighboring rulings. In a singular developable limit, the line of striction approaches the edge of regression. 

These sharp edges can be obtained from the complete integral as envelopes of the rays and are aptly called \textit{singular solutions}. We will show later that these singular solutions correspond to the sharp ridges on the flux surfaces. Thus, near a sharp ridge where the hyperbolic side of the flux surface is well approximated by a ruled surface, the ridge may be interpreted as a limiting striction/regression curve near \(K_G\to0\).

Choosing $\psi(X,Y,Z), \alpha(X,Y,Z)$ as flux surface and field line labels, we can obtain the singular solutions as envelopes of $\alpha$ curves on a fixed flux surface by solving
\begin{align}
\varphi(X,Y,Z)=u, \quad \del_\psi u=0, \quad \del_\alpha u=0, \quad \del_\alpha^2 u=0.
\label{eq:singular_sol_B0}
\end{align}
The singular solution \eqref{eq:singular_sol_B0} does not necessarily exist for all geometries since $\del_\alpha^2u$ might not vanish anywhere on the surface. However, if a solution exists, it represents a space-curve along which different field line labels $\alpha$ touch each other, forming an envelope. Clearly, $|\dl\alpha|\to \infty$ near the envelope. Since $\B=\dl\psi\times \dl \alpha$ must be finite, and $\nabla \psi$ and $\nabla \alpha$ are not parallel, $|\dl\psi|\to 0$ near the singular solutions. In optics, such singular curves are identified as the envelope of rays forming ray caustics. Since magnetic field lines can not actually touch each other, the singular solution, if it exists, represents a limiting curve, beyond which smooth solutions can not exist.  

\subsection{\textbf{Field line curvature near ridges \label{sec:OA_QS_curvature}}}

In the last Section, we showed that an important consequence of the optical analogy is the direct correlation between the field line curvature, gradients of $B$, and the Gaussian curvature of the flux surface. In QS, one can show that near the magnetic axis, where $B$ is essentially a constant, the vanishing of the magnetic axis curvature signals the breakdown of QS \citep{landreman2018a,landreman2018b,rodriguez_sengupta2023_constructing_QSspace_NAE}. Thus, the behavior of the critical points of $B$ and field line curvature deserves special attention in the study of singularities of the vacuum QA system. In the following, we discuss the constraints on field line curvature imposed by QS.

The field line curvature $\bm{\kappa}=\th\cdot \dl\th,\; \th\equiv \B/B,$  is determined completely by the gradients of $B$ since for vacuum fields \citep{helander2014}
\begin{align}
\bm{\kappa}=\frac{\dl B}{B}-\frac{\th}{B^2}\BD B.
\end{align}
By definition, $\bm{\kappa}$ has no component along $\th$. The other two components, $\kappa_n,\kappa_g$ along $\nh=\dl \psi/|\dl\psi|$ and $\bh\equiv\th\times \nh$, called the normal and the geodesic components, are given by \citep{helander2014}
\begin{align}
\bm{\kappa}= \kappa_n \nh+\kappa_g \bh, \quad \kappa_n =\frac{\dl\psi\cdot \dl B}{B|\dl\psi|}, \quad \kappa_g = \frac{\BxDpsi\cdot \dl B}{B^2 |\dl\psi|}.
\label{eq:kappa_ng_exp}
\end{align}
Using the two-term form of vacuum QS \eqref{eq:2term}, we can rewrite the geodesic curvature, which is proportional to $\B\times \dl \psi \cdot \dl B$ in terms of $\BDB$ such that
\begin{align}
\kappa_g = \frac{G_0\BDB}{B^2|\nabla \psi|}.
\label{eq:kappa_g_exp}
\end{align}
Since $|\nabla \psi|$ becomes small as we approach a sharp magnetic ridge, continuity forces $\BDB$ to be small. Thus, $B$ is close to a critical point of $B$, either a maxima $B_M(\psi)$ or minima $B_m(\psi)$ or a saddle of $B$. Away from the sharp ridge but in its immediate neighborhood where $|\nabla \psi|$ is no longer zero, continuity implies that $\kappa_g$ is still zero due to the numerator. The field lines are then locally geodesic. The local geodesic nature is consistent with the variational form or Fermat’s principle for the optical analogy, which shows that rays will minimize distances along their path. Configurations where all field lines are geodesics are either isodynamic with $B=B(\psi) $\citep{palumbo1968,bishop_Taylor1986degenerate} or QP with $B=B(\varphi_B,\psi)$ \citep{spong2001compact_drift,madan2026geometry}. For QS, the critical field lines and the field lines neighboring them are approximately isodynamic. However, if the $|\nabla \psi |$ term in the denominator simultaneously vanishes, the geodesic curvature can still be nonzero near critical points of $B$ and a more careful limit must be taken.

Even if a critical field line has zero geodesic curvature, it is not necessarily a straight line since the normal curvature of the field line at that point is, in general, nonzero, and can only vanish under further stringent restrictions imposed on $B$. It can be shown \citep{rodriguez2024maxJ} that for vacuum fields $B'_M(\psi)\geq 0$, and $B$ can only attain its maximum on the boundary. For QS, the minima and maxima are typically symmetric, hence, $B_m'(\psi)\leq 0$. Therefore, for generic QS systems, the radial derivative $\del_{\psi} B$, and hence the normal curvature $\kappa_n$, do not vanish, even if the geodesic curvature does near the critical points of $B$.

In the immediate vicinity of the sharp ridge, where $|\dl\psi|$ is not exactly zero, which is also the case with all numerically optimized compact QA configurations, we must have $B_M'(\psi)\to 0$ to have $\kappa_n\to 0$. The vanishing of $B_M'$ is in sharp contrast to the near-axis behavior where $B_M(\psi)/B_0=1+\etab \sqrt{2\psi} +O(\psi)$, where $\etab$ is a positive constant with $\etab R_a\sim 1$. Near the magnetic axis, where $|\dl\psi|\sim \sqrt{2\psi}$,  $\kappa_n \sim |\dl\psi|B_M'(\psi)/B_0\sim \etab>0$ and $\kappa_n R_a\sim 1$. Although this behavior is expected near the axis, it holds quite well in the full volume \citep{rodriguez2022_thesis,sengupta2023periodic} for 2-field period QA configurations such as the Landreman-Paul precise QA, which lacks a sharp 3D ridge. Thus, for sharp ridges to exist, $B_M$ must deviate significantly from the near-axis prediction.

Indeed from Figure \ref{fig:deviation_from_NAE} we find that all configurations with $n_{fp}>2$ show significant deviation from the near-axis behavior. On the inboard side with $B\approx B_M$ we find $B_M'(\psi)\to 0$. Thus, field lines must be approximately asymptotic curves for which $\kappa_n \to 0$. The fact that asymptotic lines must lie in the negative Gaussian curvature region of the flux surface, implies that the region of 3D ridges with $B\approx B_M$ must be contained in the overall $K_G \leq 0$ region of the flux surface. A systematic proof of these statements are given in Section \ref{sec:GradB_ridges} and in Appendix \ref{app:Diff_geo_proof}. 

We now contrast the 3D sharp ridges with the axisymmetric  X-point. Axisymmetry constrains the X point to be a planar circle with zero torsion and normal curvature typically of the order of the inverse major radius $R_a$ such that $\kappa_n R_a\sim 1$. Thus, $\kappa_n R_a$ is non-zero along a sharp axisymmetric ridge, in general, and can only vanish if the ridge is exactly on the top and the bottom of the torus, where the Gaussian curvature is identically zero on the parabolic curve. 

Finally, we point out that the appearance of these straight-line geodesic magnetic fields near the maximum $B$ is a characteristic signature of the failure of QS. From the ray equation, we find that near points of vanishing field line curvature close to $B_M$, $d^2\bm{x}/d\cT^2=0$. Therefore,
\begin{align}
\bm{x}=\bm{x}_0+\bm{x}_1 \cT, \quad \cT= \ell/B_M,
\label{eq:Straight_B}
\end{align}
where $\bm{x}_0,\bm{x}_1$ are constant vectors. Since $d\bm{x}/d\cT=\B$, $\bm{x}_1$ points in the direction of the local $\B$. The key reason for the failure of QS near these straight field lines is that as $|\dl\psi|\to 0,$ the field line $\B = G_0\dl \varphi_B$ and the QS vector $\u= \del \bm{X}/\del \varphi_B$ must align since $\B\times \u =\dl \psi$ \citep{burby2020,rodriguez2020a}. However, this alignment cannot work when the rotational transform is nonzero because $\B$ must have both toroidal and poloidal components in Boozer coordinates, whereas $\u$ is purely toroidal.

We have seen earlier in Figures \ref{fig:RIchard_typical_QA} and \ref{fig:RIchard_typical_QH} that there are points on the ridges, where simultaneously the principal curvature of the surface $\kappa_2$ is a maximum, and $\kappa_1$ is zero, such that the local Gaussian curvature is also zero. The ridges align quite well with the magnetic field lines and are local minima of $|\dl\psi|$. Furthermore, these points also align very well with the maxima of $B$. Near such points, the field lines are essentially straight lines, which we demonstrate numerically in Figure \ref{fig:straight_B_nfp346}. Essentially, the top view of any compact QA device should look like an $n_{fp}$-sided polygon on the inboard side, where $n_{fp}$ is the number of field periods, and the sharp ridges define the sides of the polygon. Clear examples of the approximately polygonal shape in the inboard side of the flux surfaces can be seen in Figure 2 from \citep{plunk2020_near_axisymmetry_MHD}, Figure 1 from \citep{Schuett_henneberg2024compactQA}, Figures 1 and 3 from \cite{nies_exploration_2024} and Figure \ref{fig:straight_B_nfp346} in this paper.

\section{Exact results on the Grad-B tensor and its eigenvalues: implications for ridges and coils}\label{sec:Exact_results}

We present two exact results that show how ridges and coils are connected to the determinant of the $\dlBT$ tensor.
In Section \ref{sec:GradB_ridges}, we formalize the optical analogy into concrete statements connecting the $\dlBT$ tensor and the Hessian of the flux surface $\dl\dl\psi$. In Section \ref{sec:GradB_coils} we prove that all filamentary coils must lie on zero level sets of the determinant of the $\dlBT$ tensor.

\subsection{\textbf{Quasisymmetric ridges must lie on zero-determinant manifold of the magnetic gradient tensor}}
\label{sec:GradB_ridges}
We have so far relied on the optical analogy and the continuity to understand the structure of ridges. In this Section, we obtain exact results that show the deep connections and the parallels between ridges and current filaments.

The only assumption we make is that a quasisymmetric vacuum field with nested flux surfaces can be found. More explicitly, we assume that $\B=\dl\Phi$ is a vacuum field with flux function $\psi$ such that $\BD\psi=0$ and level sets of $\psi$ foliate the toroidal volume. However, as discussed in Section \ref{sec:BnB_ridges} we must consider the possibility of $|\nabla \psi|$ to vanish on a sharp ridge. The vanishing on $|\nabla \psi|$ being a singular limit, care must be exercised in the description of sharp magnetic ridges. We avoid indeterminacy of the normal vector $\nh\equiv \nabla \psi/|\nabla \psi|$ and the elements of the first and second fundamental forms of the flux surface by allowing $|\nabla \psi|$ to be small but nonzero. The necessary mathematical details are provided in Appendix \ref{app:ridge_GradB_Hessian_psi}.

From the gradient of the condition $\BD\psi=0$ or equivalently $\nabla \Phi \cdot \nabla \psi=0$ we obtain
\begin{align}
\dlBT\cdot \dl \psi + \dl \dl \psi \cdot \B =0,
\label{eq:gradBDpsi}
\end{align}
which connects components of the $\dlBT$ and the $\nabla\nabla \psi$ tensors. It is convenient to use the orthonormal Darboux frame $(\nh,\bh, \th)$ where
\begin{align}
\th\equiv \frac{\B}{B}, \quad \nh\equiv \frac{\dl \psi}{|\dl\psi|}, \quad \bh \equiv \th \times \nh,
\label{eq:darboux_frame_again}
\end{align}
to rewrite \eqref{eq:gradBDpsi} as
\begin{align}
\dlBT\cdot \nh \frac{|\dl \psi|}{B}+\dl\dl\psi \cdot \th =0
\label{eq:gradBDpsi_Darboux}
\end{align}
We now dot \eqref{eq:gradBDpsi_Darboux} with $\th$, use the fact that $\dlBT$ is symmetric for vacuum fields, and utilize the curvature identities given in \eqref{eq:kappa_ng_exp}. We find the fundamental relation
\begin{align}
\kappa_n |\dl\psi| + \th \cdot \dl\dl \psi \cdot \th =0,
\label{eq:kappa_n_Hessian_psi}
\end{align}
between the field line normal curvature $\kappa_n$ and the $\th\th$ component of the Hessian $\dl\dl\psi$. Since $\kappa_n = \nh \cdot \nabla \ln{B}$ for vacuum fields, \eqref{eq:kappa_n_Hessian_psi} is a direct constraint on the radial gradient of $B$. Consequently, if $\kappa_n$ or equivalently $\nh \cdot \dl B$ vanishes, $\th \cdot \dl\dl \psi \cdot \th $ must also vanish, which means that, by definition, the field lines are asymptotic curves with $\th$ denoting the asymptotic direction on the flux surface. Note that $\th\cdot\dl \kappa_n \neq 0$ in general. However, if $\BDB=0$ and $\nh\cdot \dl B$ vanish simultaneously, then $\kappa_n=0, \th\cdot \dl \kappa_n=0$.

Next we represent $\dl\dl\psi/|\nabla \psi|$ and $\dlBT/B$ by $3\times 3$ matrices of the form $a_{ij}$ and $c_{ij}$ respectively, where $i,j=1,2,3$ correspond to the Darboux frame directions ($\nh,\bh,\th$). The following components of the $\dlBT/B$ tensor can be deduced from the definitions of the field line curvature components given in \eqref{eq:kappa_ng_exp},
\begin{align}
c_{13}=\kappa_n, \quad c_{23}=\kappa_g, \quad c_{33}= \frac{ \BDB}{B^2}.
\label{eq:cij_comps}
\end{align}
Further, $ c_{11}+ c_{22}+c_{33}=0$ from the divergence-free nature of the magnetic field.

The condition \eqref{eq:gradBDpsi_Darboux} connects $c_{1j}$ to $a_{j3}$ through
\begin{align}
c_{1j}+a_{j3}=0,  \quad j=1,2,3.
\label{eq:b1jaj3}
\end{align}
In particular, $a_{33}=-\kappa_n$ and $ c_{13}=\kappa_n$, both of which vanish in the limit of vanishing $\kappa_n$. The Gaussian curvature of the flux surface $K_G$ is equal to $(a_{22}a_{33}-a_{23}^2)$. The limit $\kappa_n\to 0$, $a_{33}\to 0$ forces
\begin{align}
K_G \to -\lbr a_{23}\rbr^2,
\label{eq:neg_Gauss_curv}
\end{align}
which shows that $K_G$ is locally negative definite. The parabolic curve $K_G=0$ is then given by $a_{23}\to 0$. Further, in this limit, $c_{12}\to 0$ from \eqref{eq:b1jaj3}.

We now turn to $\scD_3=\text{Det}\dlBT$. As discussed in Appendix \eqref{app:Diff_geo_proof}, near a sharp ridge both $K_G$ and $\kappa_n$ go to zero simultaneously. Therefore, near a sharp ridge, 
\begin{align}
(\text{Det}\dlBT)/B = -c_{11}((c_{11}+c_{33})c_{33}+c_{23}^2).
\label{eq:D3neq0}
\end{align}
Thus, $\scD_3\neq 0$ in general near ridges. This is where QS plays a crucial role. QS connects the geodesic curvature $\kappa_g$ to $\BDB$ as shown in \eqref{eq:kappa_g_exp}. Due to QS, the components $c_{23}, c_{33}$ are no longer independent but are both proportional to $\BDB$. Consequently, Det$(\dlBT)$ is also proportional to  $\BDB$ from \eqref{eq:cij_comps} and \eqref{eq:D3neq0}
\begin{align}
\frac{\text{Det}\dlBT}{B}= -c_{11}\frac{\BDB}{B^2}\lbr c_{11}+\frac{\BDB}{B^2} \lbr 1+\lbr \frac{G_0}{|\dl\psi|}\rbr^2 \rbr \rbr,
\label{eq:DetDlBT_BDB_form}
\end{align}
and vanishes near $B=B_M$ on a ridge. 

We note that several components of the $\dl\B/B$ tensor namely $c_{12}, c_{13}, c_{23}, c_{33}$ vanish near the ridge where $\kappa_n\to 0,\kappa_g\to 0$. However, $c_{11}=-c_{22}$ does not necessarily vanish. Therefore, the Frobenius norm $|\dlBT|^2_F$ expressed in terms of the metric elements as $B^2\sum_{ij}c_{ij}$, stays finite as long as $c_{11}\neq 0$. The subharmonic nature of $B$ \citep{rodriguez2024maxJ,Solovev1970} sheds light on the degeneracy of the $\dlBT$ tensor. From the Laplacian of $B^2$ in Cartesian coordinates and $\dl\cdot\B=0$, we get
\begin{align}
    (1/2)\nabla^2 B^2=\sum_i \dl B_i \cdot \dl B_i \geq 0,
    \label{eq:Subharmonic}
\end{align}
which implies that $B^2$ is subharmonic and that it can not only attain its maximum on the outer flux surface boundary and not in the interior. Near the magnetic ridge, where $B$ approaches a constant both in radial and angular directions, $\partial_{X_i} B_i$ is small. Here, $X_i$ denotes local coordinates in the $(\nh,\bh,\th)$ frame. Thus, $\dl B_i$ and the Frobenius norm vanish if $\dl X_i$ stays finite for all $i$. However, as we discussed in Section \ref{sec:geom_ridges} $\dl X_i$ need not stay finite since $|\dl \psi|\to 0$ imply divergence of $|\nabla \alpha|$ for finite $B$ near a ridge.

We now discuss the significance of the $K_G\leq 0$ condition and limits on its applicability. As discussed in Section \ref{sec:BnB_ridges} and Appendix \ref{app:Diff_geo_proof}, the vanishing of $K_G$ related to the vanishing of $\kappa_n$ holds only when the sharp ridge is not aligned with the symmetry direction. We thereby exclude axisymmetric or exactly helically symmetric ridges in exactly symmetric devices. Since a torus must have $K_G$ of both signs, \eqref{eq:neg_Gauss_curv} cannot be satisfied on the whole surface, but is locally possible only near an extremum of $B$. Near a sharp magnetic ridge, $\kappa_2$ blows up while $\kappa_1 \approx \kappa_n$ goes to zero. Thus, $K_G=\kappa_1 \kappa_2$ is zero along the ridge but varies considerably across the ridge. Since field lines must cover the whole flux surface ergodically for generic irrational rotational transform the $K_G\leq 0$ imposes a natural barrier for the field lines to cross the sharp ridge on the flux surface. We demonstrate this barrier for field lines clearly in Figure \ref{fig:nfp4_fieldline} for a $n_{fp}=4$ QA device that has a sharp ridge. In contrast, the $n_{fp}=2$ case shown in Figure \ref{fig:nfp2_fieldline} do not show a clear barrier since its ridge is not sharp. We quantify the sharpness of ridges through the eigenvalue of the $\dlBT$ tensor as discussed in Section \ref{sec:OA_LgradB}.

From the differential geometry point of view, the parabolic curve $K_G=0$ is an envelope of the asymptotic curves $\kappa_n=0$. Therefore, the simultaneous vanishing of $K_G$ and $\kappa_n$ implies that neighboring field lines with $\kappa_n=0$ must form a ray caustic or a singular  envelope of rays near $K_G=0$. From the negative definiteness of $K_G$ in \eqref{eq:neg_Gauss_curv}, we can approximate the flux surface locally as a ruled surface. The field lines that are asymptotic curves on this surface provide the natural ruling. The rulings approach each other as $K_G\to 0$ leading to a singular line of striction as anticipated from the optical analogy in Section \ref{sec:geom_ridges}. Finally, we note that from the Beltrami-Enneper theorem, $\tau^2 +K_G=0$ for asymptotic curves. Thus, \eqref{eq:neg_Gauss_curv} implies that $a_{23}$ must be equal in magnitude to the torsion $\tau$ of the field line. Since $K_G$ varies a lot across the ridge, the torsion of nearby field lines must be large.

\subsection{\textbf{Filamentary coils must lie on zero-determinant manifold of the magnetic gradient tensor } \label{sec:GradB_coils}}

For vacuum fields, the eigenvalues of the $3\times 3$, real, traceless, symmetric $\dlBT$ tensor must satisfy a cubic equation of the form
\begin{align}
\lambda^3 + \cP \lambda +\cQ=0.
\label{eq:cubic_lambda_eqn}
\end{align}
The coefficient of $\lambda^2$, given by $\mbox{Tr}(\dlBT)$, vanishes identically thanks to the divergence-free property of $\B$. For vacuum configurations, the other two quantities $\cP$ and $\cQ$, which depend on the details of the specific type of equilibrium, are related to the determinant and the $L_{\nabla B}$ metric \citep{kappel_Landreman2024magnetic}
as
\begin{align}
\cQ=-\mbox{Det}(\dlBT), \quad L_{\nabla B}= \frac{1}{\sqrt{-\cP/B^2}}.
\label{eq:Det_LDB_cPcQ}
\end{align}
The $L_{\nabla B}$ metric is defined in terms of the Frobenius norm as
\begin{equation}
L_{\nabla \B} \equiv \frac{B \sqrt{2}}{\sqrt{|\dlBT|_F^2}}. \quad \frac{\left|\dlBT\right|^2_F}{B^2} \equiv \sum_{ij}c_{ij}^2.
\end{equation}

The discriminant of the cubic $\Delta_d=-(4\cP^3 +27 \cQ^2)$ must be positive for three real roots. In the limit of $\cQ\to 0$, near the locus of vanishing 3D determinant $\scD_3=0$, it follows from \eqref{eq:cubic_lambda_eqn} that the eigenvalues are $(-\sqrt{-\cP},0,\sqrt{-\cP})$ or equivalently $(-B/L_{\nabla B},0,B/L_{\nabla B})$. Therefore, if the current source is such that $\scD_3=0$, which coincides with the zero eigenvalue, the magnitude of the non-zero eigenvalues is directly related to the $L_{\nabla B}$ metric \citep{kappel_Landreman2024magnetic}.

We now show that indeed the singularities of vacuum fields originating from filamentary coils are always tangential to surfaces defined by $\scD_3(x,y,z)=0$ with necessary details in Appendix \ref{app:Det0_coils_NCE}.
The essential idea is to use a Frenet-Serret frame along the filament and approximate the Biot-Savart integral via \textit{local induction approximation} (LIA), which is used extensively in the theory of vortex filaments \citep{saffman1992vortex,callegariTing1978motion}. 

We represent the filamentary coil by a space curve $\bm{r}_c(\ell)$ parameterized by its arclength $\ell$, and employ the standard orthogonal Mercier coordinate system $(\rho,\omega,\ell)$ \citep{Mercier1964,Solovev1970,sengupta2021NSE,jorge_sengupta2020near} equipped with the orthonormal unit vectors $(\th=\bm{r}_c'(\ell),\rhoh=\dl\rho,\wh=\dl\omega)$. We use the following notation: $C=\mu_0 I/2\pi$ is the strength of the current filament, $\rho$ is the distance from the filament, $\theta=\omega-\int\tau_c d\ell$ is the angle between the normal $\nh$ and the radial unit vector $\rhoh$, and $\kappa_c(\ell), \tau_c(\ell)$ are the curvature and torsion of the filament. LIA in these coordinates is a formal expansion in $\kappa_c \rho$, identical to the near-axis expansion with the magnetic axis replaced by the coil. 

Within the LIA, the field near the current filament is
\begin{align}
&\B= B_\rho \rhoh+ B_\omega\wh +O((\kappa_c \rho)^0), \label{eq:LIA_meets_NAE}\\
& B_\rho = -\frac{C}{2\rho}\lbr \kappa_c \rho \ln{\kappa_c \rho }\rbr\sin\theta,\quad B_\omega = \frac{C}{\rho}\lbr 1+\frac{\kappa_c \rho}{2}\cos\theta \lbr -\ln{\kappa_c\rho}+1\rbr\rbr \nonumber.
\end{align}
The $\dlBT$ tensor in the $(\th,\rhoh,\wh)$ basis is of the form
\begin{align}
\dlBT \equiv
\begin{pmatrix}
-\kappa_c B_n \quad 0\; \\
\quad 0 \qquad  \scH_2 \;
\end{pmatrix} \quad\scH_2\equiv
\begin{pmatrix}
\del_\rho B_\rho \qquad \qquad \del_\rho B_\omega\\
\frac{1}{\rho}(-B_\omega +\del_\omega B_\rho) \quad \frac{1}{\rho}(\del_\omega B_\omega + B_\rho)
\end{pmatrix},
\label{eq:dlBT_coil}
\end{align}
where the normal component in the Frenet frame,  $B_n = \B \cdot \nh$, reads
\begin{align}
B_n \equiv \B \cdot \nh = -C\frac{\sin\theta}{\rho}\lbr 1+ \frac{\kappa_c\rho}{2}\cos\theta\rbr.
\label{eq:Bn_exp}
\end{align}

The traceless condition of $\dlBT$ implies
\begin{align}
\del_\rho B_\rho +\frac{1}{\rho}(\del_\omega B_\omega + B_\rho)-\kappa_c B_n =0.
\label{eq:divB_coil}
\end{align}
The largest terms in the left side of \eqref{eq:divB_coil} from \eqref{eq:LIA_meets_NAE} are $O(1/\rho \ln{\kappa_c\rho})$, which can be seen to cancel out, leaving only $O((\kappa_c\rho)^0)$ terms. Thus, $\B$ is divergenceless to the order of accuracy of \eqref{eq:LIA_meets_NAE}.

The determinant $\scD_3=\mbox{Det}(\dlBT)$ obtained from \eqref{eq:dlBT_coil} is
\begin{align}
\scD_3 = -\kappa_c B_n \scD_2 , \quad \scD_2\equiv \mbox{Det}\scH_2=\frac{1}{\rho}\lbr \del_\rho B_\rho\lbr \del_\omega B_\omega +B_\rho \rbr  - \del_\rho B_\omega \lbr -B_\omega +\del_\omega B_\rho\rbr \rbr. \label{eq:Det_D3_coil}
\end{align}
The term represented by $\scD_2$ is the determinant of the 2D Hessian $\scH_2$.

At a distance $\rho>0$, we observe that $\scD_3$ vanishes if $\kappa_c=0$ or $B_n=0$. $\scD_2$ is negative definite due to the traceless condition and is $O(1/\rho^2)$. Since $\kappa_c\neq 0$ for a generic filament, $\scD_3=0$ and $B_n=0$ are satisfied when $\sin\theta=0$. Thus, for $\theta=0$ and $\theta=\pi$, $\dlBT$ is rank 2, with the null direction coinciding with the direction of the coil. In the limit $\rho\to 0^+$ with $\sin\theta$ fixed at zero, the null direction aligns everywhere with the direction of the tangent to the coil, implying that the coil itself must lie on a $\scD_3=0$ surface. We note here that as we approach the $\scD_3\to 0$ surface, the two non-zero eigenvalues of the $\dlBT$ tensor grow large as the inverse of $\rho$. For a rigorous $\rho\to 0$ limit we show in Appendix \ref{app:Det0_coils_NCE} that a finite and normalized determinant, $\bar{\scD_3}$, can be defined instead of $\scD_3$, which does not alter any of our conclusions. Finally, we note that due to the large $1/\rho$ terms in the expression for $\B$ near a given filament, the LIA with a single coil is still an excellent approximation for multiple coils.

\subsection{\textbf{Ridges, coils and the $L_{\nabla B}$ metric \label{sec:OA_LgradB}}}

The framework we have developed provides crucial insight not only for coils but also for flux surface geometry and properties of the magnetic field lines near the regions of vanishing $|\dl\psi|$. The key to understanding the geometry dependence is through the behavior of the $\dlBT$ tensor and its eigenvalues. For coils, the eigenvalues determine the $L_{\nabla B}$ metric for the coil-plasma distance. For flux surfaces with sharp features like X points or ridges that can potentially serve as divertors,  the eigenvalues control the local behavior of the magnetic field lines. In particular, we can quantify the sharpness of the ridge in terms of the eigenvalues of the $\dlBT$ tensor. 

We now discuss the $\dlBT$ tensor and its eigenvalues in three cases of practical importance: the axisymmetric tokamak X point, the O and X points of an island divertor of a stellarator, and, finally, the 3D sharp ridges we have discussed thus far. The details are provided in Appendix \ref{app:GradB_tensor_forms}. In each of these cases, we restrict ourselves to regions near O or X points or near ridges, with a common feature that $|\dl\psi|$ is locally a minimum and can tend to zero. We show that the eigenvalues of $\dlBT$ in these three cases can be markedly different. 

Since we compare the eigenvalues from different geometries, it is convenient to normalize the equilibrium quantities with the major radius $R_a$ and a reference magnetic field strength $B_a$. From here on, we assume normalized quantities, such that a regular eigenvalue corresponds to $\lambda_0 B_a/R_a$, where $\lambda_0$ is order unity constant.

We start with the axisymmetric case. As shown in Appendix \ref{app:GradB_tensor_AS}, the normalized eigenvalues are regular if the second derivatives of the flux surface label $\psi$ are bounded. However, if the second derivatives of $\psi$ are large, which can happen due to strong axisymmetric shaping, both $\cP$ and $\cQ$ can be large, leading to large eigenvalues. Thus, $L_{\nabla B}$ can be small near areas of strong flux surface shaping. 

Next, we consider the case of a regular O or X point generated by a closed field line of an island chain. We show in Appendix \ref{app:GradB_tensor_closedB} that in this case, the normalized eigenvalues are always regular. Locally, the flux surface is always a rotating ellipse or hyperbola whose rotation rate and elongation are determined by their values near the closed field line. Thus, the second-order derivatives of $\psi$ can not be arbitrarily large. As a result, regular X points do not enforce any restrictions on $L_{\nabla B}$.

Finally, we consider the 3D sharp ridges. As shown in Appendix \ref{app:GradB_near_ridges}, the normalized $\cQ$ goes to zero near the ridge, which is generally not true in either the axisymmetric or the closed field line cases. The normalized $\cP$ is related to the second-order derivatives of $\Phi$ through its 2D Hessian. 
Thus, $L_{\nabla B}=1/\sqrt{|K_G|}$ near the ridges. Just like the axisymmetric case, large second-order derivatives of $\Phi$ lead to large values of $K_G$ and hence small $L_{\nabla B}$. However, on the ridge where $K_G=0$, $L_{\nabla B}$ can diverge. We can obtain the same result based on the subharmonic property of $B$ discussed in Section \ref{sec:GradB_ridges}. It is indeed possible for the $\dlBT$ tensor to become degenerate in the vicinity of the ridge, when $B$ approaches a constant on the outer flux surface in both angles and radial variables. In this degenerate case, the Frobenius norm is small, which implies that $L_{\nabla B}$ diverges. Therefore, the $L_{\nabla B}$ metric must be handled with care near the ridges.

The three cases discussed above illustrate the interesting role played by the second-order derivatives of the flux surface label and Gaussian curvature in the case of axisymmetry with strong axisymmetric flux surface shaping and 3D ridges, but not in the case of regular O and X points.

We have shown that $\cQ=-\mbox{Det}(\dlBT)$ goes to zero both near the 3D ridges and coils. A quantity analogous to $\cQ$ has been recently investigated in the fluid dynamics context, generalizing the Okubo-Weiss criterion by investigating the $\mathsfbi{\dl\bm{v}}$ tensor for a flow field $\bm{v}$ \citep{bachman2021}. Associated with $\cQ=0$ is a null direction of the  $\dlBT$ tensor, which aligns with ridges and coils. In both these cases, the nonzero eigenvalues are given by $\pm \sqrt{-\cP}=\pm B/L_{\nabla B}$. Thus, while using the $L_{\nabla B}$ metric to find nearby coils, one needs to be careful near regions with significant flux surface shaping.

We conclude by noting that the regular O and X points and the regular tokamak X points are periodic and are uninterrupted on the torus. We can also include the $n_{fp}=2$ QA configurations in this group, characterized by a lack of strong flux surface shaping. Our analysis shows that the normalized eigenvalues of the $\dlBT$ are $O(1)$ and so coils can be further out. Now we consider 3D sharp ridges that are highly localized, being restricted to mostly lie on the inboard $K_G\leq 0$ region. Localized sharp ridges naturally imply localized flux surface shaping, which, from our analysis, would require complex coils in the vicinity of the strong shaping on the inboard side. This is also correlated with large normalized eigenvalues.  

\section{Numerical verification \label{sec:Numerical}}

Using Parker’s optical analogy, we have constructed an analytical theory of near-axisymmetric ridges. Now we present numerical evidence in support of the main results discussed in Section \ref{sec:Prop_near_ridges} and \ref{sec:Exact_results}.

In \ref{sec:Rogerio_numerics} we present a suite of numerically optimized compact QA with field periods $n_{fp}$ varying between $2$ to $7$. Using these configurations, we then show that sharp ridges tend to form on the inboard side.

\subsection{Optimized near-axisymmetric configurations with QS in a volume}
\label{sec:Rogerio_numerics}
We perform a numerical optimization of ideal MHD fields to obtain quasi-axisymmetry using the SIMSOPT code \citep{landreman2021simsopt}.
The data and scripts used in this section are available on \emph{github} \footnote{\url{https://github.com/rogeriojorge/QA_nfpX}}.
The magnetic field equilibrium is obtained using the VMEC code \citep{VMEChirshman1983steepest}.
The objective function $f$ targeted here for minimization follows \cite{landremanPaul2021} and is given
\begin{equation}\label{eq:objective}
f = f_{\iota}+f_{A}+f_{QS},
\end{equation}
where $f_\iota=10^2 (\overline \iota - \overline \iota_*)^2$ with $\overline \iota = \int_0^1 ds\; \iota$ the mean rotational transform $\iota$ over the normalized toroidal flux $s$ and $\overline \iota_*$ is a specified target value taken here as 0.21, and $f_A=(A-A_*)^2$ where $A$ is the aspect ratio of the boundary surface as computed by the VMEC code and $A_*$ a specified target value taken here as 5.
In Eq. (\ref{eq:objective}), $f_{QS}$ is the objective function used to achieve QS
\begin{equation}
f_{QS} = -\sum_{s_j} \left<\left(\frac{1}{B^3}[\iota \mathbf B \times \nabla B \cdot \nabla \psi_T + G_0 \mathbf B \cdot \nabla B \right)^2\right>,
\end{equation}
where  $\left\langle \dots \right\rangle$ is a flux-surface average.
The sum is over a set of flux surfaces $s_j$, where a uniform grid 0, 0.1, …, 1 was used.
The independent variables used here are the Fourier coefficients ${R_{mn},Z_{mn}}$ of the boundary surface, defined in cylindrical coordinates $(R, \phi, Z)$ as a function of the VMEC poloidal $\theta$ and toroidal $\phi$ angles
\begin{align}
R(\theta,\phi)=\sum_{m,n}R_{mn}\cos(m\theta-n_{fp}n\phi), \quad
Z(\theta,\phi)=\sum_{m,n}Z_{mn}\sin(m\theta-n_{fp}n\phi),
\end{align}
where stellarator symmetry has been assumed.
Here, $R_{00}$ is excluded from the parameter space, being equal to 1 meter, therefore fixing the spatial scale.
The initial condition is an axisymmetric torus, and the parameter space is expanded in a series of 4 steps, with modes $|m|,|n| \le j$ optimized in step $j$.
Using SIMSOPT, the optimization is performed using a nonlinear least-squares minimization (trust region reflective) algorithm, with the gradients computed using finite differences.

\begin{figure}
\centering
\includegraphics[trim={0.01cm 2.1cm 0.01cm 0.01cm},clip,width=0.3\linewidth]{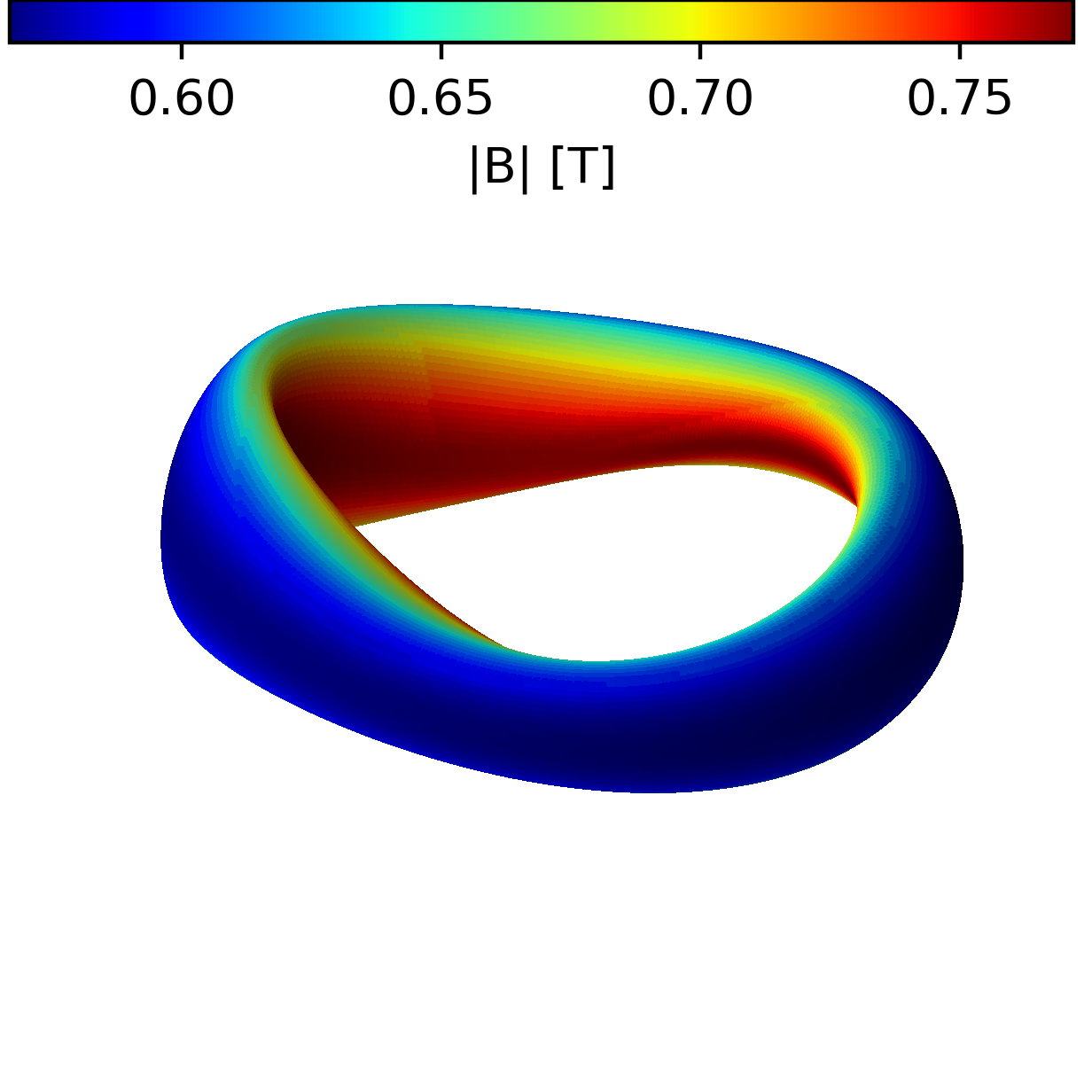}
\includegraphics[width=0.3\linewidth]{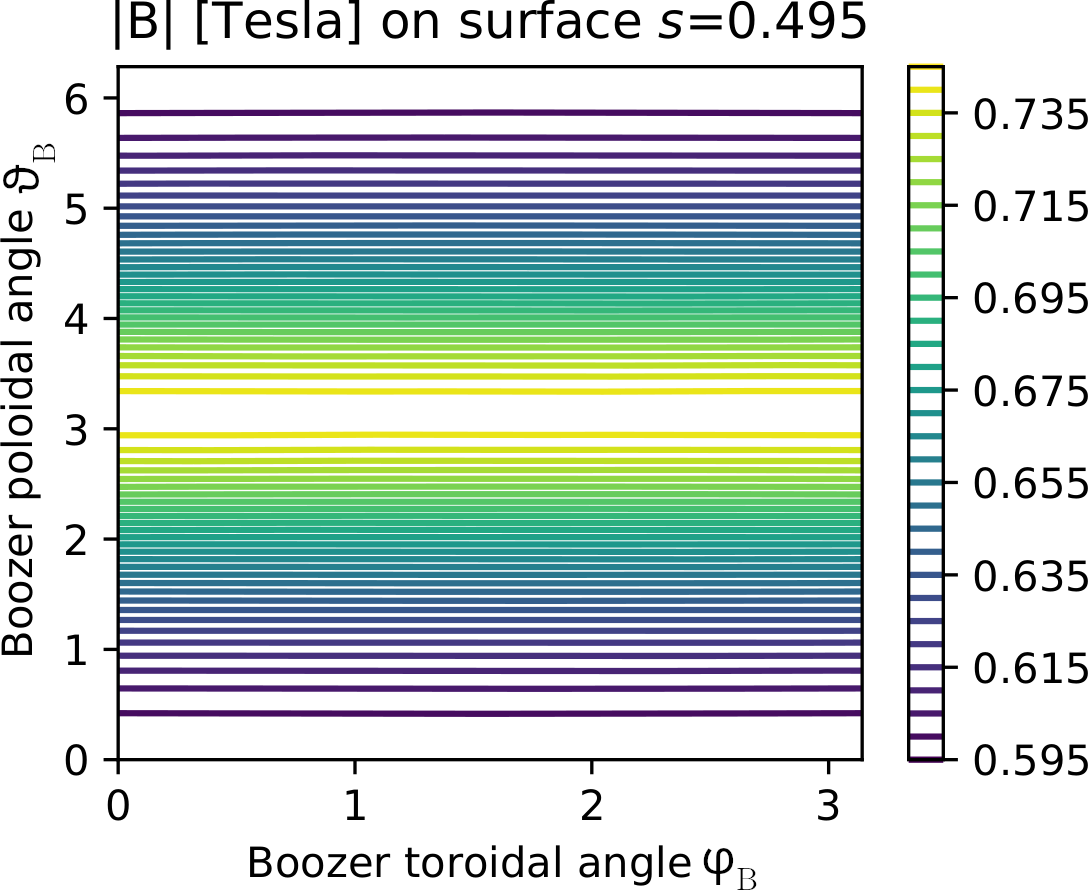}
\includegraphics[trim={0.01cm 11.9cm 23.4cm 0.01cm},clip,width=0.35\linewidth]{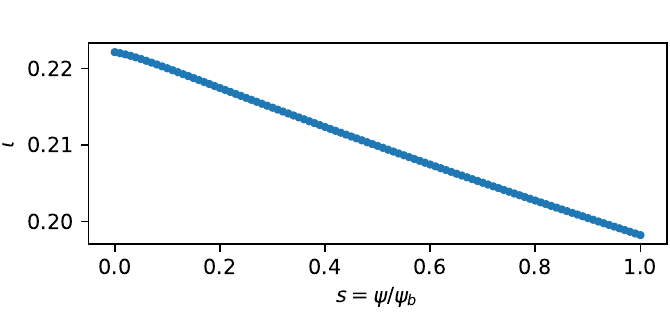}
\caption{Magnetic field configuration with precise quasi-axisymmetry and $n_{fp}=2$. Left: plasma boundary and the corresponding strength of the magnetic field. Middle: contours of magnetic field strength at mid-radius in Boozer coordinates showcasing precise QS. Right: rotational transform profile with a mean of $\overline \iota=0.21$.}
\label{fig:nfp2}

\includegraphics[trim={0.01cm 2.1cm 0.01cm 0.01cm},clip,width=0.3\linewidth]{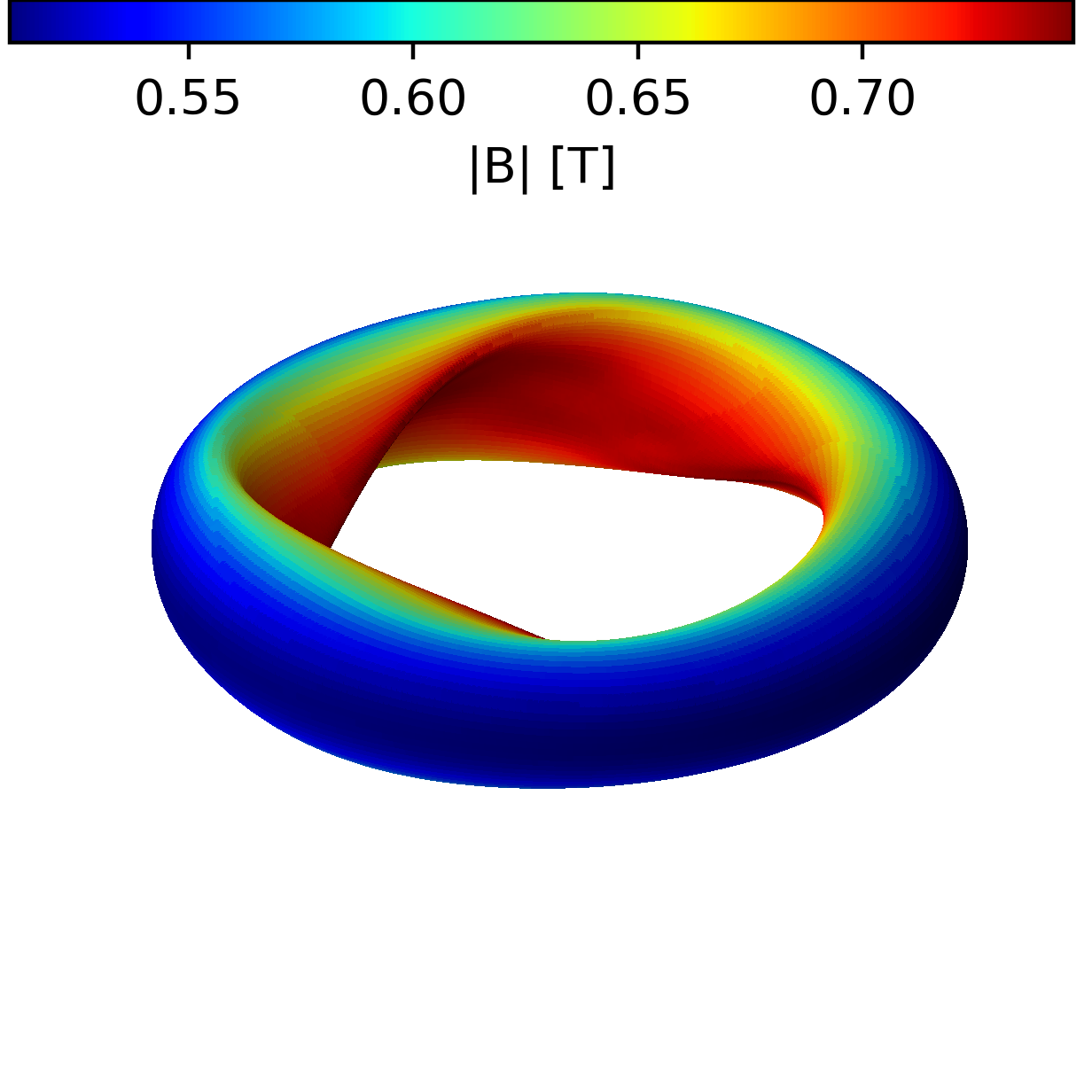}
\includegraphics[width=0.3\linewidth]{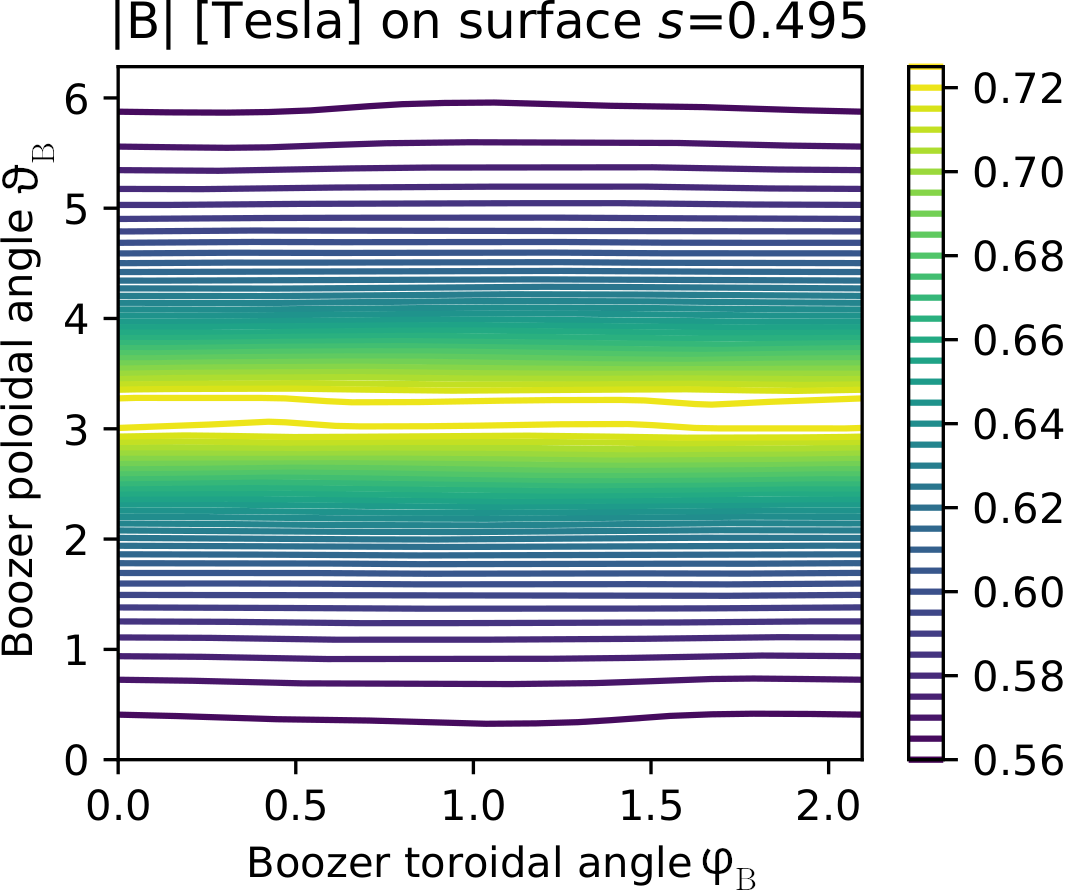}
\includegraphics[trim={0.01cm 11.9cm 23.4cm 0.01cm},clip,width=0.35\linewidth]{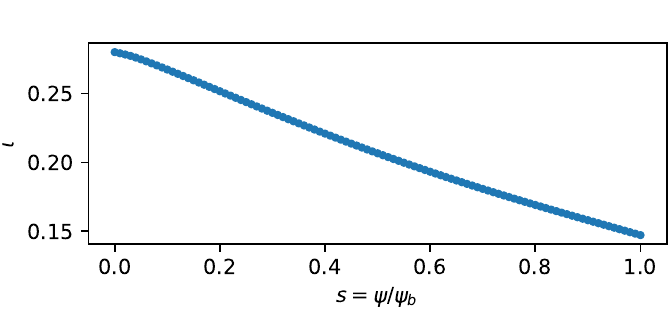}
\caption{Configuration with precise quasi-axisymmetry and $n_{fp}=3$.  See \autoref{fig:nfp2} for details.}
\label{fig:nfp3}

\includegraphics[trim={0.01cm 2.1cm 0.01cm 0.01cm},clip,width=0.3\linewidth]{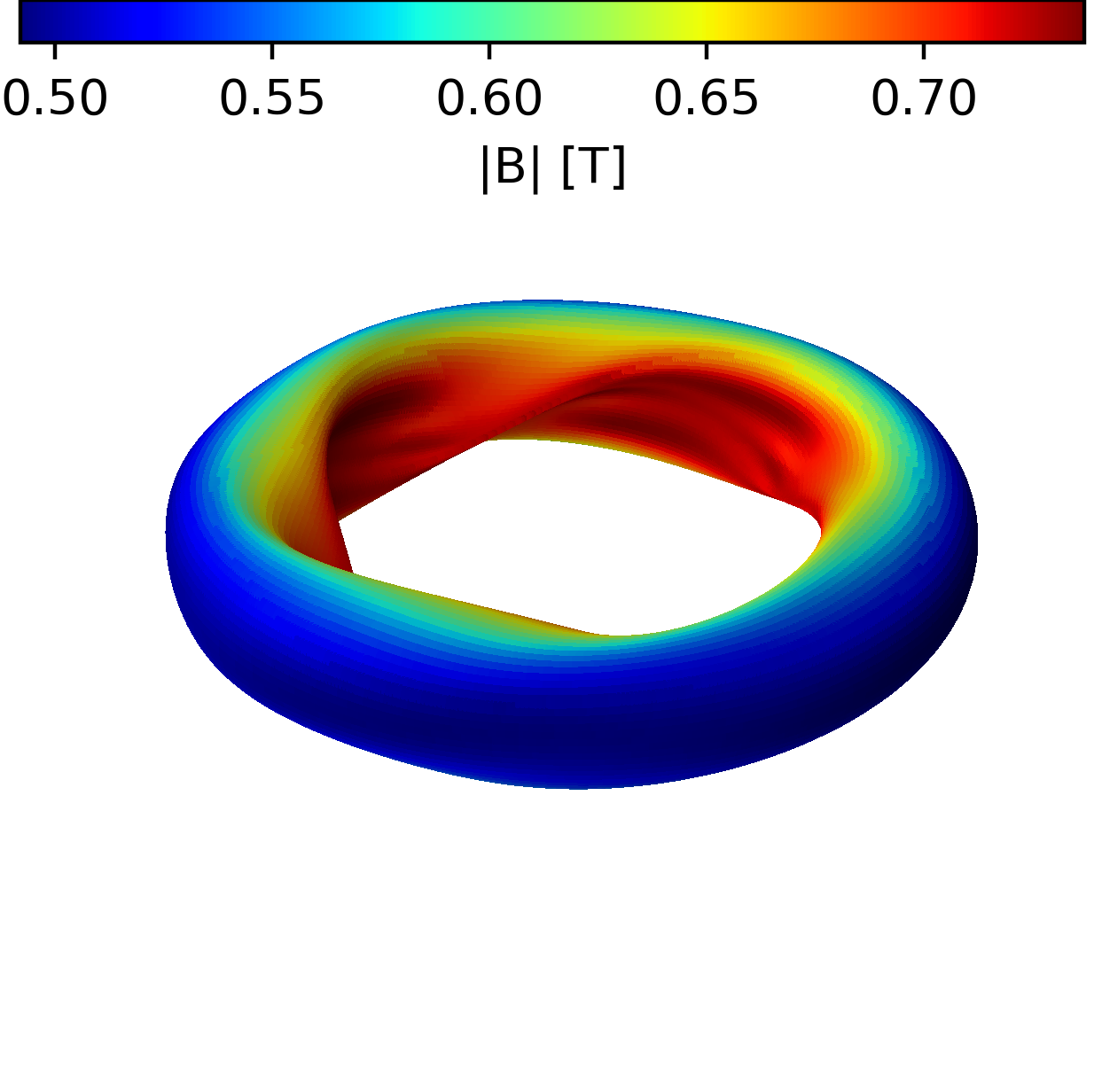}
\includegraphics[width=0.3\linewidth]{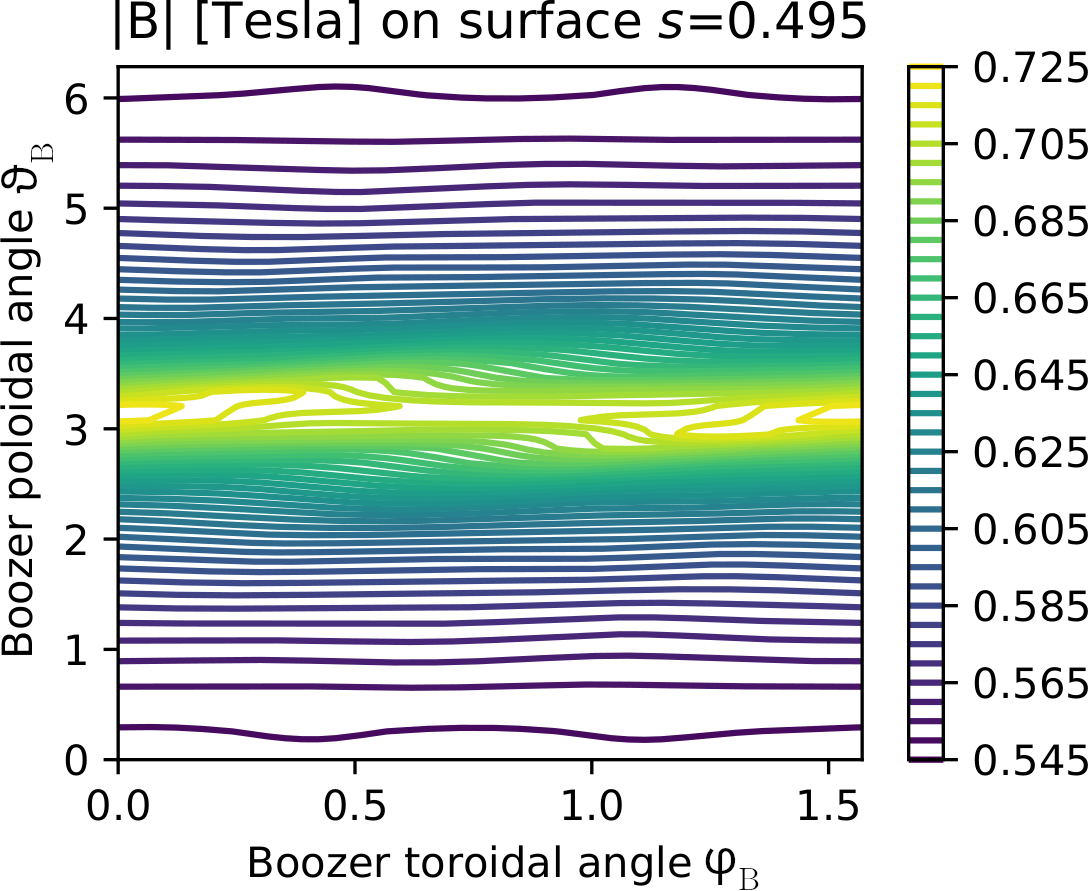}
\includegraphics[trim={0.01cm 11.9cm 23.4cm 0.01cm},clip,width=0.35\linewidth]{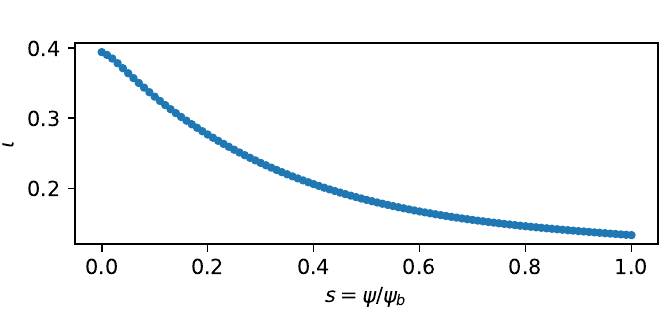}
\caption{Configuration with precise quasi-axisymmetry and $n_{fp}=4$.  See \autoref{fig:nfp2} for details.}
\label{fig:nfp4}

\includegraphics[trim={0.01cm 2.1cm 0.01cm 0.01cm},clip,width=0.3\linewidth]{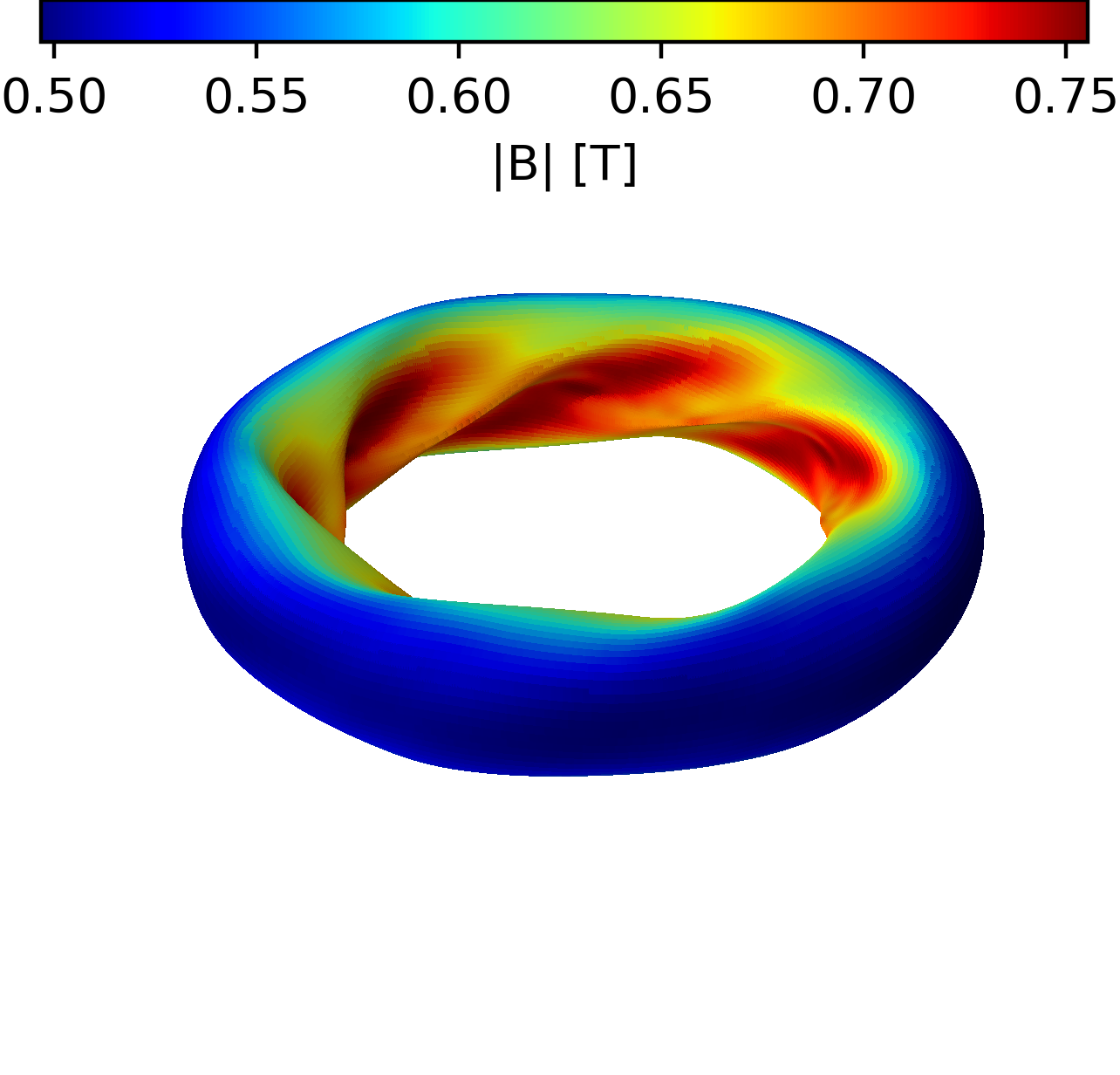}
\includegraphics[width=0.3\linewidth]{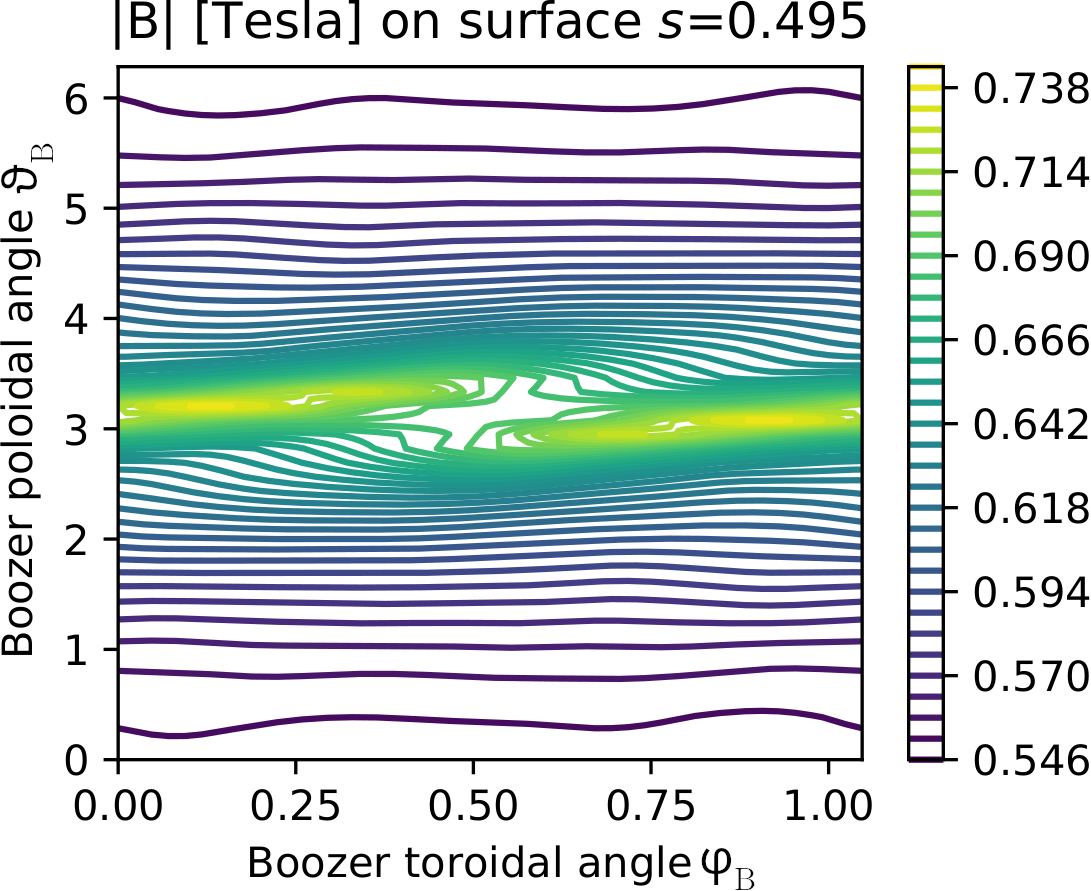}
\includegraphics[trim={0.01cm 11.9cm 22.4cm 0.01cm},clip,width=0.35\linewidth]{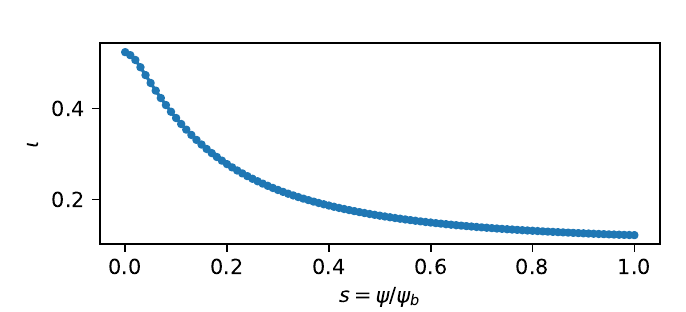}
\caption{Configuration with precise quasi-axisymmetry and $n_{fp}=6$. See \autoref{fig:nfp2} for details.}
\label{fig:nfp6}
\end{figure}

In order to benchmark our optimization method against established QA equilibria, we first show the result of an optimization using an $n_{fp}=2$ equilibrium in Figure \ref{fig:nfp2}.
Here, we present the three-dimensional shape of the plasma boundary, along with the magnetic field strength at both the boundary and at mid-radius in Boozer coordinates.
This shows that this algorithm is able to find precise quasi-axisymmetry when $n_{fp}=2$ is used.
We also show in Figure \ref{fig:nfp2} is the resulting rotational transform profile with a mean of $\overline \iota = 0.21$.

\begin{figure}
\centering
\includegraphics[width=0.315\linewidth]{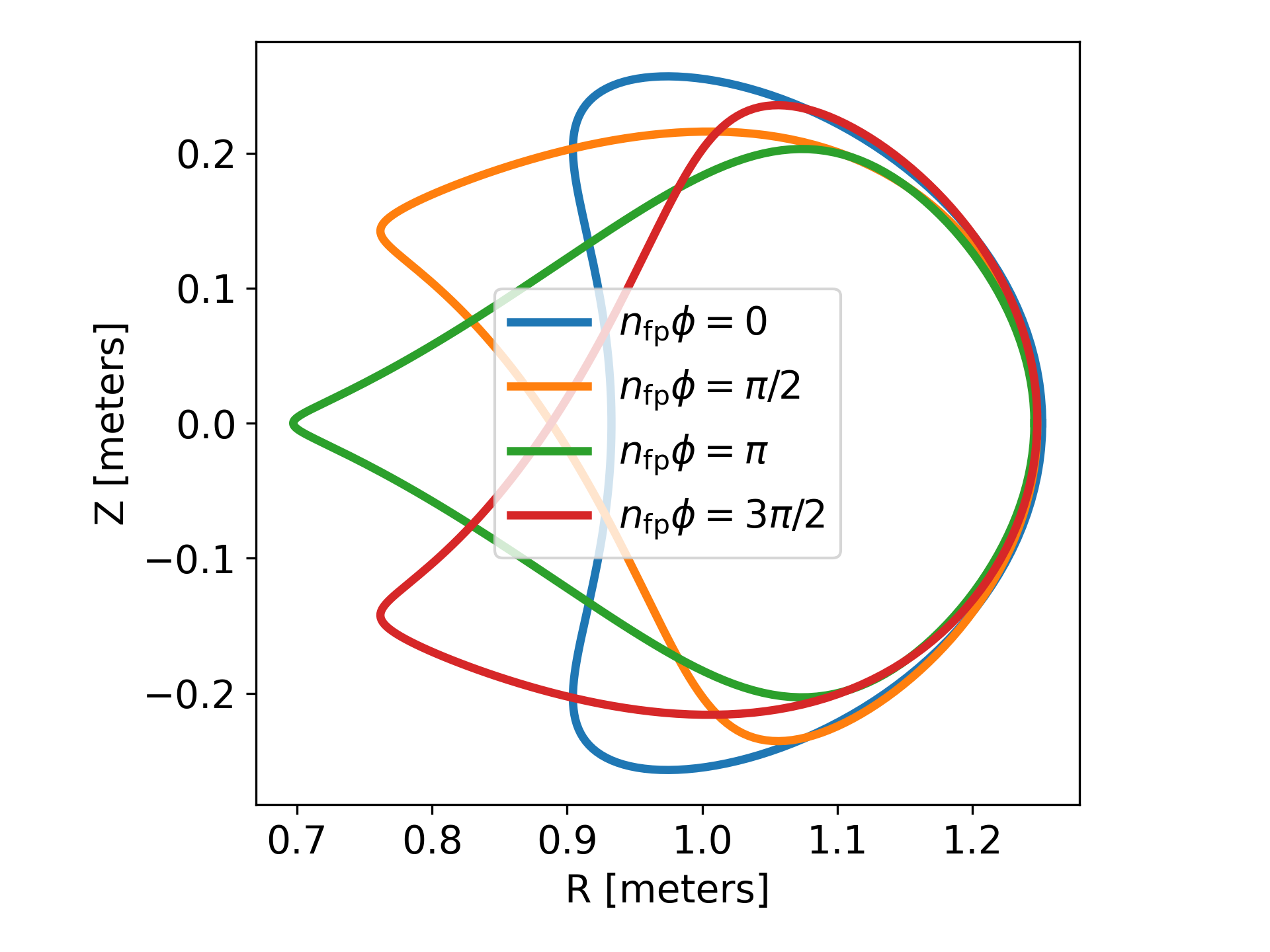}
\includegraphics[width=0.315\linewidth]{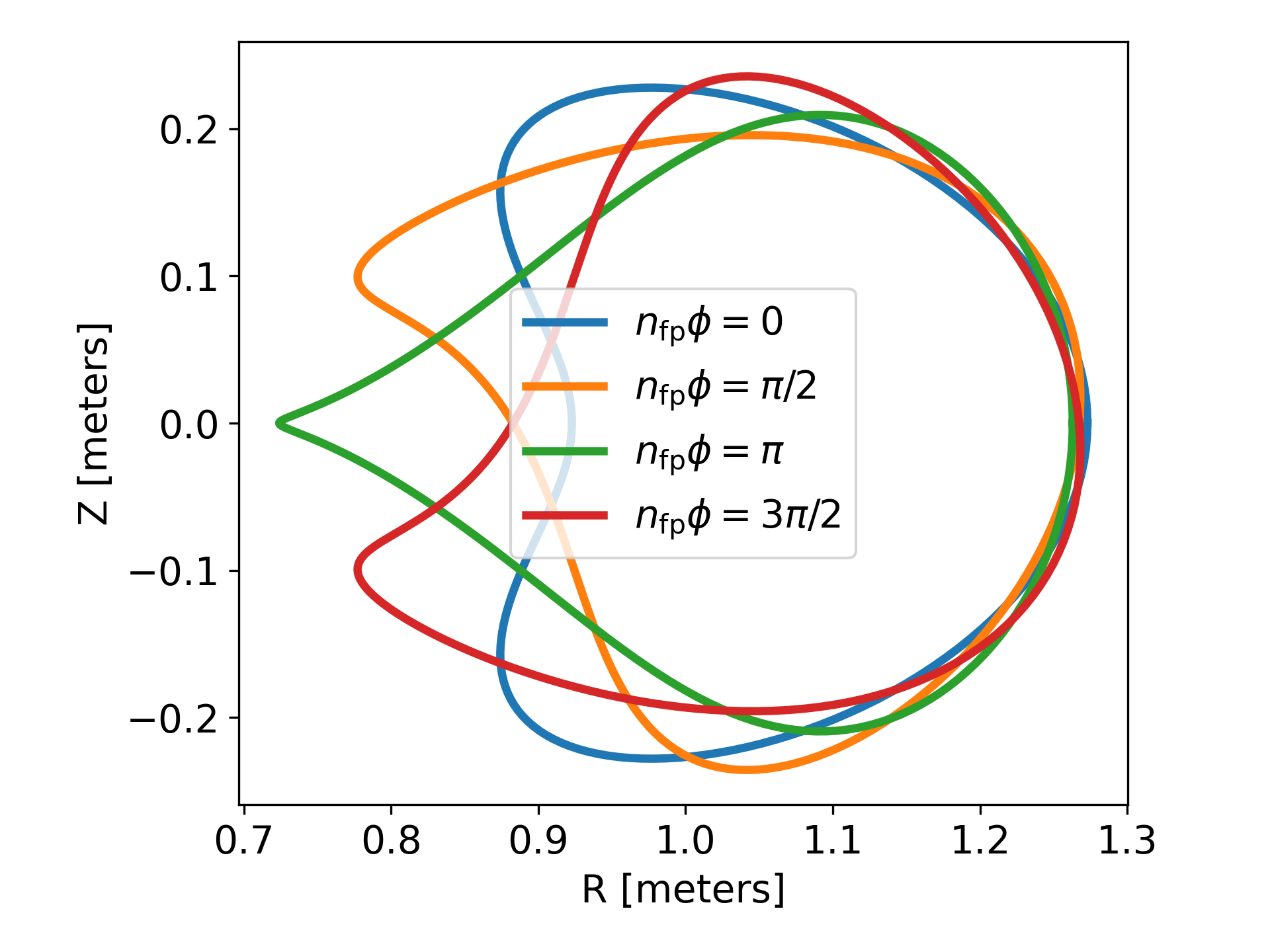}
\includegraphics[width=0.315\linewidth]{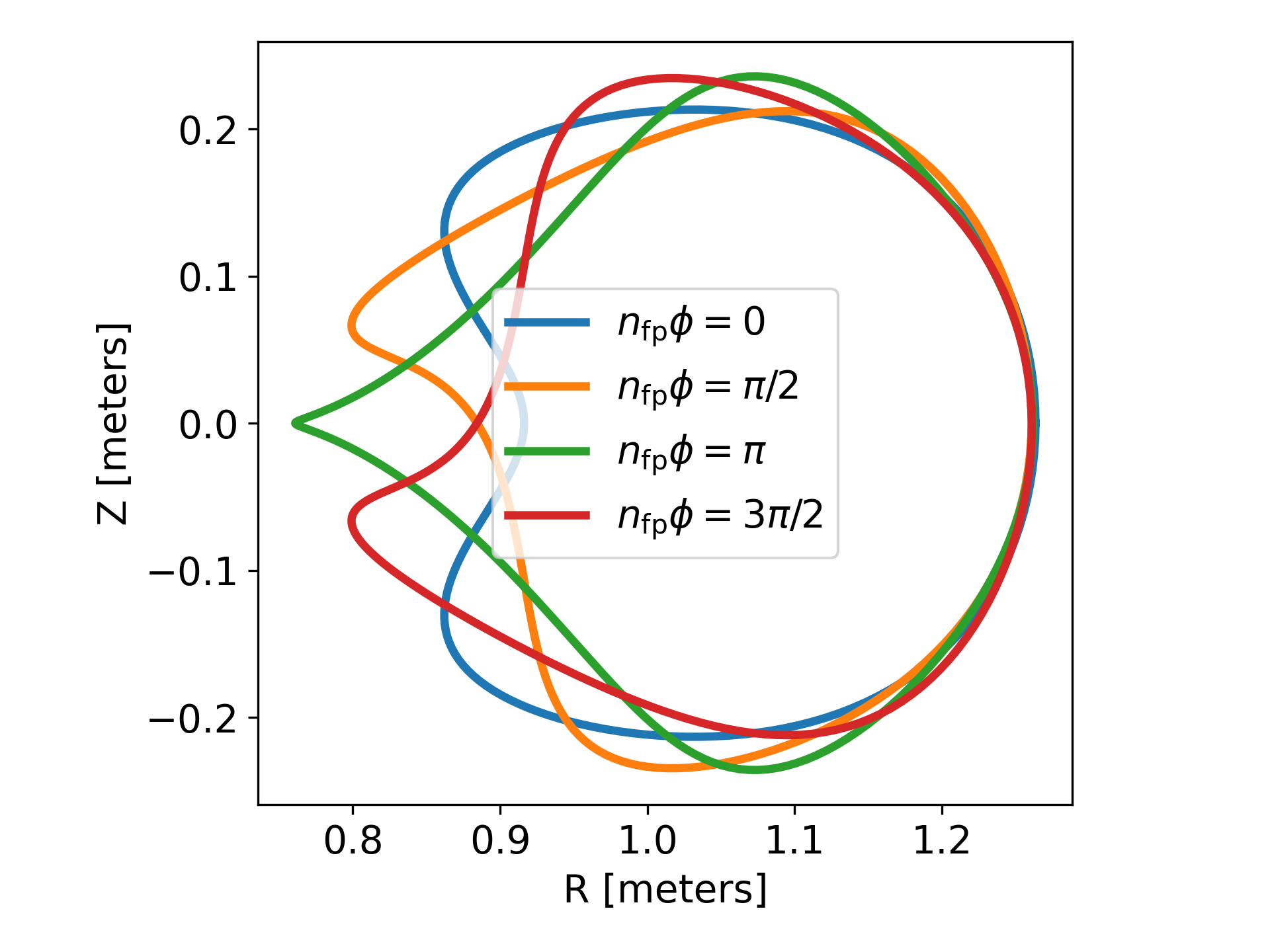}
\caption{Poloidal cross-sections of the plasma boundary of quasi-axisymmetric configurations at toroidal angles $n_{fp}\phi=0, \pi/2, \pi$ and $3\pi/2$ with $n_{fp}=3$ (left), $n_{fp}=4$ (middle) and $n_{fp}=6$ (right).}
\label{fig:poloidal_nfp346}
\end{figure}

We now apply the same optimization algorithm that produced  the $n_{fp}=2$ solution in Figure \ref{fig:nfp2} to configurations with $n_{fp}=3, 4, 5, 6$ and $7$.
The configurations found for $n_{fp}=3, 4$ and $6$ are shown in Figs. \ref{fig:nfp3}, \ref{fig:nfp4} and \ref{fig:nfp6}, respectively.
Here, we find that, using the same optimization method, the quasisymmetry residual $f_{QS}$ increases with the number of field periods, making the contour plots of the magnetic field strength in Boozer coordinates depart from horizontal straight lines at higher $n_{fp}$.
Despite the differences in $n_{fp}$, Figure \ref{fig:poloidal_nfp346} shows remarkable similarity in the poloidal cross sections of these configurations when plotted in cylindrical coordinates $R,Z$ at different cylindrical toroidal angles $\phi$.
In particular, increasingly sharper ridges appear on the inboard side of the stellarator, while the outboard side displays a more axisymmetric geometry, with a nearly circular cross-section around the mid-board plane at poloidal angle $\theta=0$.
Such features were also observed in \cite{henneberg2024compact}.

\begin{figure}
\centering
\includegraphics[width=0.8\linewidth]{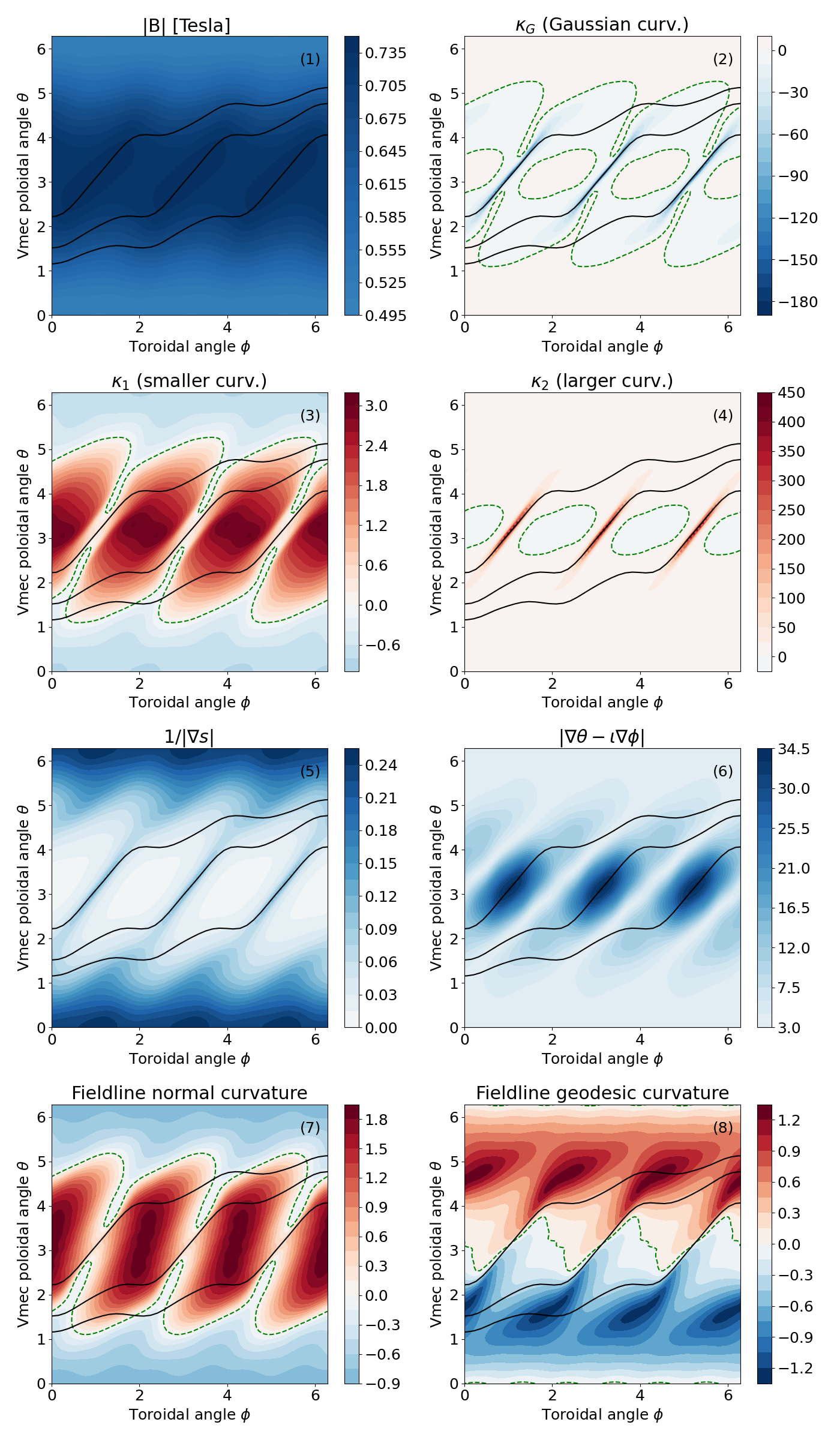}
\caption{From top-left to bottom right: (1) $|B|$, (2) $K_G$, (3) $\kappa_1$ (4) $\kappa_2$, (5) $1/|\nabla s|$, (6) $|\nabla \theta - \iota \nabla \phi|$, (7) $\kappa_n$,  $\kappa_g$. For the boundary of the $n_{\text{fp}}=3$ configuration in \autoref{fig:nfp3}. The black lines overlaid on each plot depict the field lines passing through the center of the ridges.}
\label{fig:nfp3_allplot}
\end{figure}

\begin{figure}
\centering
\includegraphics[width=0.3\linewidth]{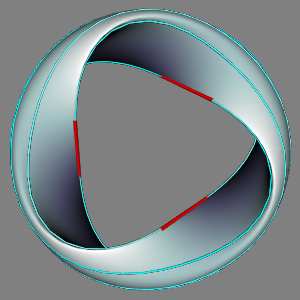}
\includegraphics[width=0.3\linewidth]{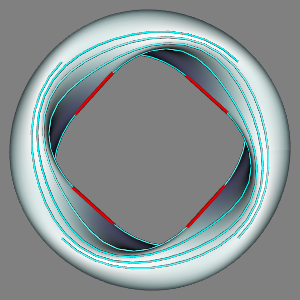}
\includegraphics[width=0.3\linewidth]{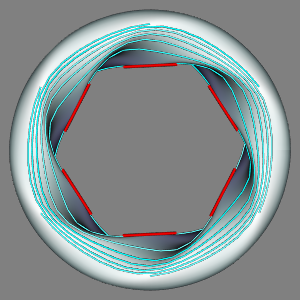}
\caption{Demonstration of straightness of field lines on the ridges near the maximum $B$ for $n_{fp}=3$ (left), $n_{fp}=4$ (middle), and $n_{fp}=6$ (right). The red lines are perfectly straight lines centered at the mid-point of ridge and oriented along the ridge. The cyan lines are the field lines.}
\label{fig:straight_B_nfp346}
\end{figure}

To confirm that these are true ridges in the mathematical sense, we plot the principal curvature $\kappa_1,\kappa_2$. This is shown in subplot 3 and 4 of Fig.~\ref{fig:nfp3_allplot} for the QA $n_{\text{fp}} =3$ configuration. The differential geometry definition of the sharp ridge are collection of points of maximum of $\kappa_2$. We see that the $\kappa_2$ attains a large positive value highly localized near a curve in the $(\theta,\phi)$ plane that is almost straight, which we call the magnetic ridge. We note that $\kappa_1$ goes to zero and stays zero along the ridge. The fieldlines (black curves in the plot) pass through the ridge, confirming the predictions in Section~\ref{sec:BnB_ridges}. As a consequence of the vanishing principal curvature on the ridge, the fieldlines passing through the ridge are straight lines in real space, which is confirmed by the vanishing normal and geodesic curvatures of the fieldlines in subplot 7 and 8. Straight fieldlines and flux conservation imply that $B$ must be constant along the ridge, which is confirmed in subplot 1. Finally, the fact that $B$ is constant while $\dl \psi$ becomes small on the ridge (subplot 5) implies that the parts of $\dl \alpha$ that are orthogonal to $\dl \psi$ must become large, since $|\B| = |\dl \psi \times \dl \alpha|$. This is confirmed in subplot 6.

The fieldlines being straight is also shown in Fig.~\ref{fig:straight_B_nfp346}, where we plot the boundary of the $n_{\text{fp}}= 3$, $n_{\text{fp}}= 4$, and $n_{\text{fp}}= 6$ together with a straight line (red) centered on the mid-point of the ridge and oriented in the ridge direction at the point.

\begin{figure}
\centering
\includegraphics[width=0.45\linewidth]{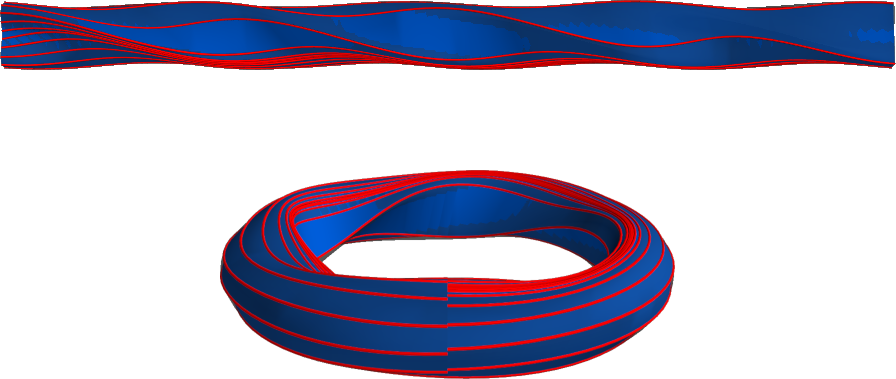}\includegraphics[width=0.45\linewidth]{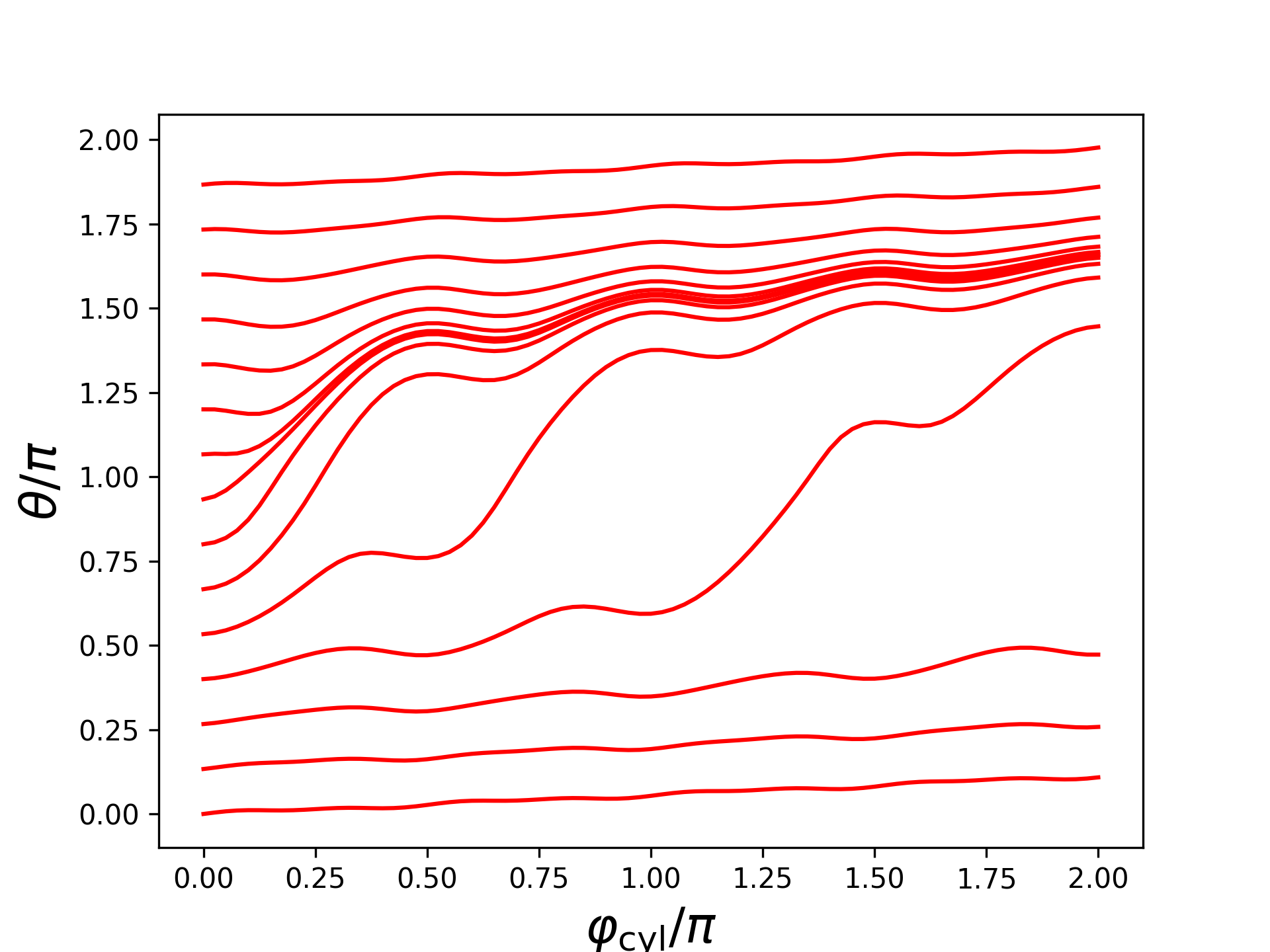}
\caption{Field lines in the $n_{fp}=4$ configuration depicted in Figure~\ref{fig:nfp4}. Left: Field lines on the torus and a cut-open torus where the toroidal angle increases to the right. Right: $\theta$ and $\phi_{\text{cyl}}$ coordinates of the field lines. Note how field lines get focused as they approach the ridge.}
\label{fig:nfp4_fieldline}
\end{figure}

\begin{figure}
\centering
\includegraphics[width=0.45\linewidth]{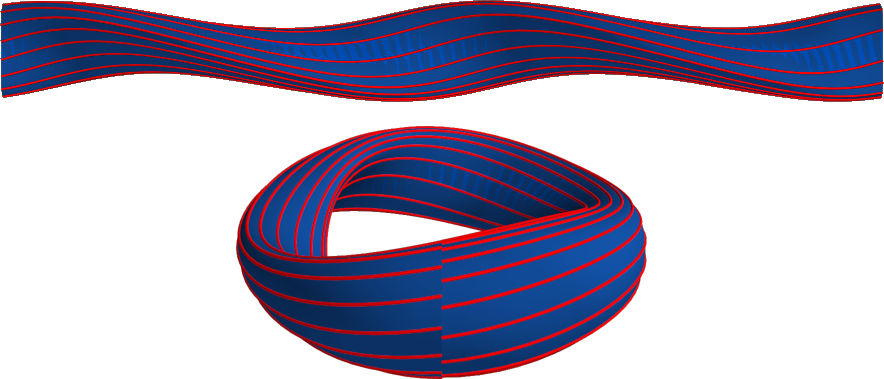}\includegraphics[width=0.45\linewidth]{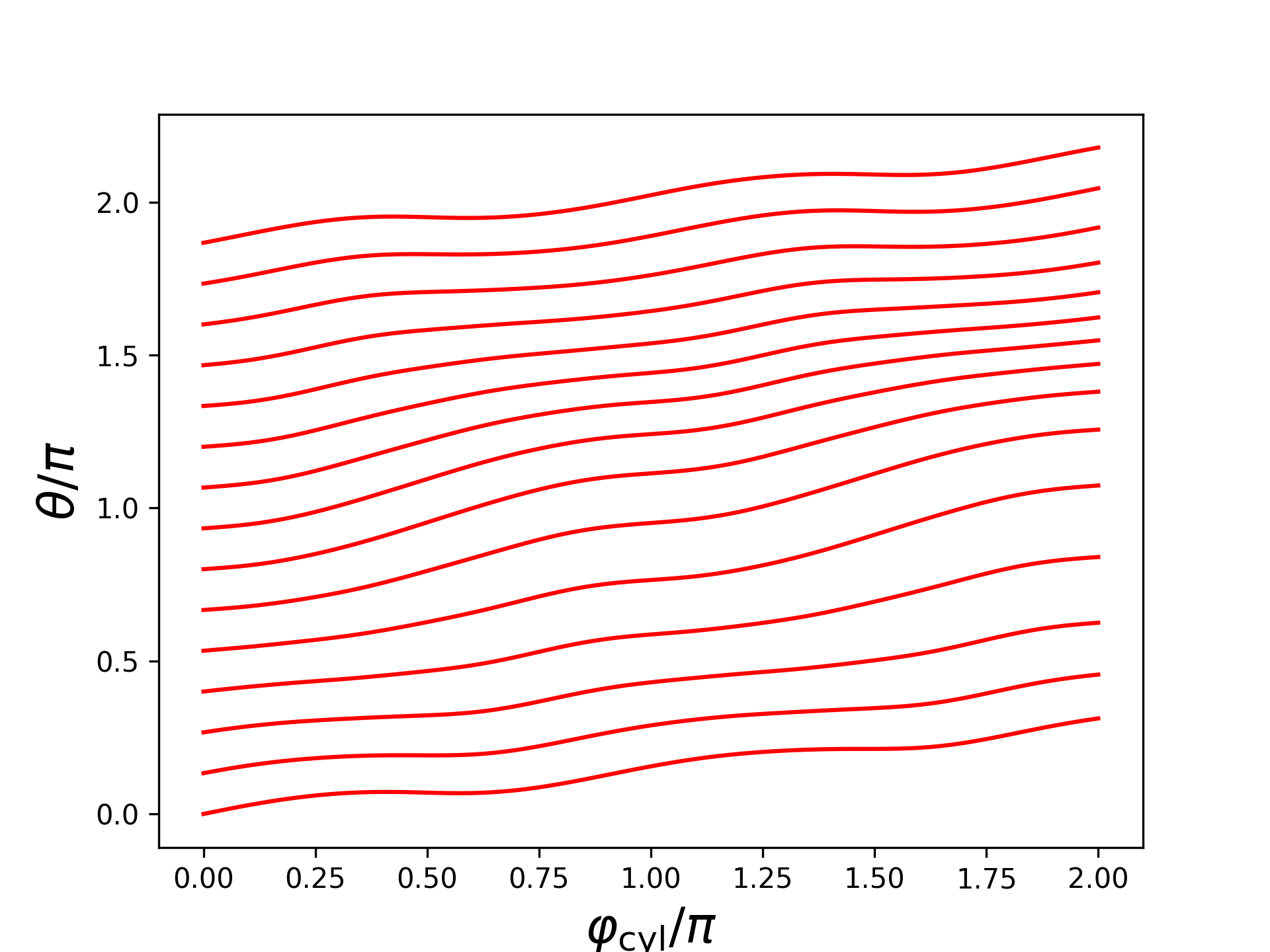}
\caption{Figure corresponding to Figure~\ref{fig:nfp4_fieldline} but for the  $n_{fp}=2$ configuration depicted in Figure~\ref{fig:nfp2}. Notice the absence of ridges and how the field lines thus do not get focused.}
\label{fig:nfp2_fieldline}
\end{figure}

\begin{figure}
\centering
\includegraphics[width=0.45\linewidth]{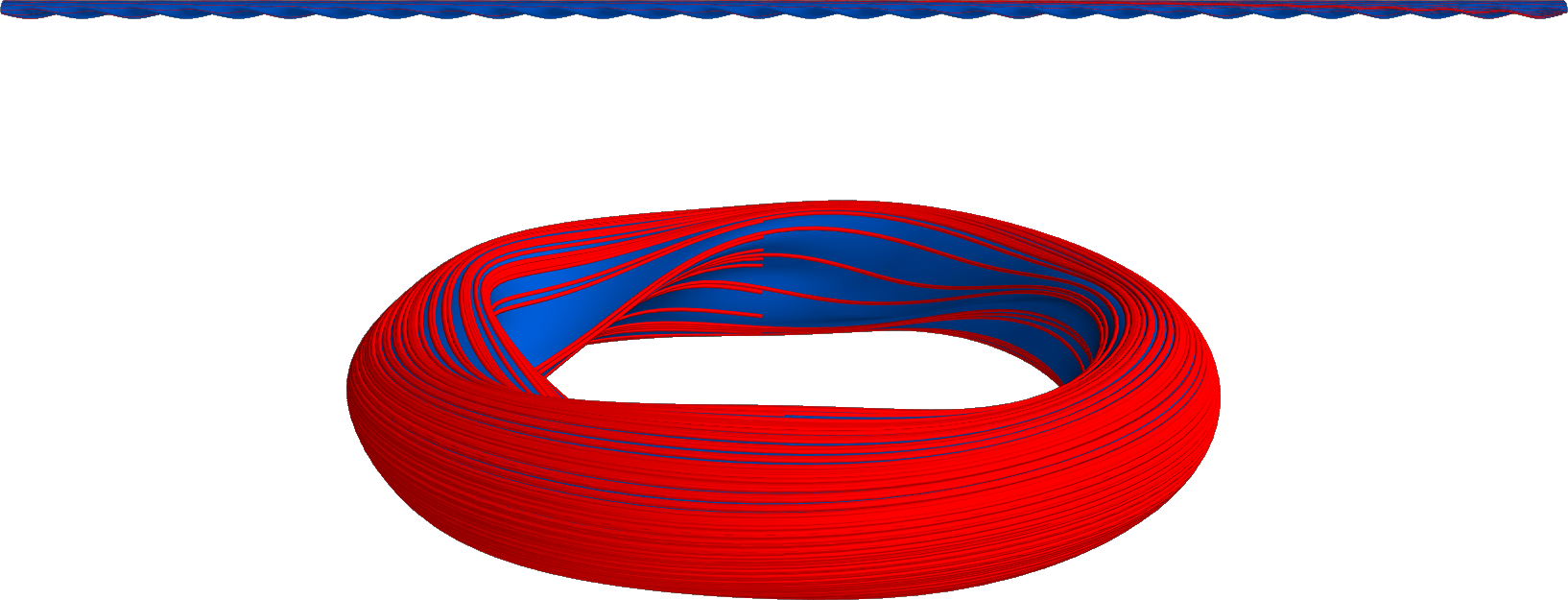}\includegraphics[width=0.45\linewidth]{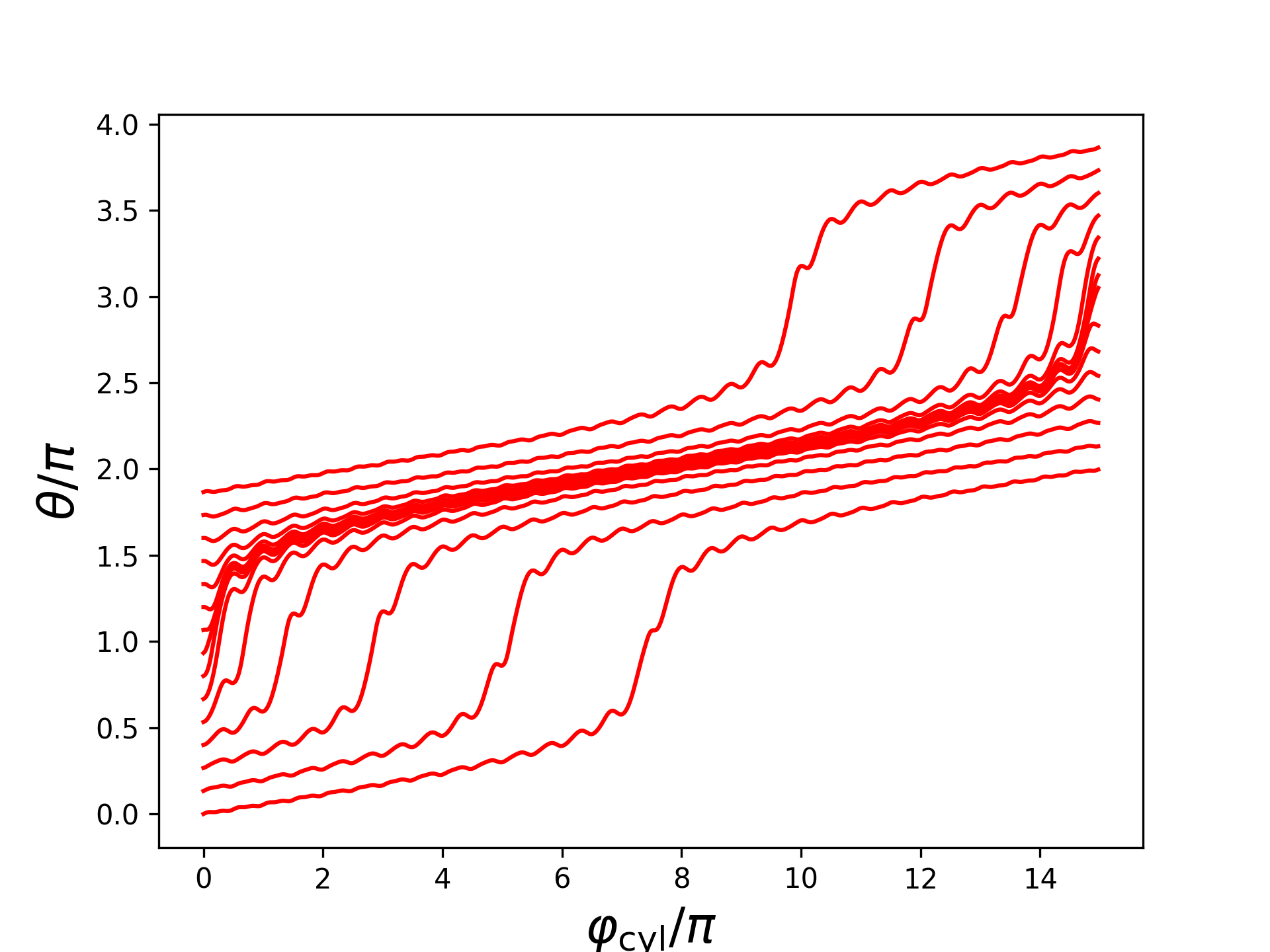}
\caption{Figure corresponding to Figure~\ref{fig:nfp4_fieldline} but for for field lines traced a toroidal distance of $2\pi/\iota$. At this point, the field lines have defocused and are again uniformly distributed.}
\label{fig:nfp4_fieldline_long}
\end{figure}

The fact that $|\dl \alpha|$ becomes large near the ridges means that a small physical displacement results in a large change in $\alpha$, resulting in field lines becoming compressed.
In Figure~\ref{fig:nfp4_fieldline}, we see how the ridges affect the field lines in the $n_{fp}=4$ configuration: an ensemble of field lines that is initially homogeneous in $\theta$ becomes focused approaching the ridge, as field lines are forced to go along the ridge. For comparison, the   $n_{fp}=2$ configuration, which lacks sharp ridges, is depicted in Figure~\ref{fig:nfp2_fieldline}.  After traversing a toroidal distance of $2\pi/\iota$, the field lines return to their originally uniform $\theta$ distribution, as can be seen in Figure~\ref{fig:nfp4_fieldline_long}.
The strong focusing along the ridge causes the field line to not sample the torus uniformly, with parts of the inboard side not being covered as densely. This could have major implications for the particle transport in devices with ridges, as well as present practical problems for gyrokinetic flux-tube simulations.
Note that the toroidal distance $2\pi/\iota$  corresponds to about $7.487$ full toroidal turns, which is not an integer. Thus, the field lines have not returned to their original position.

\begin{figure}
\centering
\includegraphics[width=\linewidth]{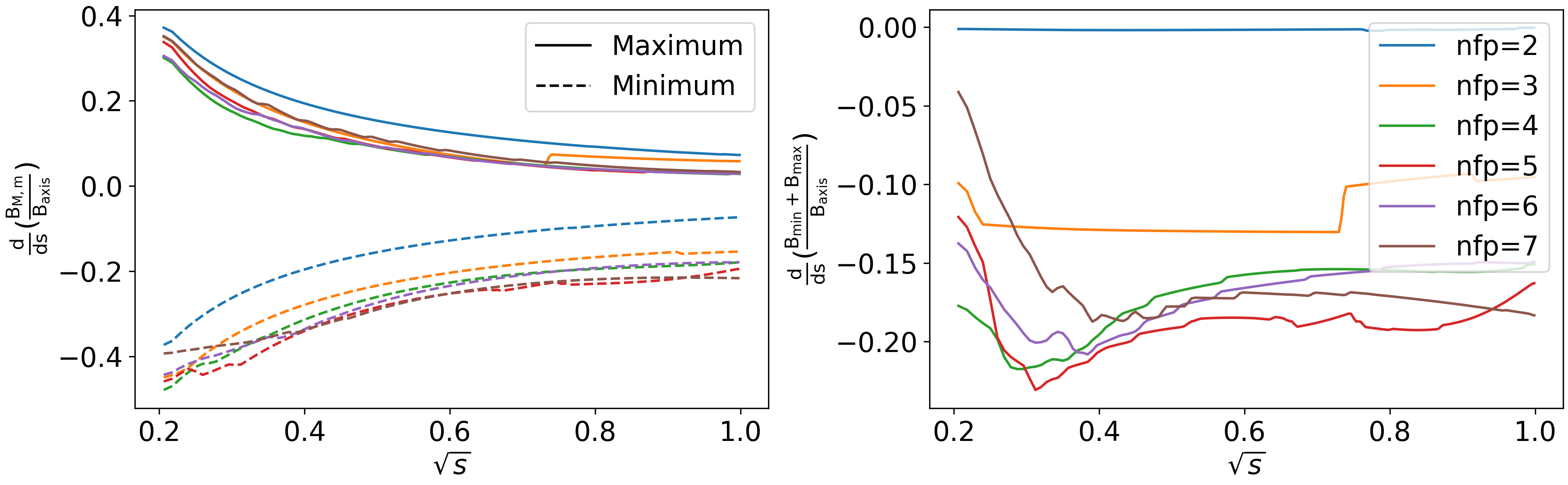}
\caption{We show how the derivatives of the maximum $B_M$, the $B_m$, and their sum vary with the square root of the normalized flux $\sqrt{s}$. We find that the magnitudes of $B_M'$ and $B_m'$ become small as we approach the outermost surface where the ridges are. Their sum, which must approximately vanish in the near-axis limit, is finite and negative for all $n_{fp}\geq 3$. The $n_{fp}=2$ retains the near-axis behavior. Thus, on the inboard side of the outermost surface near the ridges, $B\approx B_M$ approaches a constant.}
\label{fig:deviation_from_NAE}
\end{figure}

We demonstrate in figure \ref{fig:deviation_from_NAE} the significant deviation of the higher $n_{\text{fp}}\geq 3$ from the near-axis behavior where the sum of the maximum and the minimum of $B$ on a surface, $B_M+B_m$, is a constant to first order. For $n_{\text{fp}}\geq 3$ we find that, $B_M' + B_m' < 0$.
In contrast, $n_{fp}=2$ has a very near-axis-like behavior. Our findings for $n_{fp}\geq 3$ support the claim \citep{plunk2018,plunk2020_near_axisymmetry_MHD} that the perturbed near-axisymmetric QA devices fall outside the standard near-axis class of equilibrium, whereas $n_{fp}=2$ is in the near-axis class.

\begin{figure}
\centering
\includegraphics[width=0.6\linewidth]{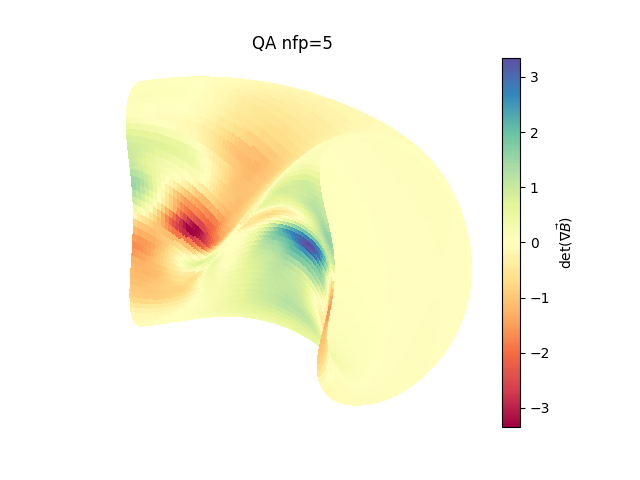}
\caption{A field period of the $n_{fp}=5$ configuration, where the surface is colored based on the local value of $\mbox{Det} (\dlBT)$. Near the ridge, this determinant goes to zero, confirming the predictions in Section.~\ref{sec:GradB_ridges}.}
\label{fig:cdet_nfp5_boundary}
\end{figure}

\begin{figure}
\centering
\includegraphics[width=0.6\linewidth]{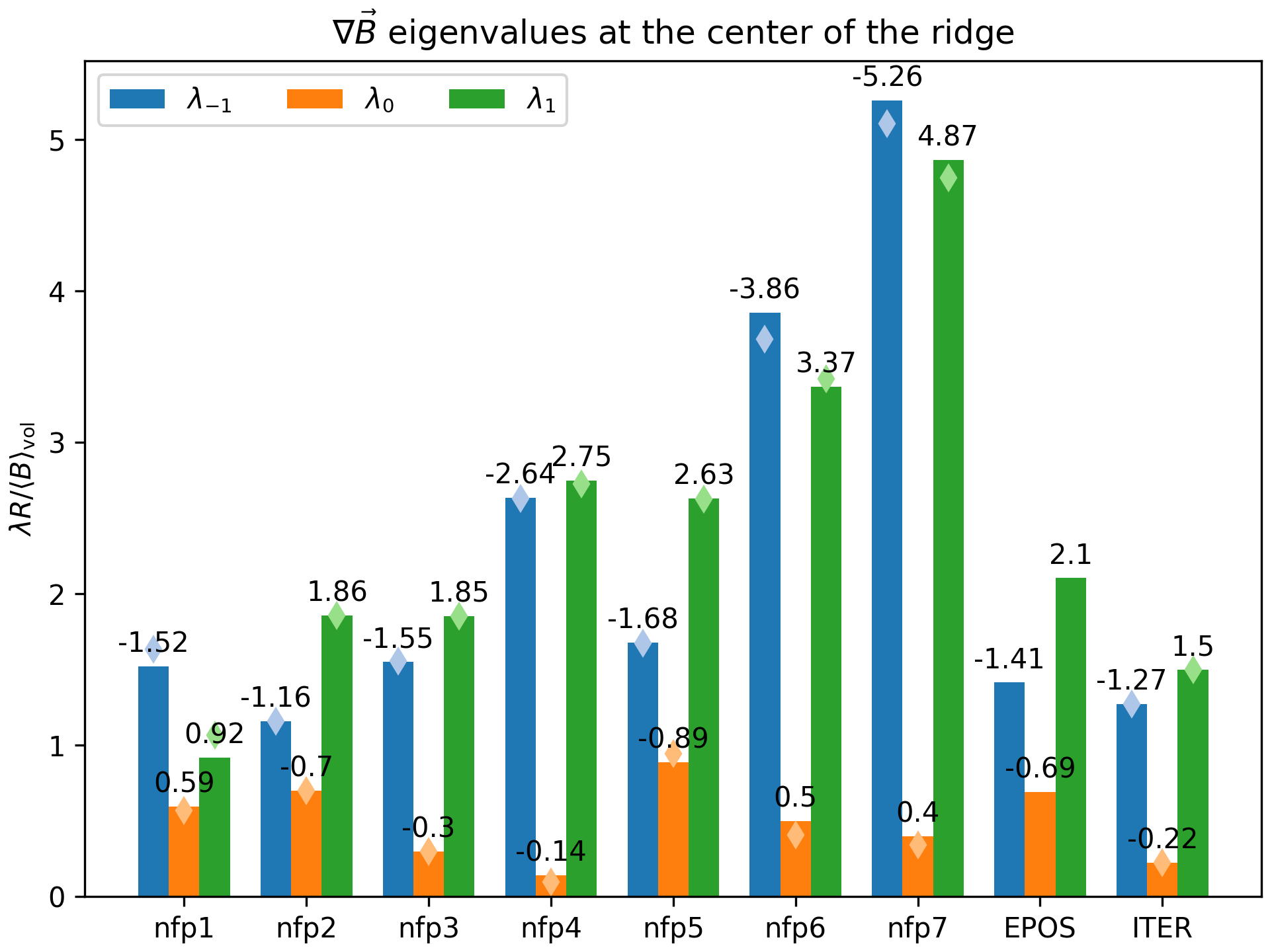}
\caption{Eigenvalues of the matrix $\dlBT$ for various configurations, calculated at the point at the center of the ridge. We used a high-resolution version of VMEC to obtain the bar-plot. Typical resolution used in optimization produces even larger eigenvalues for higher $n_{fp}$ values. Values calculated from even higher resolution equilibria are shown with diamonds when available, to verify convergence.}
\label{fig:barplot}
\end{figure}

Additionally, the quantity $\mbox{Det} (\dlBT)$ goes to zero on the ridges, as shown in Section.~\ref{sec:GradB_ridges} and in Figure~\ref{fig:cdet_nfp5_boundary}.
This can be rephrased in terms of the eigenvalues of the matrix $\dlBT$ becoming zero. The remaining eigenvalues become large due to the sharp real-space variations on ridges, with one eigenvalue being equal to the negative of the other eigenvalue since the sum of eigenvalues must be zero due to $\B$ being divergence-free. Accordingly, we will label the eigenvalues $\lambda_{-1}$, $\lambda_0$ and $\lambda_{1}$. These eigenvalues are shown in Fig.~\ref{fig:barplot}, for the various configurations presented in this paper.
For each configuration, the value is calculated at the point at the center of the ridge.
Note that eigenvalues are normalized to $R/\langle B\rangle_{\text{vol}}$ for each configuration, where $\langle B\rangle_{\text{vol}}$ is the volume-averaged magnetic field. Despite this normalization, there is a large variation in the size of the eigenvalues, since the curvatures of the ridges in stellarators are generally not related to the size of the major radius. This is different from the X-point of a tokamak, where toroidal symmetry forces one of the curvatures to scale with the major radius. This picture does not change if the eigenvalues are calculated on neighboring points along the ridge, i.e.~the choice of the central point on the ridge does not notably affect the results.

The same eigenvalue behavior also occurs near electromagnetic coils, which are used in the $L_{\nabla B}$ metric to quantify distance to coils. Since ridges and coils are not easily distinguishable by this metric, we expect the correlation between $L_{\nabla B}$ and coil-plasma distance to be poorer for configurations with ridges.

\begin{figure}
\centering
\includegraphics[width=0.6\linewidth]{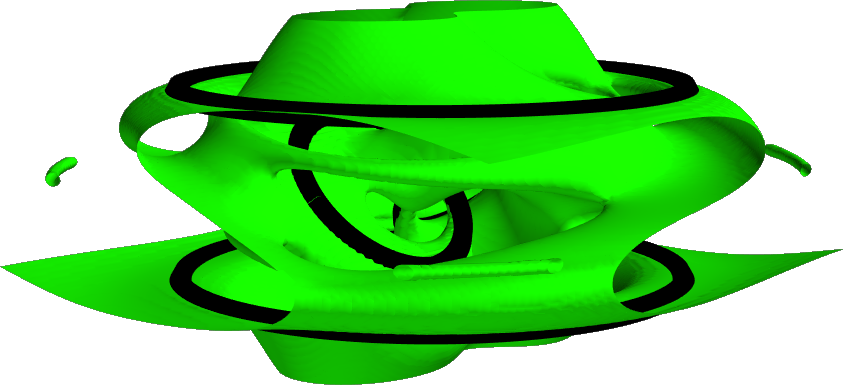}
\caption{Surface at which Det$(\dlBT)=0$ for a stellarator with 4 planar coils (the CNT experiment \citep{CNT2002}). In general, the Det$(\dlBT)=0$ surface is very complicated, but all the coils must lie on the surface.}
\label{fig:cabbage}
\end{figure}

In Fig.~\ref{fig:cabbage}, we show a set of electromagnetic coils, plotted on top of the $\mbox{Det} (\dlBT)=0$ surface, indicating where one of the eigenvalues is zero. As expected, the coil currents lie on the $\mbox{Det} (\dlBT)=0$ region. Based on this hypothosis, it might be possible to derive a relation between ridge and coil locations. However, as seen in the figure, the $\mbox{Det} (\dlBT)=0$ region is extremely complicated, spanning basically all space. In light of this complication, deriving such a relation will be left for future work.

\begin{figure}
\centering
\includegraphics[width=0.60\linewidth]{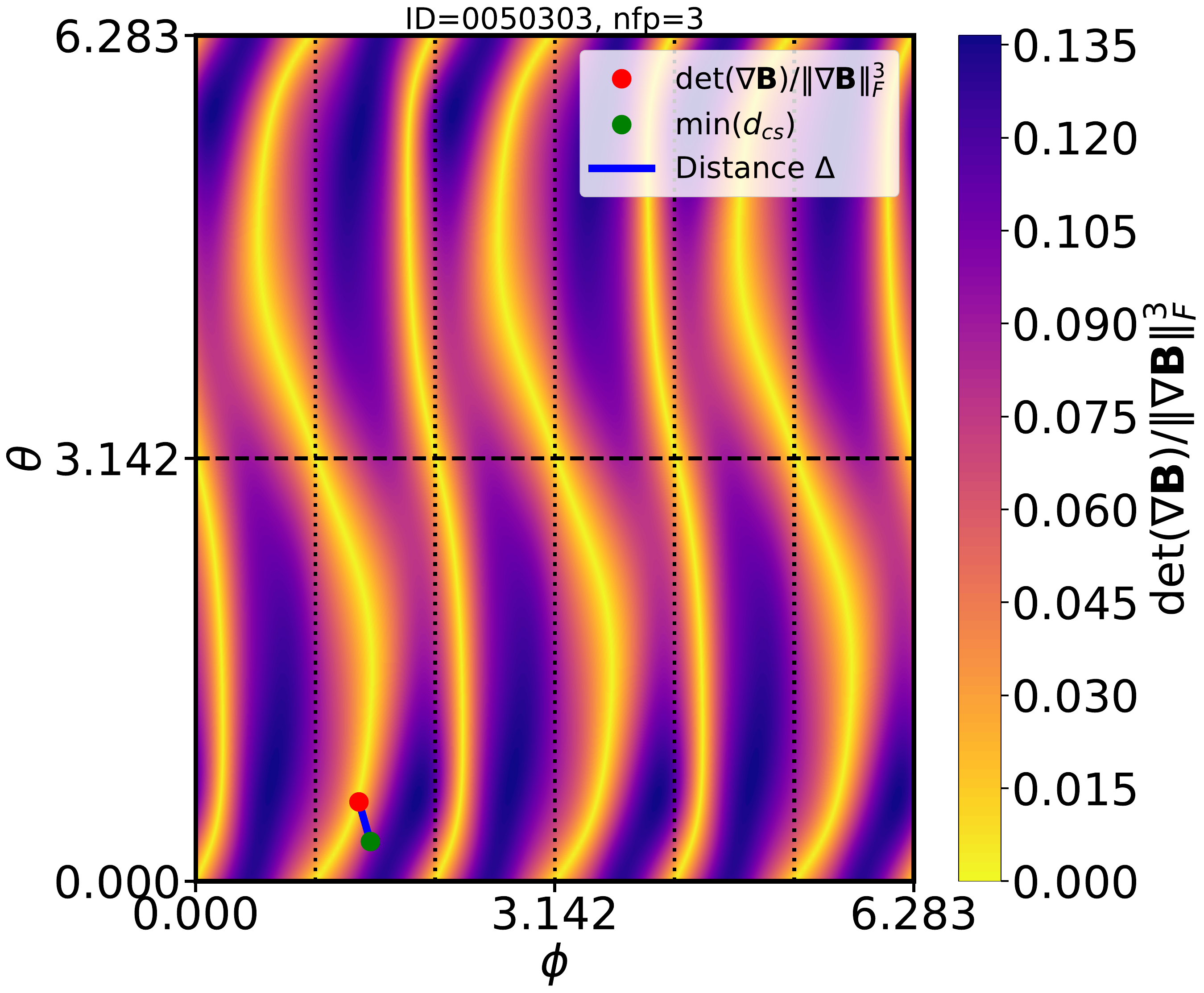}
\caption{Effective distance $\Delta$ evaluated on the last closed flux surface for an example equilibrium, which is ID=0050303 in the QUASR dataset. The location of the minimum coil surface distance is depicted with a green dot. The red dot is the closest point to the green dot where $\mbox{Det} (\dlBT)/\|\nabla{\mathbf{B}}\|_F^3$ is in the bottom 0.5\%. The blue line depicts the distance $\Delta$, which is 0.095 for this equilibrium.}
\label{fig:kappel_Delta}
\end{figure}

\begin{figure}
\centering
\includegraphics[width=0.6\linewidth]{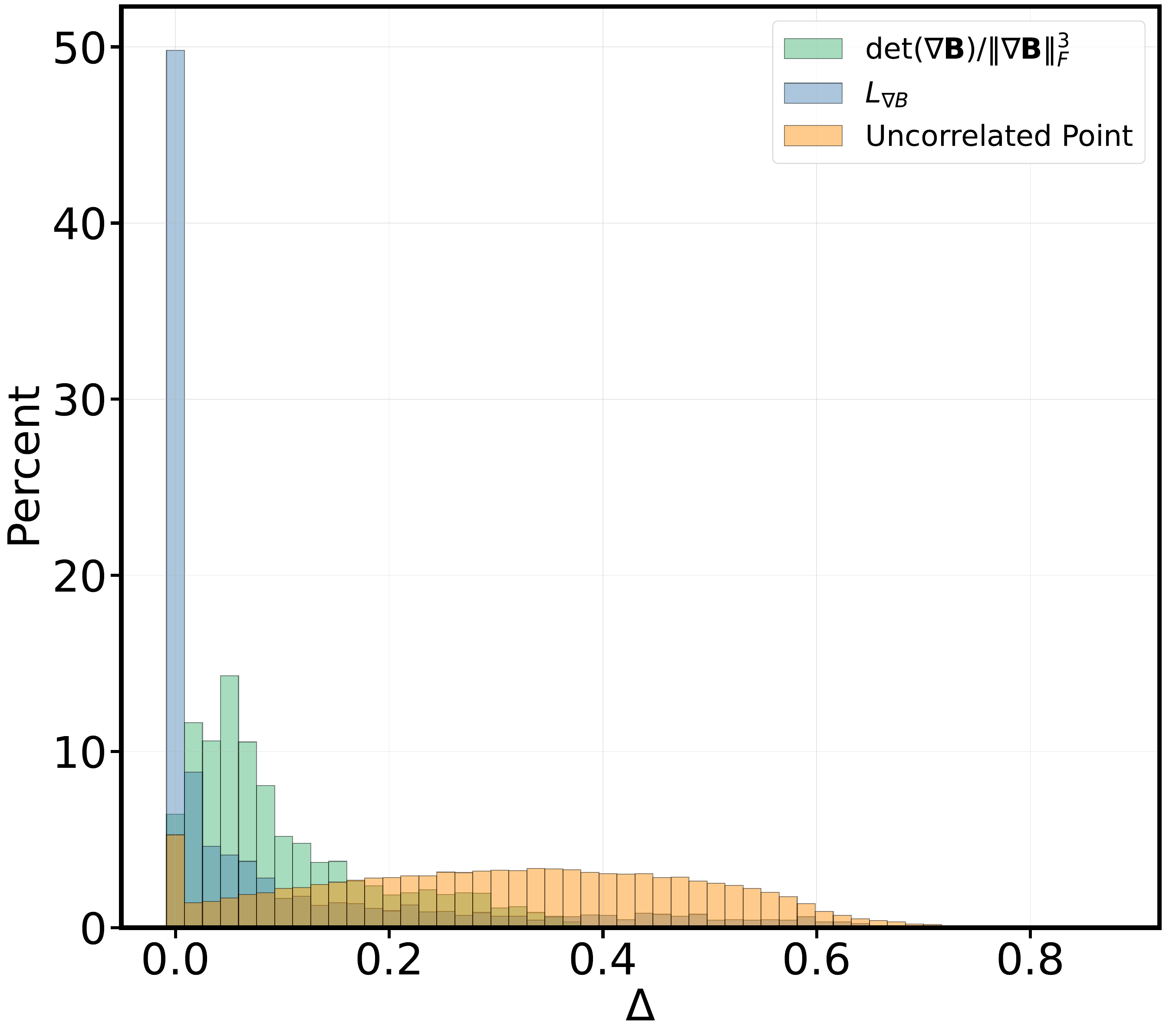}
\caption{A histogram depicting the effective distance $\Delta$ for $L_{\nabla B}$,  $\mbox{Det} (\dlBT)/\|\nabla{\mathbf{B}}\|_F^3$, and an uncorrelated point.}
\label{fig:kappel_hist}
\end{figure}

Since $\mbox{Det} (\dlBT) = 0$ where the coils lie, $\mbox{Det} (\dlBT)$ it may be possible to use $\mbox{Det} (\dlBT)$ as an alternative to the $L_{\nabla B}$ metric.
To test this, we analyzed a dataset of 3027 equilibria and coils from the QUAsisymmetric Stellarator Repository (QUASR) \citep{giuliani_quasr_2024}. 
This subset of QUASR is the same as the one used by \citet{kappelHowDoesMagnetic2026}.

For each configuration in this dataset,
we calculate the coil surface distance and $\normdet{} \equiv \mbox{Det} (\dlBT)/\|\nabla{\mathbf{B}}\|_F^3$, for each point on the last closed-flux surfaces. The normalization of  $\mbox{Det} (\dlBT)$ makes the quantity dimensionless and removes the dependence of the size of $|B|$.
We also calculated the $L_{\nabla B}$-metric in order to compare the results.

We then extract the $(\theta,\phi)$-locations corresponding to the bottom 0.5\% range of $L_{\nabla B}$ and $\normdet{}$, respectively. Denote these sets of points as $\{\normdet{} < 0.5\%\}$ and $\{L_{\nabla B} < 0.5\%\}$.
From these sets of points, we calculate $\Delta =  \sqrt{\delta\theta ^2+ (n_{fp}\delta\phi)^2}/(\sqrt{2}\pi)$, where $\delta\theta$ and $\delta \phi$ are the minimum differences in $\theta$ and $\phi$ between the location of the minimum coil-surface and any location in $\{\normdet{} < 0.5\%\}$ or $\{L_{\nabla B} < 0.5\%\}$. The smaller $\Delta$ is, the closer our set of points is to the point with the minimum coil-surface distance, with $\Delta = 0$ for configurations where this point is within the sets.
The $\Delta$ metric is normalized to account for field period and stellarator symmetry so that the farthest possible distance between two points is $\Delta=1$.  An example of this metric is shown in Figure~\ref{fig:kappel_Delta}. 


For comparison, we also evaluated the $\Delta$ distance between uncorrelated points on the last-closed flux surface. Specifically, we pick two random points on the last-closed flux surface and calculate the minimum distance between any point with a small radius around point 1 and any point within a radius of point 2, where the radii are chosen to cover 0.5\% of the last-closed flux surface.
This was done 100000 times. This is an adaptation of the method used in \citep{kappelHowDoesMagnetic2026}.


The results of this analysis are shown in Figure \ref{fig:kappel_hist}. We find that the effective distance $\Delta$ for $\mbox{Det} (\dlBT)/\|\nabla{\mathbf{B}}\|_F^3$ (which has a median of 0.078 and a mean of 0.108), which is larger than the $\Delta$ for $L_{\nabla B}$ (which has a median of 0.017 and a mean of 0.095). Thus, the new metric performs slightly worse than $L_{\nabla B}$ .
However, both outperform $\Delta$ calculated from the randomized points (which has a median of 0.434 and a mean of 0.428), meaning both are encoding information about the location of where the minimum coil surface distance is located on the last closed flux surface. This is not surprising; because $\mbox{Det} (\dlBT)/\|\nabla{\mathbf{B}}\|_F^3$ is small both near the coils and near the ridge, it is a necessary but not sufficient condition for determining coil location. The same is also true $L_{\nabla B}$, and it is not clear why it ends up being a better predictor of coil location.

\section{Conclusions}
\label{sec:conclusions}

In conclusion, we have analyzed the behavior of magnetic field lines and the field strength of vacuum quasisymmetric fields on a surface far from the magnetic axis. Formulating quasisymmetry as an eikonal equation, we have shown that in compact QA, sharp ridges tend to form on the inboard side where the Gaussian curvature is typically negative. The magnetic field lines in the vicinity of the ridges are essentially straight, with far-reaching consequences, discussed below. 

Firstly, the straightness of the magnetic field lines immediately puts constraints on the field strength.  Since the field strength is dictated by vacuum QA, the implications of straight field lines have observable consequences, such as the radial derivative of the maximum $B$ is approximately zero on the outer surface. Also, straight lines tangential to a surface can only exist on the negative Gaussian curvature side. We have obtained an exact relation between the 3D $\dlBT$ tensor and the 3D Hessian of the flux surface $\dl\dl\psi$ and demonstrated how quasisymmetry plays a crucial role in ensuring Det$\dlBT=0$ near ridges.  

Secondly, we have argued that 3D ridges are very different from regular X points associated with rational surfaces. We have quantified the difference in terms of the eigenvalues of the $\dlBT$ tensor normalized by the major radius. For regular X points, we have shown that the eigenvalues of $\dlBT$ normalized by the major radius are $O(1)$, whereas for sharp 3D ridges, they are either zero or much larger in magnitude than unity. The largeness of the eigenvalue corresponds to the variation of the Gaussian curvature across the ridge. We note here that the large change in curvature across a ridge is not the mathematically accurate definition of a ridge, since the ridge point is defined by an extremum of a principal curvature along the same principal direction. For our case, this translates to $\del_\ell \kappa_n=0$ since the principal direction and the principal curvature of the surface locally coincide with the field line direction and the normal curvature $\kappa_n$ close to $B=B_M$. It happens that a large variation in the other principal curvature $\kappa_2$ across the ridge is beneficial for a better alignment of the ridge with the magnetic field, but not strictly necessary. However, in terms of numerical computation, detecting the large $\kappa_2$ might be easier for a ridge detection.

Thirdly, we have shown that filamentary coils must lie on $\mbox{Det}\dlBT=0$ surfaces. The implication of this is that the direction of the coil filament must align with the null direction of the $\dlBT$ tensor. The other eigenvalues are large and of opposite signs. The similarity with the case of the 3D ridges is striking and indeed shows that ridges and coils are both associated with the null space of the $\dlBT$ tensor. Thus, it would be harder to generate coils for configurations with sharp 3D ridges due to the strong surface shaping.  The coil-to-surface distance and coil non-planarity are both affected by surface shaping. We have defined a new coil-to-surface metric, $\mbox{Det} (\dlBT)/\|\nabla{\mathbf{B}}\|_F^3$, which approximates where (but not to what extent) the coils are close to the plasma. While only slightly less accurate than the $L_{\nabla B}$ metric, it significantly outperforms the null case. 

Finally, we have shown that field line torsion varies rapidly near the ridges. The ridge itself is almost straight with zero torsion but the neighboring field lines have large torsion, essentially proportional to the square root of the negative Gaussian curvature. In the simple model where the winding surface is a flux surface \citep{rodriguezSengupta2026WindingCoil}, the non-zero torsion or non-planarity of $\B$ directly constrains the non-planarity of the coils since the current potential $\bm{K}=\nh \times \B$. Further, in the light of recent work on QI by \cite{warmer_pavone2026Nonplanarity}, the large variation in the torsion of the field lines due to the large variation of the Gaussian curvature near the ridge has a direct relevance for coil non-planarity. We showed that near the sharp ridges, the field lines locally align with the principal direction with the smaller principle curvature. Since the field lines must also lie on flux surfaces, the principal directions must twist sharply in the vicinity of the ridges. Thus, the the flux surface itself need to bend at the ridges and then twist sharply away from the ridge to accommodate the field lines with large torsion. The sharp twist can contribute to large non-planarity of coils in QS much like in QI. Further work along these lines is ongoing.

\section*{Acknowledgements}
The authors would like to thank E. Rodriguez, M. Landreman, F. P. Diaz, A. Bader, J. W. Burby, N. Cao, S. Naik, T. Klotz, J. Meiss, J. L. Velasco, S. Patil, D. Seidita, C. B. Smiet, R. Davies, S. Henneberg, G. Plunk, J. Loizu , and E. J. Paul for stimulating discussions and helpful suggestions.

\section*{Funding}

This research was supported by a grant from the Simons Foundation/SFARI (560651, AB) and the Department of Energy Award No. DE-SC0024548 (until March 31, 2025). Some computations were performed on the HPC system Viper at the Max Planck Computing and Data Facility (MPCDF).

\section*{Declaration of interests}
The authors report no conflict of interest.

\section*{Data availability}
The data that support the findings of this study are available from the corresponding author upon reasonable request.

\appendix

\section{ The formal mathematics underlying ridges \label{app:ridge_caustics}}
We now discuss the formal mathematical definition and several important geometrical properties of the ridges. We also emphasize its connection with caustics and envelopes that play a fundamental role in geometrical optics. 

\subsection{Brief summary of the relevant ideas from differential geometry of surfaces  \label{app:surf_theory}}

We need a few essential definitions from the theory of surfaces in differential geometry \citep{struik2012lectures,koenderink1990solid,bruceGiblin1992curves} before we can proceed. 

On a smooth surface (say of constant flux $\psi$), parametrized by local coordinates $\bm{r}=(u,v)$ at each point $\bm{P}$, we can define a unit normal $\bm{N}$, which coincides with $\dl\psi/|\dl\psi|$ at a non-critical point where $|\dl\psi|\neq 0$. The map from a point to its normal is the \textit{Gauss map}. The first and second fundamental forms of the surface are defined as
\begin{align}
    \mathrm{I}\equiv d\bm{r}\cdot d\bm{r}=g_{ij}dr^i dr^j, \quad \mathrm{II}\equiv -d\bm{r}\cdot d\bm{N}.
    \label{eq:first_second_fund_form}
\end{align}

The \textit{shape operator} can be obtained through the $(u,v)$ derivatives of $\bm{N}$ as $S(\bm{r}_{,u})=-\bm{N}_{,u}$ and $ S(\bm{r}_{,v})=-\bm{N}_{,v}$. The eigenvalues and eigenvectors of the shape operator gives the \textit{principal curvatures} $(\kappa_1,\kappa_2)$ and \textit{principal directions} $(\bm{e}_1,\bm{e}_2)$.
The \textit{Gaussian curvature} $K_G\equiv\kappa_1\kappa_2$, and the \textit{mean curvature} $H=(\kappa_1+\kappa_2)/2$ are the determinant and half the trace of the shape operator. 

The \textit{parabolic line} with $K_G=0$ when either $\kappa_1=0$ or $\kappa_2=0$,  is the surface analogs of inflection points. It separates the hyperbolic $K_G<0$ region and the elliptic $K_G>0$ region. The Gauss map has singularities on the parabolic line as the determinant of the shape operator vanish along the line.  

The \textit{asymptotic direction} consists of directions along which $\mathrm{II}$ vanish identically. \textit{Asymptotic curves} are curves with tangent in the asymptotic direction, for which the normal curvature is always zero. Asymptotic curves exist only in $K_G\leq 0$ regions. For $K_G<0$, a pair of asymptotic curves exist, while on the parabolic line $K_G=0$, there is only one. An \textit{asymptotic inflection point} is a point where the asymptotic curve has zero geodesic curvature. Since by definition, asymptotic curves have zero normal curvatures, at the inflection points, they are straight lines. Thus, any straight line on a surface must be an asymptotic curve with degenerate inflection points.  

At points called \textit{umbilical points}, the two principal curvatures are equal. The ratio of the two principal curvatures typically scales as the aspect ratio of the device. Since in a stellarator, the aspect ratio is always larger than one, even for the most compact designs, the principal curvatures are almost always unequal. Thus, umbilical points are rare and will be ignored henceforth. 

\subsection{Formal definition of ridges and ridge points
\label{app:ridge_def_n_caustics}}
Formally, ridges are defined as a collection of ridge points at which one of the principal curvatures attains an extremum along the corresponding principal direction \citep{porteous2001geometric,koenderink1990solid}. Thus, curves along which $\bm{e}_1\cdot\dl \kappa_1=0$ or $\bm{e}_2\cdot\dl \kappa_2=0$ defines ridges. Being an extrema of a principal curvature, ridge points are analogs of \textit{vertices} of planar curves \citep{bruceGiblin1992curves}, where the curvature has a local extremum. 

Closely associated with planar vertices is the concept of an \textit{evolute} \citep{arnold1990huygens}, which is defined as the envelope to the normals to the curve. In the case of a planar curve, the evolute is obtained by joining all points that are located at normal distances from the curve equal to the radii of curvature. Evolutes have cusp singularities near the vertices. In other words, on a planar curve, the vertices are also points  that are closest to the cusp singularity of the evolute to the curve.

The surface analog of an evolute is obtained from the loci of the two principal curvatures. Hence, there are typically two of these surfaces, which are also called caustic surfaces or focal surfaces \citep{porteous2001geometric,koenderink1990solid}. 

The caustic surfaces can have singularities such as cusp points or cuspidal edges called \textit{ribs} \citep{porteous2001geometric}. At these singularities, two or more normals meet. Ridge points are the surface analogs of planar curve vertices in the sense that they are the closest points on the surface to the cusp singularities of the caustic surface, the distance between them scaling inversely with the curvature. 

\subsection{Connection with caustics}

 Caustics play an important role in optics \citep{berry1976waves_and_Thom,berry1981Les_houches_singularities}, and are defined as surfaces where two or more rays meet such that the formal geometrical optics breakdown leads to formally infinite intensity of light. The meeting of rays in a caustic signals a characteristic degeneracy that play a crucial role in the theory of rainbows and ship wakes \citep{BerryHowlsNIST}. 
 
 We illustrate the underlying mechanism for the formation of caustics in optics with the example of asymptotic approximations of oscillatory integrals typical in WKB theory. The method of stationary phase allows one to expand the oscillatory integral near a critical point up to quadratic terms and evaluate the resulting Gaussian integral analytically. As a result, we obtain the square root of the second derivative of the phase in the denominator. However, if the critical point is degenerate, the second derivative of the phase goes to zero. This happens precisely near the turning points in WKB theory where incident, reflected and transmitted ``rays" can coexist. Near such degenerate critical points, the amplitude of the Gaussian approximation formally blows up, which is the essence of a caustic formation. Near turning points, we need the third order terms in the phase to obtain a finite approximation of the integral in terms of Airy functions.

The mathematical machinery needed for treating caustics self-consistently is catastrophe theory \citep{arnold1986catastrophe, berry1976waves_and_Thom,berryUpstill1980iv,postonStewart2014catastrophe}. Due to the association with caustics, the description of ridges requires similar tools \citep{porteous2001geometric,koenderink1990solid}, which we discuss in further details in our forthcoming paper. In the following, we briefly summarize relevant known results from \citet{Banchoff_cusp_of_gauss}. 

The singularities of the Gauss map that happens on the parabolic curves have deep connection with ridges. Banchoff, Gaffney, and McCrory \citet{Banchoff_cusp_of_gauss} shows that the generic singularities of the Gauss map are \textit{folds and cusps}. Typically, lines of curvature $\bm{e}_i$ cross asymptotic lines transversely at folds. However, at the cusp they are tangential. Gaussian cusps occur precisely where the ridge lines cross the parabolic curve with vanishing principal curvature. The inflection points on the asymptotic lines also accumulate near the cusps of Gauss, essentially requiring the asymptotic lines to be straight. Asymptotic lines approach the parabolic curve and are tangent to it at the cusp. The parabolic curves thus serves as a local envelope to all the asymptotic curves near the cusp of Gauss.

\subsection{Implications for stellarators \label{app:surf_theory_ridges_B}}

We now emphasize certain key differences between regular X point of a tokamak or a resonant island and the 3D ridges in stellarators that arises due to caustics. 

Magnetic field lines are essentially straight and therefore asymptotic lines in the vicinity of the ridges, as we have shown using the optical analogy. They approach the parabolic line but can not cross since $K_G$ mus be negative and so stay tangential. Thus, the parabolic curves form an envelope along the ridge. As we saw in the last Section, this configuration is the signature of the cusp of Gauss. The essential point to note here is that the vanishing of the field line normal on the inboard side with $B=B_M, K_G<0$ is highly nontrivial and does not happen in regular X points.  

Next, we note that in both these cases, $|\dl\psi|\to 0$, implying that the X-point or the ridge point is a fixed point. For regular X-points, the fixed point is not degenerate since the Jacobian of the second order derivatives of $\psi$ does not vanish. For instance, from the normal form of $\psi$ near a regular X point, $\psi=B_0(\ell)(y^2-x^2)/2$, yields a non-vanishing Jacobian $|\del(\psi_{,x},\psi_{,y})/\del(x,y)|=B_0^2\neq 0$. However, the form $\psi=B_0(y^2-x^3)$ that can occur near cusps \citep{brown2025Palumbo} is degenerate.

\section{Differential geometric characteristics of the sharp ridges}\label{app:Diff_geo_proof}

We explore the differential geometric constraints of having a sharp ridge on a flux surface in this Appendix. We start with a parametric representation of the flux surface such that the parametric curves are orthogonal (p. 130 \cite{struik2012lectures}). The first fundamental form is
\begin{align}
    \mathrm{I} = E\,du^2 + G\,dv^2, \qquad (F = 0),
\end{align}
with arc-length parameters
\begin{align}
  s_1 = \int\!\sqrt{E}\,du, \qquad s_2 = \int\!\sqrt{G}\,dv.
\end{align}
The geodesic curvatures of the coordinate curves are
\begin{subequations}
    \begin{align}
  \kappa_{g_1} &= \bigl(\kappa_g\bigr)_{v=\mathrm{const}}
    = -\frac{1}{2}\frac{E_v}{E\sqrt{G}}
    = -\frac{d}{ds_2}\ln\sqrt{E}, \label{eq:kappag1} \\
  \kappa_{g_2} &= \bigl(\kappa_g\bigr)_{u=\mathrm{const}}
    = +\frac{1}{2}\frac{G_u}{G\sqrt{E}}
    = +\frac{d}{ds_1}\ln\sqrt{G}.\label{eq:kappag2}
\end{align}
\label{eq:kappag1_2}
\end{subequations}

Taking lines of curvature as coordinates ($F = f = 0$), both fundamental forms
can be diagonalized (p. 113 \cite{struik2012lectures})
\begin{align}
    \mathrm{I} = E\,du^2 + G\,dv^2, \qquad
  \mathrm{II} = e\,du^2 + g\,dv^2.
\end{align}
The principal curvatures are
\begin{align}
    \kappa_1 = \frac{e}{E}, \qquad \kappa_2 = \frac{g}{G}.
\end{align}
In these coordinates, the Codazzi--Mainardi Equations are
\begin{equation*}
  \dfrac{1}{\sqrt{G}}\dfrac{\partial\kappa_1}{\partial v}
    = +\dfrac{(\kappa_2-\kappa_1)}{2}\dfrac{E_v}{E\sqrt{G}}, 
  \qquad
  \dfrac{1}{\sqrt{E}}\dfrac{\partial\kappa_2}{\partial u}
    = -\dfrac{(\kappa_2-\kappa_1)}{2}\dfrac{G_u}{G\sqrt{E}}.
\end{equation*}
Rewriting in arc-length coordinates ($\partial_{s_1} = \tfrac{1}{\sqrt{E}}\,\partial_{u}, \partial_{s_2} = \tfrac{1}{\sqrt{G}}\,\partial_{v}$\,), we obtain
\begin{align}
  \frac{\partial\kappa_1}{\partial s_2}
    = -(\kappa_2-\kappa_1)\,\kappa_{g_1}, \quad
  \frac{\partial\kappa_2}{\partial s_1}
    = -(\kappa_2-\kappa_1)\,\kappa_{g_2}.
    \label{eq:Dskappa12}
\end{align}
The Gaussian Curvature ($K_G$) for orthogonal coordinates is
\begin{align}
  K_G = -\frac{1}{\sqrt{EG}}\!\left[
    \frac{\partial}{\partial u}\!\left(\frac{\partial\sqrt{G}}{\partial s_1}\right)
    + \frac{\partial}{\partial v}\!\left(\frac{\partial\sqrt{E}}{\partial s_2}\right)
  \right].
\end{align}
Using $\partial_u = \sqrt{E}\,\partial_{s_1}$,\; $\partial_v = \sqrt{G}\,\partial_{s_2}$ and \eqref{eq:kappag1_2}, the expression for $K=\kappa_1 \kappa_2$ reduces to
\begin{align}
   K_G
  &= -\frac{1}{\sqrt{G}}\,\partial_{s_1}\!\left(\sqrt{G}\,\kappa_{g_2}\right)
    + \frac{1}{\sqrt{E}}\,\partial_{s_2}\!\left(\sqrt{E}\,\kappa_{g_1}\right)\\
   &=\partial_{s_2}\kappa_{g_1} - \kappa_{g_1}^2
       - \bigl(\partial_{s_1}\kappa_{g_2} + \kappa_{g_2}^2\bigr)
       \label{eq:KG_exp}
\end{align}
 
Until now, everything follows from basic differential geometry in local principal curvature coordinates. Now, we focus on the ridge conditions. Ridges of the principal curvatures are defined by
\begin{align}
    \dfrac{\partial\kappa_i}{\partial s_i} = 0, \quad \dfrac{\partial^2\kappa_i}{\partial s_i^2}\neq 0, \quad i=1,2.
\end{align}

We now utilize the sharpness of the ridge by assuming $\kappa_2\gg \kappa_1$. We proved (using continuity) that field lines must
align with the ridges. This implies that as $\kappa_2\to\infty$, the field lines must align with the $u$-lines, which are lines of principal curvature $\kappa_1$. Thus, $s_1$ is along $\th=\B/B$ and $s_2$ is along $\bh=\th\times \nh, \nh =\nabla \psi/|\nabla \psi|$. The alignment of the field lines with the sharp ridges continue as long as the ridge stays sharp.
 
Since $\kappa_2\gg\kappa_1$, equations \eqref{eq:Dskappa12} simplify to

\begin{align}
  \frac{\partial\kappa_1}{\partial s_2} = -\kappa_2\,\kappa_{g_1},
    \quad
  \frac{\partial\kappa_2}{\partial s_1} = -\kappa_2\,\kappa_{g_2}.
   \label{eq:dsikappai}
\end{align}
Since $\kappa_2\gg\kappa_1$, 
\begin{align}
  |\kappa_{g_1}| \to 0,
  \label{eq:kappag1_0}
\end{align}
as long as $\del_{s_2}\kappa_1$ is bounded. The boundedness assumption of $\del_{s_2}\kappa_1$ will be later demonstrated to be self-consistent. The vanishing geodesic curvature of $\B$ means the field lines are geodesics on the flux. Therefore,
surface:
\begin{align}
  \kappa_{g_1}\to 0
  \;\Longrightarrow\; \BDB = 0
  \;\Longrightarrow\; B\;\text{is constant along }s_1.
\end{align}
Combined with quasi-axisymmetry (QA), $\BD B = 0$ implies $B$ is constant along two directions. 
 
Next, we analyze the relation between $\kappa_{g_2}$ and $\kappa_2$,
\begin{align}
  -\kappa_{g_2}
  = \frac{1}{\kappa_2}\frac{\partial\kappa_2}{\partial s_1}
  = \partial_{s_1}\ln\kappa_2.
  \label{eq:Ds1logkappa2}
\end{align}
In general for $\kappa_{g_2}\neq 0$, we can integrate \eqref{eq:Ds1logkappa2} to get 
\begin{align}
  \kappa_2 = \kappa_{20}\exp\!\left(-\!\int\!\kappa_{g_2}\,ds_1\right).
\end{align}

 
We mow make a local approximation near a localized maximum of a $\kappa_2$-ridge, such that both first derivatives vanish of $\kappa_2$
\begin{align}
  \frac{\partial\kappa_2}{\partial s_1} = 0, \qquad
  \frac{\partial\kappa_2}{\partial s_2} = 0.
\end{align}
From \eqref{eq:Dskappa12} we find that the local condition $\partial_{s_1}\kappa_2 = 0$ implies that $\kappa_{g_2} = 0$.  With both $\kappa_{g_1} = \kappa_{g_2} = 0$, \eqref{eq:KG_exp} reduces to $K_G = -\partial_{s_1}\kappa_{g_2}$. Differentiating \eqref{eq:Ds1logkappa2} with $s_1$ we find
\begin{align}
  -\partial_{s_1}\kappa_{g_2}
  = \frac{1}{\kappa_2}\frac{\partial^2\kappa_2}{\partial s_1^2}
    - \left(\frac{1}{\kappa_2}\frac{\partial\kappa_2}{\partial s_1}\right)^{\!2}.
    \label{eq:Ds1kappag2}
\end{align}
Substitution of \eqref{eq:Ds1kappag2} into \eqref{eq:KG_exp} and the restriction to the vicinity of the local maximum ($\partial_{s_1}\kappa_2 = 0$, $\;\partial_{s_1}^2\kappa_2 < 0$) yields a considerably simplified expression for $K_G$,  
\begin{align}
  K_G = \frac{1}{\kappa_2}\frac{\partial^2\kappa_2}{\partial s_1^2} \leq 0.
\end{align}
A parabolic profile of $\kappa_2$ in $s_1$ near the local maximum of $\kappa_2$ further implies that 
\begin{align}
  \frac{\partial^2\kappa_2}{\partial s_1^2} \sim -c_2^2\,\kappa_2,
\end{align}
with a bounded $c_2$, which physically correspond to the local value of $K_G$. Since $K_G = \kappa_1\kappa_2$, we get
\begin{align}
  \kappa_1 = \frac{1}{\kappa_2^2}\frac{\partial^2\kappa_2}{\partial s_1^2}= -\frac{c_2^2}{\kappa_2}.
  \label{eq:kappa1_from_kappa2}
\end{align}
Further, in the light of the ridge condition $\del_{s_2}\kappa_2$, the boundedness assumption of $\del_{s_2}\kappa_1$ is satisfied by the $\kappa_1$ solution in \eqref{eq:kappa1_from_kappa2}. 

Therefore, as $\kappa_2\to\infty$, $\kappa_1\to 0$.  Since $\kappa_1$ is the normal curvature of $\B$,
$\kappa_n \to 0$ near the ridge.  Since both the normal $\kappa_n$ and the geodesic curvature, $\kappa_{g_1}$, of the magnetic field vanishes near the ridge, $\B$ must be locally straight.

\section{ Magnetic field lines approach asymptotic lines on the outermost surface as $B$ approaches a constant \label{app:ridge_GradB_Hessian_psi}}

We provide the necessary mathematical details for the Section \ref{sec:GradB_ridges} in this Appendix. We use the Darboux frame $(\th=\B/B, \nh=\dl\psi/|\dl\psi|, \bh = \th \times \nh)$ consistently and start with the derivative of the $\BD\psi=0$ condition,
\begin{align}
    \dlBT\cdot \nh \frac{|\dl \psi|}{B}+\dl\dl\psi \cdot \th =0.
    \label{eq:gradBDpsi_Darboux_again}
\end{align}
Dotting \eqref{eq:gradBDpsi_Darboux_again} with $\th$ and using the expression for the field line curvature,
\begin{align}
    \bm{\kappa} = \frac{\th \cdot \dl \B}{B}-\frac{\th \cdot \dl B}{B}\th,
    \label{eq:kappa_expr}
\end{align}
we find that
\begin{align}
    \bm{\kappa} \cdot \dl \psi + \th \cdot \dl \dl \psi \cdot \th =0,
    \label{eq:norm_curv_Hessian_psi}
\end{align}
which can be recast in terms of normal curvature as in \eqref{eq:kappa_n_Hessian_psi}.

We now prove that field lines are asymptotic curves on a constant flux surface with $\th$ in the asymptotic direction when $\th\cdot \dl B\to 0$. A curve on a surface is an asymptotic curve if the second fundamental form, II, vanishes along the curve. On a flux surface $\psi$ the second fundamental form 
\begin{align}
    \text{II}=
    \begin{pmatrix}
        \cL \quad \cM\\
        \cM \quad \cN
    \end{pmatrix},
\end{align}
where
\begin{align}
    \cL = -\bh\cdot \frac{\dl \dl \psi}{|\dl\psi|} \cdot \bh, \quad \cM = -\bh\cdot\frac{\dl \dl \psi}{|\dl\psi|}  \cdot \th, \quad \cN = -\th \cdot \frac{\dl \dl \psi}{|\dl\psi|}  \cdot \th.
    \label{eq:II_form}
\end{align}
Thus, \eqref{eq:norm_curv_Hessian_psi} implies that
\begin{align}
    \cN = \kappa_n .
    \label{eq:cNn}
\end{align}
It is clear that when $\kappa_n \to 0$, II vanishes along $\th$. Thus, field lines are asymptotic curves. We can express $\dl\dl\psi$ and $\dlBT$ in terms of $3\times 3$ matrices in the $(\nh,\bh,\th)$ frame such that
\begin{align}
    \frac{1}{|\nabla \psi|}\dl\dl\psi = 
    \begin{pmatrix}
        a_{11} \quad a_{12} \quad a_{13}\\
        a_{12}\quad a_{22} \quad a_{23}\\
        a_{13} \quad a_{23}\quad a_{33}
    \end{pmatrix} \qquad 
    \frac{\dlBT}{B} = 
    \begin{pmatrix}
        c_{11} \quad c_{12} \quad c_{13}\\
        c_{12}\quad c_{22} \quad c_{23}\\
        c_{13} \quad c_{23}\quad c_{33}
    \end{pmatrix}.
    \label{eq:ab_mat}
\end{align}
Here, we have used the fact that both tensors are symmetric. The second fundamental form \eqref{eq:II_form} from \eqref{eq:ab_mat} is
\begin{align}
   \text{II}=-
   \begin{pmatrix}
        a_{22} \quad a_{23}\\
        a_{23}\quad a_{33}
    \end{pmatrix}, \quad \text{Det(II)}=a_{22}a_{33}-a_{23}^2.
    \label{eq:II_form2}
\end{align}

To connect the components of these tensors to physically meaningful quantities, we start from the expressions for $\kappa_n,\kappa_g, \BDB$ given in \eqref{eq:kappa_ng_exp}. It follows that
\begin{align}
    \kappa_n = \frac{\th \cdot \dlBT \cdot \nh }{B}= c_{13}, \quad \kappa_g = \frac{\th \cdot \dlBT \cdot \bh }{B}= c_{23}, \quad \frac{\th\cdot \dl B}{B}= \frac{\th \cdot \dlBT \cdot \th}{B}=c_{33}.
    \label{eq:bi3_expr}
\end{align}
Next, $\dl\cdot \B$ implies that $c_{11}+c_{22}+c_{33}=0$. Finally, employing \eqref{eq:norm_curv_Hessian_psi}, we find that
\begin{align}
    (c_{11} \quad c_{12} \quad c_{13})+(a_{13} \quad a_{23} \quad a_{33})=0.
\end{align}
As $\kappa_n\to 0$ and the field line tends to an asymptotic curve, $a_{33}\to 0$. Now,
\begin{align}
    K_G= \frac{\text{Det}(II)}{\text{Det}(I)},
\end{align}
where I,II are the first and second fundamental forms on the $\psi$ surface. From \eqref{eq:II_form2}, we get $K_G \propto (a_{22}a_{33}-a_{23}^2)$. Now, as $\kappa_n\to 0, a_{33}\to 0$, we have
\begin{align}
   \text{II}\to -
   \begin{pmatrix}
        a_{22} \quad a_{23}\\
        a_{23}\quad \;\; 0
    \end{pmatrix}, \quad \text{Det(II)}\to -\lbr a_{23}\rbr^2.
    \label{eq:II_form3}
\end{align}

We now focus on the case $\kappa_n\to 0$, which is appropriate near ridges as discussed in Section \ref{sec:BnB_ridges} and Appendix \ref{app:Diff_geo_proof}. The vanishing $\kappa_n$ near $B\approx B_M$ is also deduced from $B_M'\to 0$ in Fig \ref{fig:deviation_from_NAE}. In the $\kappa_n\to 0$ limit, we examine the scenario situation where $K_G$ also vanish. It is clear that $a_{33}, a_{23}$ both must go to zero in this limit, resulting in a highly degenerate fundamental form 
\begin{align}
   \text{II}\to -
   \begin{pmatrix}
        a_{22} \quad 0\\
        0\quad \;\; 0
    \end{pmatrix}, \quad \text{Det(II)}\to 0.
    \label{eq:II_form4}
\end{align}
The diagonal nature of II shows that the limiting $(\bh,\th)$ are also principal curvature directions on the flux surface. Moreover, in this limit $\dlBT$ takes the following form in the $(\nh,\bh,\th)$ basis
\begin{align}
    \frac{\dlBT}{B}= 
    \begin{pmatrix}
        c_{11} \quad \;  0 \qquad 0\\
       \;\; 0 \qquad c_{22}\quad \; c_{23}\\
        \;\;0 \qquad c_{23}\quad \; c_{33}
    \end{pmatrix}.
    \label{eq:dlB_form_darboux_ridge}
\end{align}
Thus, the Det$(\dlBT)/B=c_{11}(c_{22}c_{33}-c_{23}^2)$ is clearly nonzero in general. However, because in QS, $\kappa_g = (G_0/|\dl\psi|)(\BDB/B^2)$ as shown in \eqref{eq:kappa_g_exp}, $c_{23}=c_{33}G_0/|\dl\psi|$. Therefore, 
\begin{align}
    \frac{\text{Det}\dlBT}{B}= -c_{11}c_{33}\lbr c_{11}+c_{33} \lbr 1+\lbr \frac{G_0}{|\dl\psi|}\rbr^2 \rbr \rbr,
\end{align}
which vanishes when $\BDB \to 0$. This happens on the ridge where $\BDB$ and $\nh\cdot \dl B$ both vanish forcing $B$ to be almost a constant. Thus, we have come full circle to where we started: near almost constant $B$, which implies
\begin{align}
    \kappa_n=0 , \quad \th\cdot \dl\kappa_n=0.
    \label{eq:ridge_kappan}
\end{align}
Since $\th$ is the principal direction and $\kappa_n$ the curvature associated with it, \eqref{eq:ridge_kappan} defines a ridge point, in the true mathematical sense. Moreover, it is a cusp of Gauss since $\kappa_n=0$.

\section{$\dlBT$ tensor in axisymmetry, near closed field lines, and near ridges \label{app:GradB_tensor_forms}}

\subsection{Eigenvalues of the Grad-B tensor \label{app:GradB_tensor_eig} }
The $\dlBT$ tensor is a $3\times 3$ matrix with real coefficients. It is traceless owing to divergence-free nature of $\B$. Since current $\mu_0\J=\dl\times \B$ is related to the antisymmetric components of the $\dlBT$

Using Cayley-Hamilton theorem \citep{strang2006linearAlgebra}, the characteristic equation of $\dlBT$ is
\begin{align}
    \mbox{Det}(\lambda \mathbb{I} -\dlBT)= \lambda^3-\mbox{Tr}(\dlBT) \lambda^2 +\frac{1}{2}\lbr \lbr \mbox{Tr}(\dlBT) \rbr^2 - \mbox{Tr}((\dlBT)^2) \rbr \lambda- \mbox{Det}(\dlBT) =0.
\end{align}
Since $\mbox{Tr}(\dlBT)=\dl\cdot\B=0$, we have
\begin{align}
    \mbox{Det}(\lambda \mathbb{I} -\dlBT)=\lambda^3+\cP \lambda +\cQ=0, \quad \cP= -\frac{1}{2} \mbox{Tr}\lbr(\dlBT)^2 \rbr, \quad \cQ=-\mbox{Det}(\dlBT).
\end{align}
The eigenvalues $\lambda_i, i=1,2,3$ of the $\dlBT$ tensor satisfy $\prod_i(\lambda-\lambda_i)=0$ such that
\begin{align}
   \lambda_1+\lambda_2+\lambda_3=0, \quad 
    \cP= \lambda_1 \lambda_2 + \lambda_2 \lambda_3 +\lambda_3 \lambda_1, \quad \cQ=- \lambda_1 \lambda_2 \lambda_3.  \label{eq:Char_Det_dlB}
\end{align}
It follows from the traceless condition that $2\sum_{i\neq j}\lambda_i \lambda_j+\sum_i \lambda_i^2=0$ for $i,j=1,2,3$. Thus, $\cP$ can also be written as
\begin{align}
    \cP = -\frac{1}{2}(\lambda_1^2+\lambda_3^2+\lambda_3^2). 
\end{align}
Now, from the definition of the Frobenius norm, $| \mathsfbi{M}|_F^2 = \mbox{Tr} (\mathsfbi{M} \mathsfbi{M}^\top)$ and $L_{\nabla B}$ of an arbitrary symmetric tensor $\mathsfbi{M}$ we find that
\begin{align}
   | \dlBT|_F^2  = (\lambda_1^2+\lambda_3^2+\lambda_3^2)=-2\cP \quad \Rightarrow \quad L_{\nabla B}=\frac{\sqrt{2}B}{\sqrt{| \dlBT|_F^2 }}=\frac{1}{\sqrt{|\cP|/B^2}}.
\end{align}

\subsection{Grad-B tensor in axisymmetry \label{app:GradB_tensor_AS} }
We first analyze the axisymmetric case relevant to tokamak-like divertors. We employ the standard cylindrical coordinates $(R,\zeta, z)$ with orthonormal vectors $(\Rh=\dl R,\zth=R \dl\zeta,\zh=\dl z)$ \citep{haeseleer_flux_coordinates,freidberg2014idealMHD}, and consider a general axisymmetric toroidal magnetic field of the form
\begin{align}
    \B= I(\psi)\dl\zeta + \dl\zeta \times \dl \psi, \quad \psi=\psi(R,z).
    \label{eq:ASB}
\end{align}
Here, $I(\psi)$ is the current, $\psi(R,z)$ is the poloidal flux surface label that satisfies the Grad-Shafranov equation,
\begin{align}
    \psi_{,zz}+\psi_{,RR}-\frac{1}{R}\psi_{,R}+ R^2p'(\psi) + I I'(\psi)=0.
    \label{eq:GS_eqn}
\end{align}
The gradient of $\B$ from \eqref{eq:ASB}, taking into account
\begin{align}
    \dl\Rh =\frac{\zth\zth}{R}, \quad \dl\zth=-\frac{\zth\Rh}{R},
    \label{eq:AS_derivs}
\end{align}
is given by
\begin{align}
\dlBT =& \frac{1}{R}\lbr \Rh \dl \psi_{,z}-\zh \dl \psi_{,R} \rbr -\frac{I}{R^2}\lbr \zth \Rh +\Rh \zth \rbr
\nonumber\\
&+
\lbr \frac{I'(\psi)}{R}\dl\psi \zth - \frac{\Rh}{R}\frac{\zth\times \dl\psi}{R} +\frac{\psi_{,Z}}{R}\frac{\zth\zth}{R}\rbr.
    \label{eq:AS_gradB_general}
\end{align}
Near O and X points, where $|\dl\psi|^2=(\psi_{,R})^2+(\psi_{,z})^2\to 0$, the last term in \eqref{eq:AS_gradB_general} drops out, and $\dlBT$ takes the simple form
\begin{align}
\dlBT=\frac{1}{R}\lbr \psi_{,Rz}\lbr \Rh\Rh - \zh\zh\rbr +\lbr \zh \Rh\; \psi_{,zz} - \Rh \zh\; \psi_{,RR}\rbr  \rbr -\frac{I}{R^2}\lbr \zth \Rh +\Rh \zth \rbr.
    \label{eq:AS_gradB}
\end{align}
In the alternative matrix form in the $(\zth,\zh,\Rh)$ basis
\begin{align}
    \dlBT= \frac{1}{R}\begin{pmatrix}
       \;\; 0 \qquad \quad \; 0 \qquad -I/R\\
        \quad 0 \qquad -\psi_{,Rz} \qquad \;\psi_{,zz}\\
        -I/R \quad -\psi_{,RR} \qquad \psi_{,Rz}
    \end{pmatrix},
    \label{eq:dlBT_AS_form}
\end{align}
with trace and determinant given by
\begin{align}
   \mbox{Tr}{\lbr \dlBT\rbr}=0, \quad  \mbox{Det}{\lbr \dlBT \rbr}=\lbr \frac{I}{R^2}\rbr^2 \psi_{,Rz}.
   \label{eq:TrDet_AS}
\end{align}
Normalizing the axisymmetric quantities with major radius $R_a$ and field strength $B_a$, and denoting normalized quantities by a bar,
\begin{align}
    \bar{R}=\frac{R}{R_a},\quad \zb=\frac{z}{R_a},\quad \Ib = \frac{I}{B_a R_a},\quad \psibar=\frac{\psi}{B_a R_a^2},
\end{align}
we find that
\begin{align}
    \frac{\cP}{(B_a/R_a)^2}=\frac{1}{\Rb^2}\lbr \overline{\scH}_2[\psibar]-\lbr\frac{\Ib}{\Rb}\rbr^2\rbr , \quad \frac{\cQ}{(B_a/R_a)^3}= 
     -\lbr \frac{\Ib}{\Rb^2}\rbr^2 \psibar_{,\Rb\zb},
    \label{eq:cPcQ_AS}
\end{align}
The normalized eigenvalues $(R_a/B_a)\lambda_i$ will be $O(1)$ if $\overline{\cP},\overline{\cQ}$ are both $O(1)$. 

\subsection{Grad-B tensor near closed field lines \label{app:GradB_tensor_closedB}}
Next, we consider the case of closed field lines, which has been
has been discussed extensively by \citet{Solovev1970,Mercier1964,burby2021_NAE_normal_forms} and others. Avoiding re-derivation, we simply quote relevant formulas from Section 3 and Appendix C of \citet{guinchard2025application}. The magnetic axis is assumed to be a smooth curve $\bm{r}_0(\ell)$ parametrized by the arclength $\ell$. A Frenet-Serret frame $(\th(\ell)=\bm{r}_0'(\ell),\nh(\ell),\bh(\ell))$ is fitted to the axis such that 
\begin{align}
  \frac{d}{d\ell}  
  \begin{pmatrix}
        \th\\ \nh\\\bh
    \end{pmatrix}
    = \begin{pmatrix}
        \; 0 \quad +\kappa \quad \;\; 0\\ 
        -\kappa\qquad 0\;\; +\tau\\ 
        \; 0\quad -\tau\quad \;\;0
    \end{pmatrix}
    \begin{pmatrix}
        \th\\ \nh\\\bh
    \end{pmatrix}.
    \label{eq:Frenet_frame}
\end{align}
Following \citep{Solovev1970}, a set of ``rotating frame" $(\th,\bm{\cN}, \bm{\cB})$ with a rotation angle $\delta(\ell)$ can then be constructed such that
\begin{align}
\begin{pmatrix}
    \bm{\cN} \\ \bm{\cB} 
\end{pmatrix}
  = \begin{pmatrix}
    \cos{\delta(\ell)} \quad  -\sin{\delta(\ell)}\\
    \sin{\delta(\ell)} \quad +\cos{\delta(\ell)}
\end{pmatrix}
  \begin{pmatrix}
    \nh \\ \bh
\end{pmatrix}.
\end{align}
The position vector of any point can now be expressed as
\begin{align}
    \bm{r}=\bm{r}_0(\ell)+x\bm{\cN}+y\bm{\cB}.
    \label{eq:pos_vec_rot}
\end{align}
We obtain the following tangent and the reciprocal basis vectors \citep{haeseleer_flux_coordinates} from \eqref{eq:pos_vec_rot}
\begin{align}
     &\bm{e}_1\equiv \del_x \bm{r}=\bm{\cN}, \quad \bm{e}_2\equiv \del_y \bm{r}=\bm{\cB}, \quad \bm{e}_3\equiv \del_\ell \bm{r}=h \th-u'(\ell)(x\bm{\cB}-y\bm{\cN}), \label{eq:basis_vec_rot}\\
    &\bm{e}^1\equiv \dl x = \bm{\cN}-\frac{u' y}{h}\th, \quad \bm{e}^2\equiv \dl y = \bm{\cB}+\frac{u' x}{h}\th, \quad \bm{e}^3\equiv \dl \ell = \frac{\bm{t}}{h},
     \nonumber\\
     &\sqrt{g}\equiv \bm{e}_1 \times \bm{e}_2 \cdot \bm{e}_3=h, \quad h=1-\kappa (x \cos\delta -y \sin{\delta}), \quad u'(\ell)=\delta'(\ell)-\tau(\ell).
\end{align}
To obtain the equation of the field lines and $\BD$ we need $\B$ in the form
\begin{align}
    \sqrt{g}\B = \sqrt{g}B^1 \bm{e}_1 +\sqrt{g}B^2 \bm{e}_2 +\sqrt{g}B^3 \bm{e}_3, \quad \BD=B^3\del_\ell+B^1\del_x + B^2\del_y.
    \label{eq:sqrtgB^i_BD_closed}
\end{align}
Analyticity near the closed field line then implies that we can carry out regular Taylor expansions in $(x,y)$
\begin{align}
    \sqrt{g}B^1=a_1 x+ a_2 y , \quad \sqrt{g}B^2=b_1 x+ b_2 y, \quad \sqrt{g}B^3=B_0+c_1 x+ c_2 y.  
    \label{eq:sqrtgB_Taylor_closed}
\end{align}
Ensuring $\dl\cdot\B=0$, $\dl\times \B =J_0\th $ using \eqref{eq:sqrtgB_Taylor_closed}, we get to linear order, 
\begin{align}
    &\B=B_t \th + B_\cN \bm{\cN} + B_\cB \bm{\cB}, \label{eq:B_form_closed}\\
    &B_t=B_0(\ell) +c_1 x +c_2 y, \;\; B_\cN=a_1 x +(a_2+u' B_0)y,\;\; B_\cB=b_2 y +(b_1-u' B_0)x ,\nonumber\\
   &\frac{a_1}{B_0}=-\frac{1}{2}\lbr \frac{B_0'}{B_0}+\eta' \rbr, \;\; \frac{b_2}{B_0}=-\frac{1}{2}\lbr \frac{B_0'}{B_0}-\eta' \rbr,\;\; \frac{a_2}{B_0}=-\Omega_0 e^{-\eta},\;\; \frac{b_1}{B_0}=\Omega_0 e^{\eta}, \nonumber\\
   &\frac{c_1}{B_0}= \kappa \cos\delta ,\quad \frac{c_2}{B_0}= \kappa \sin\delta, \quad \Omega_0=\frac{J_0/(2B_0)+u'}{\cosh{\eta}},\quad \Omega_V=\Omega_0 \sinh\eta. \nonumber
\end{align}
The flux surface label $\psi= B_0(\exp{\eta(\ell)}x^2 \pm \exp{-\eta(\ell)}y^2 )/2$ is a rotating ellipse or a hyperbola depending on whether we are at an O point or an X point. Whereas $\delta'(\ell)$ is the rotation rate of the conic, the elongation is controlled by $\eta'$.

Thus, equation of the field lines and $\BD$ to the lowest order in the distance to the closed field line are 
\begin{align}
    \frac{d\ell}{B_0}=\frac{dx}{a_1 x+a_2 y}= \frac{dy}{b_1 x+b_2 y}, \label{eq:B_eqn_closed}\\
    \BD =B_0 \del_\ell +(a_1 x+a_2 y)\del_x +(b_1 x+b_2 y)\del_y, \nonumber
\end{align}
where $a_1,b_1,a_2,b_2$ are given by \eqref{eq:B_form_closed}.

Finally, the lowest order $\dlBT$ in the $(\th,\bm{\cN},\bm{\cB})$ basis takes the form
\begin{align}
    \dlBT=B_0 
    \begin{pmatrix}
        B_0'/B_0 \qquad \quad \;\; c_1/B_0 \qquad \qquad c_2/B_0 \quad \\
        \qquad c_1/B_0 \qquad \qquad a_1/B_0 \qquad \lbr\Omega_V +\frac{J_0}{2B_0}\rbr \quad \\
    c_2/B_0 \qquad \lbr\Omega_V -\frac{J_0}{2 B_0}\rbr \qquad b_2/B_0 \quad
    \end{pmatrix}.
    \label{eq:dlBT_closed}
\end{align}
In the vacuum limit ($J_0=0$), 
\begin{align}
\Phi_{,x}&=a_1 x +\Omega_V y, \quad \Phi_{,y}=b_2 y +\Omega_V x,
    \label{eq:vacuum_closed}\\
 \Phi_{,xx}&=a_1, \quad \Phi_{,xy}=\Omega_V, \quad \Phi_{,yy}=b_2, \nonumber
\end{align}
such that the Laplace equation implies $\Phi_{,xx}+\Phi_{,yy}+B_0'=0$. The other quantities of interest are given by
\begin{align}
    \scH_2[\Phi]&=\frac{1}{4}\lbr \lbr\frac{B_0'}{B_0}\rbr^2 -{\eta'}^2\rbr -\Omega_V^2, \quad \Omega_V = \frac{\delta'-\tau}{\cosh{\eta}}.\label{eq:Hess2_cPcQ_closed}\\
    -\frac{\cP}{B_0^2} &=\lbr \kappa^2 +\frac{\eta'}{4}^2 +\Omega_V^2\rbr +\lbr \frac{3B_0'}{4 B_0}\rbr^2, \nonumber\\
    -\frac{\cQ}{B_0^3} &= \frac{B_0'}{B_0}\lbr \frac{\kappa^2}{2} +\frac{1}{4}\lbr \lbr\frac{B_0'}{B_0}\rbr^2 -{\eta'}^2\rbr -\Omega_V^2\rbr
    +\kappa^2\lbr \Omega_0 \sinh{\eta} \sin{2\delta}-\frac{\eta'}{2}\cos{2\delta} \rbr
    \nonumber.
\end{align}
The eigenvalues $\lambda_i$ normalized to the major radius $R_a$ are $O(1)$, since $\kappa,\tau$ and all $\ell$ derivatives scale as $1/R_a$. Thus,
\begin{align}
    R_a^2 \scH_2[\Phi]\sim  \frac{\cP}{(B_0/R_a)^2} \sim  \frac{\cQ}{(B_0/R_a)^3}\sim O(1)\quad \Rightarrow (R_a/B_0)\lambda_{i} \sim O(1).
    \label{eq:R^2scaling_closed}
\end{align}.

\subsection{Grad-B tensor near ridges \label{app:GradB_near_ridges}}

We now focus on generic 3D sharp ridges. We use a local signed Frenet-Serret frame $(\bm{t},\bm{n},\bm{b})$ \eqref{eq:Frenet_frame} near the ridges together with local coordinates $(\ell,x,y)$. We orient the signed Frenet frame such that the tangent vector $\bm{t}$ is aligned with the ridge, which is assumed to be a smooth space curve. The arclength along the ridge is measured by $\ell$. Since we expect the field line curvature and torsion to vanish exactly on the ridge, we choose the normal $\bm{n}$ and therefore the binormal $\bm{b}=\bm{t}\times \bm{n}$ directions such that continuity is maintained in the vicinity of the ridge. The distances along normal and binormal is measured by local variables $x$ and $y$. Thus, we can define the position vector as
\begin{align}
    \bm{r}=\bm{r}_0(\ell)+x \nh(\ell)+y \bh(\ell).
    \label{eq:pos_vec_ridge}
\end{align}
Upon differentiation \eqref{eq:pos_vec_ridge} yields the following tangent basic vectors \citep{haeseleer_flux_coordinates}
\begin{align}
    \bm{e}_1\equiv \del_x \bm{r}=\nh, \quad \bm{e}_2\equiv \del_y \bm{r}=\bh, \quad \bm{e}_3\equiv \del_\ell \bm{r}=h \th+\tau(x\bm{b}-y\bm{n}),
    \label{eq:tangent_basis_vecs}
\end{align}
where $h$ is the Jacobian 
\begin{align}
    h\equiv \bm{e}_1 \times \bm{e}_2 \cdot \bm{e}_3= 1-\kappa x. \label{eq:hisJ}
\end{align}
The reciprocal basis vectors, given by
\begin{align}
    \bm{e}^1\equiv \dl x = \nh+\frac{\tau y}{h}\th, \quad \bm{e}^2\equiv \dl y = \bh-\frac{\tau x}{h}\th, \quad \bm{e}^3\equiv \dl \ell = \frac{\bm{t}}{h}.
    \label{eq:reciprocal_basis_vecs}
\end{align}
The set-up so far is identical to the closed field line case \ref{app:GradB_tensor_closedB} with $\delta=0$. However, the key difference is that a regular Taylor series is not warranted near the ridges. However, we can still define a $B_0(\ell)$ analogous to the closed field line case by defining $B_0\equiv B_t (x=0,y=0,\ell)$. In the following, we restrict ourselves to vacuum fields $\B=\dl\Phi$ such that
\begin{align}
    \B&=B_t \th+B_n \nh +B_b \bh,\\
    B_t&=\frac{1}{h}\lbr \Phi_{,\ell} +\tau\lbr y \Phi_{,x}-x \Phi_{,y} \rbr \rbr, \quad B_n = \Phi_{,x}, \quad B_b =\Phi_{,y}. \nonumber\\
    \sqrt{g}B^1&=h\B\cdot \bm{e}^1=h B_n + \tau y B_t,\;\;\sqrt{g}B^2=h B_b-\tau x B_t, \;\; \sqrt{g}B^3= B_t\nonumber.
\end{align}
Thus, equation of the field lines and $\BD$ to the lowest order in the distance to the closed field line are 
\begin{align}
    \frac{d\ell}{B_0}=\frac{dx}{\Phi_{,x}+B_0\tau y}= \frac{dy}{\Phi_{,y}-B_0\tau x}, \label{eq:B_eqn_ridges}\\
    \BD =B_0 \del_\ell +(\Phi_{,x}+B_0\tau y)\del_x +(\Phi_{,y}-B_0\tau x)\del_y, \nonumber
\end{align}

Implementing the same procedure as in the closed field line case, we find the following form for the lowest order $\dlBT$ in the $(\th,\nh,\bh)$ basis
\begin{align}
    \dlBT=
    \begin{pmatrix}
        B_0' \qquad \kappa B_0 \qquad  \quad 0 \;\; \\
        \;\; \kappa B_0\qquad \Phi_{,xx} \qquad \Phi_{,xy} \;\; \\
    \quad 0 \qquad \;\;\; \Phi_{,xy} \qquad \;\Phi_{,yy} \;\;
    \end{pmatrix}.
    \label{eq:dlBT_ridges}
\end{align} 
The quantities $\cP,\cQ$ from \eqref{eq:dlBT_ridges} are
\begin{align}
    -\frac{\cP}{B_0^2}=\kappa^2+\lbr \frac{B_0'}{B_0}\rbr^2 -\scH_2[\varphi], \quad -\frac{\cQ}{B_0^3}=\frac{B_0'}{B_0}\scH_2[\varphi]-\kappa^2 \varphi_{,yy}, \quad \varphi=\frac{\Phi}{B_0}.
\end{align}

Near the ridges, $B_0'\to 0, \kappa\to 0$. Thus, $\cQ\to 0, \cP/B_0^2\to \scH_2[\varphi]$. Now, if $R_a^2\scH_2[\varphi]\sim O(1)$, $R_a \lambda_i\sim O(1)$. However, we showed $\scH_2= K_G$, which can be very large. Thus, normalized eigenvalues are not necessarily $O(1)$.

\section{Expansion of the Biot-Savart expression near a filamentary coil \label{app:Det0_coils_NCE}}
It is convenient to use the orthogonal Mercier coordinate system $(\rho,\omega,\ell)$ \citep{Mercier1964,Solovev1970,sengupta2021NSE,jorge_sengupta2020near}, which is exactly what \citet{callegariTing1978motion} used to carry out the local induction approximation (LIA) without explicitly calling it the Mercier coordinates. 

Starting with the current filament defined by the space curve $\bm{r}_0(\ell)$ parameterized by the arclength $\ell$, we construct an orthonormal Frenet-Serret frame $(\th(\ell)=\bm{r}_0'(\ell),\nh(\ell),\bh(\ell))$ that satisfies the frame equations \eqref{eq:Frenet_frame} with curvature and torsion given by $\kappa_c,\tau_c$. Here, the subscript $c$ will be used to denote coil filaments. 

Next, we define a position vector $\bm{r}=\bm{r}_0(\ell)+\rho \rhoh$, such that
\begin{align}
\rhoh &= \nh \cos\theta + \bh \sin \theta, \quad \wh= -\nh \sin\theta + \bh \cos \theta, \quad \cos{\theta}=\bm{\nh}\cdot \rhoh, \quad \omega = \theta +\int \tau d\ell, \nonumber\\
    \del_\rho \bm{r} &=\rhoh, \quad \del_\omega \bm{r}=\rho\; \wh, \quad \del_\ell \bm{r} = h \bm{\th}, \quad h\equiv 1-\kappa \rho \cos\theta, \label{eq:Fil_basis_vecs}\\
    \del_\ell \rhoh &= -\kappa \cos\theta \th, \quad \del_\ell \wh = \kappa \sin\theta \th. \nonumber
\end{align}
In addition we can show that
\begin{align}
    \dl \rhoh = \frac{1}{\rho}\wh\wh -\frac{\kappa\cos\theta}{h}\th\th, \;\;  \dl \wh = -\frac{1}{\rho}\wh\rhoh +\frac{\kappa\sin\theta}{h}\th\th, \;\; \dl\th= \frac{\kappa}{h}\th\nh.
\end{align}
From $\B= B_\rho \rhoh+ B_\omega\wh +O(\rho)$, with both $B_\rho, B_\omega$ having $1/\rho, \ln{\rho}$ terms, we find that $\del_\ell B_\rho, \del_\ell B_\omega$ are $O(\rho)$ smaller compared to terms such as $\del_\rho B_\rho, (1/\rho)\del_\omega B_\rho$. Thus, we naturally end up with 
\begin{align}
    \dlBT= 
    \begin{centering}
    \begin{pmatrix}
        -\kappa_c B_n \qquad \qquad 0 \qquad \qquad  \qquad 0\qquad\\
       0\qquad \qquad   \del_\rho B_\rho \qquad  \qquad \del_\rho B_\omega\\
     \qquad \quad 0\qquad   \frac{1}{\rho}(-B_\omega + \del_\omega B_\rho)\quad  \frac{1}{\rho}(B_\rho + \del_\omega B_\omega)
        \label{eq:DLBT_coils}
        \end{pmatrix},
    \end{centering}
\end{align}
where
\begin{align}
    B_n \equiv\B\cdot \nh =B_\rho \cos\theta -B_\omega \sin\theta.
    \label{eq:Bn_exp2}
\end{align}

We now define a reduced determinant surface $\scD=0 $ corresponding to $B_n=0$. This surface can be given a natural interpretation of an effective flux surface such that the normal component of $\B$ given by $B_n$ vanishes on the surface. Thus, it is natural to parametrize the determinant $\scD=0 $ by
\begin{align}
    \bm{r}=\bm{r}_0(\ell) + \rho_\scD \nh \label{eq:parametrize_Det_surf}.
\end{align}
We have the following tangent vectors on the determinant surface
\begin{align}
    \del_{\rho_\scD}\bm{r}=\nh, \quad \del_\ell \bm{r}= h_c\th +\rho_\scD\tau_c \bh, \quad h_c=1-\kappa_c \rho_\scD \cos\theta.
    \label{eq:tan_vecs_Det_surf}
\end{align}
In the limit of $\rho_\scD\to 0$, the tangent vectors are $\nh$ and $\th$
The unit normal vector to the determinant surface is 
\begin{align}
    \hat{\bm{N}}=\frac{\del_{\ell}\bm{r} \times \del_{\rho_\scD}\bm{r}}{|\del_{\ell}\bm{r} \times \del_{\rho_\scD}\bm{r}|} = \bh
\end{align}
Thus, the binormal vector $\bh$ is the normal to the surface $\hat{\bm{N}}$, which is a textbook \citep{struik2012lectures} example of asymptotic curves on a ruled surface. For asymptotic curves Gaussian curvature is non-positive and can be related to the torsion through $K_{G\scD}=-\tau_c^2$. 

Finally, for the sake of rigor, we now define a normalized determinant $\bar{\scD}_3$ that stays finite as we approach the Det$\dlBT$ surface. The essential idea is to use the Cayley - Hamilton invariants for a traceless symmetric matrix such as $\dlBT$. As discussed in Appendix \ref{app:GradB_tensor_forms}, the characteristic equation Det$(\lambda \mathbb{I}-\dlBT)$ leads to the invariants $\cP=-(1/2)\text{Tr}((\dlBT)^2)$ and $\cQ=-\text{Det}\dlBT$. In terms of the three real eigenvalues $(\lambda_0,\lambda_\pm)$, $\cP =(-1/2)(\lambda_0^2 +\lambda_+^2 +\lambda_-^2), \cQ= -\lambda_0 \lambda_+\lambda_-$. As we approach the current filament, $\lambda_0\to 0$ while $\lambda_\pm$ blow up. In this limit $\lambda_-\to -\lambda_+$, such that $-\cP\to \lambda_+^2, \cQ\to \lambda_0 \lambda_+^2$. To isolate the null eigenvalue surface defined by $\lambda_0$, we now define $\normdet{}$ such that
\begin{align}
    \normdet{} \equiv \frac{\cQ}{(-\cP)^{3/2}}.
    \label{eq:normalized_Det}
\end{align}
The $3/2$ power ensures a smooth continuity $\normdet{} \to 0$ as we take the limit $\rho\to 0$. Other choices are also possible, but we do not dwell on that here. The normalized determinant is useful for numerical purposes as it is designed to be smooth near coils. However, it does not change the conclusions obtained from $\scD_3\to 0$. The reason is that $\scD_3\to 0$ not because of any power of $\rho$ but because of the $\sin\theta$ factor that goes to zero near the coils.

\bibliographystyle{jpp}
\bibliography{plasmalit}
\end{document}